\documentclass[a4paper,11pt]{article}
\pdfoutput=1

\usepackage[height=23.6truecm,width=16truecm,top=3.05truecm]{geometry}

\usepackage[skip=6pt,indent]{parskip}

\usepackage[utf8]{inputenc}
\usepackage{graphicx}
\usepackage{amsmath}
\usepackage{amssymb}
\usepackage{hyperref}
\hypersetup{colorlinks,citecolor=TokiwaIro,linkcolor=RuriIro,urlcolor=RuriIro,linktocpage}
\usepackage{cite}
\usepackage{braket}
\usepackage{bm}
\usepackage{bbm}
\usepackage{color}
\usepackage[svgnames,table]{xcolor}
\usepackage{float}
\usepackage{hhline}
\usepackage{multirow}
\usepackage{indentfirst}
\usepackage{latexsym}
\usepackage{mathtools}
\usepackage{cancel}
\usepackage{slashed}
\usepackage{colortbl}
\usepackage[mathscr]{euscript}
\usepackage{enumitem}
\usepackage{subcaption}
\usepackage{comment}
\usepackage{amsthm}

\numberwithin{equation}{section}

\allowdisplaybreaks

\usepackage[whole]{bxcjkjatype} %

\definecolor{RuriIro}{HTML}{1E50A2}
\definecolor{TokiwaIro}{HTML}{007B43}

\definecolor{dred}{rgb}{0.7,0.15,0.09}
\definecolor{dblue}{rgb}{0,0.12,0.64}
\definecolor{dgreen}{rgb}{0.2,0.51,0.19}
\definecolor{pegn}{rgb}{0.33,0.51,0.14}

\definecolor{kblue}{rgb}{0,0.48,0.73}
\definecolor{kred}{rgb}{0.73,0.25,0}
\definecolor{kgreen}{rgb}{0.48,0.73,0}

\definecolor{rgreen}{HTML}{7BAA17}
\definecolor{rred}{HTML}{AB1732}
\definecolor{rblue}{HTML}{007BBB}

\definecolor{rnpurple}{HTML}{8B2BB0}

\theoremstyle{plain}

\theoremstyle{remark}

\newcommand{\nn}{\nonumber}

\newcommand{\mc}{\mathcal}
\newcommand{\mr}{\mathrm}

\newcommand{\mbb}{\mathbb}

\newcommand{\msc}{\mathscr}

\newcommand{\mt}{\mathtt}
\newcommand{\del}{\partial}

\newcommand{\dd}{\mathrm{d}}
\newcommand{\ee}{\mathrm{e}}
\newcommand{\iu}{\mathrm{i}}
\renewcommand{\I}{\mathrm{i}}

\renewcommand{\tt}{\mathrm{t}}

\newcommand{\Det}{\mathop{\mathrm{Det}}\nolimits}

\renewcommand{\Im}{\mathop{\mathrm{Im}}}

\newcommand{\diag}{\mathop{\mathrm{diag}}}

\newcommand{\ketbra}[2]{\ket{#1}\hspace{-4pt}\bra{#2}}

\newcommand{\HAMT}{\mathtt{HAM\text{-}T}}

\newcommand{\grad}{\mathop{\mathrm{grad}}\nolimits}
\newcommand{\eff}{\mathrm{eff}}
\newcommand{\Hess}{\mathop{\mathrm{Hess}}\nolimits}

\def\l{\left}
\def\r{\right} 

\begin{document}

\begin{titlepage}

\begin{flushright}
\end{flushright}

\vspace{1cm}

\begin{center}

{\LARGE \bfseries
Quantum Riemannian Hamiltonian Descent
}

\vspace{1cm}

\renewcommand{\thefootnote}{\fnsymbol{footnote}}
{%
\hypersetup{linkcolor=black}
Yoshihiko Abe$^{1,2,3}$\footnote[1]{yabe3@keio.jp}
\
and
Ryo Nagai$^{3,4}$\footnote[2]{ryo.nagai.jd@hitachi.com},
}%
\vspace{8mm}

{\itshape%
$^1${Graduate School of Science and Technology, Keio University, Yokohama, Kanagawa 223-8522, Japan}\\
$^2${Keio University Sustainable Quantum Artificial Intelligence Center (KSQAIC), Keio University, Tokyo 108-8345, Japan}\\
$^3${Quantum Computing Center, Keio University, 3-14-1 Hiyoshi, Kohoku-ku, Yokohama, Kanagawa, 223-8522, Japan}\\
$^4${Research and Development Group, 
Hitachi, Ltd., Kokubunji, Tokyo 185-8601, Japan
}
}%

\vspace{8mm}

\end{center}

\abstract{%
We propose Quantum Riemannian Hamiltonian Descent (QRHD), a quantum algorithm for continuous optimization on Riemannian manifolds that extends Quantum Hamiltonian Descent (QHD) by incorporating geometric structure of the parameter space via a position-dependent metric in the kinetic term. 
We formulate QRHD in both operator and path integral formalisms and derive the corresponding quantum equations of motion, showing that quantum corrections appear in the action integral but they are suppressed at late times by the time-dependent dissipation factor.
This implies that convergence near optimal points is controlled by the classical potential while quantum effects influence early-time dynamics.
By analyzing the semiclassical equation, we estimate a lower bound on the convergence time and numerically demonstrate whether QRHD works as a quantum optimization algorithm in some examples.
A quantum circuit implementation based on time-dependent Hamiltonian simulation is also discussed and the query complexity is estimated.
}%

\end{titlepage}

\renewcommand{\thefootnote}{\arabic{footnote}}
\setcounter{footnote}{0}
\setcounter{page}{1}

\tableofcontents

\section{Introduction}
\label{sec:introduction}

\begin{figure}[t]
    \centering
    \includegraphics[width=15cm]{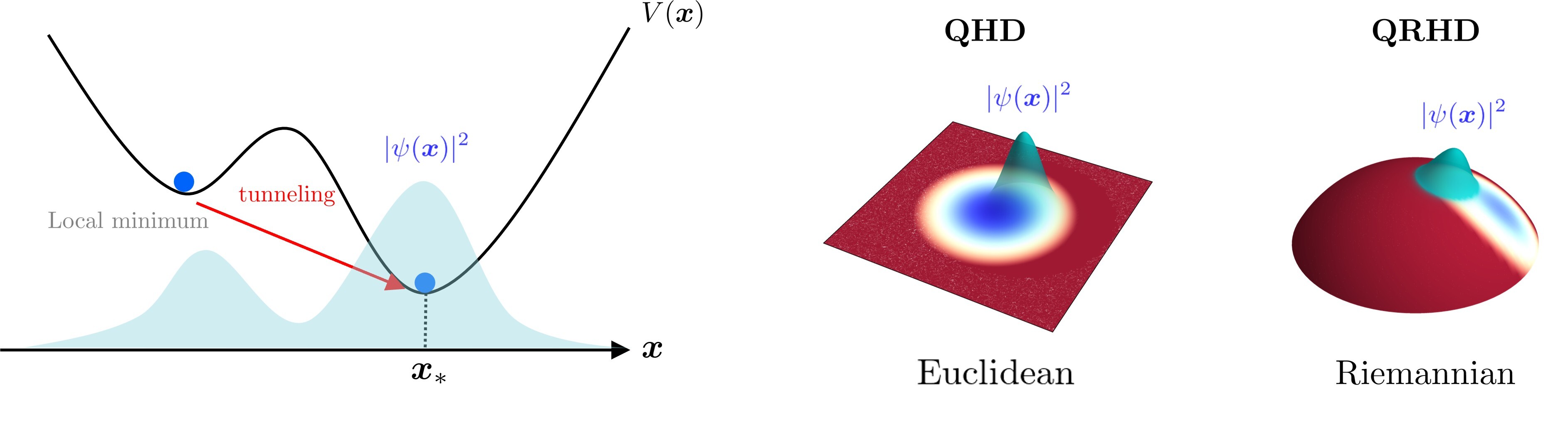}
    \caption{(Left) Schematic illustration of quantum optimization algorithms using quantum dynamics.
    The blue point and black curve represent the parameters to be optimized and the landscape of the loss function, respectively.
    The light blue region denotes the probability density of the parameters, which evolves in time according to the Schr\"odinger equation.
    Quantum tunneling enables the parameters to escape from local minima.
    (Right) Comparison between the conventional idea, QHD, and our proposal, QRHD.
    While QHD performs parameter search in a Cartesian coordinate system on a flat space, QRHD enables parameter search on a specified Riemannian manifold.}
    \label{fig:summary}
\end{figure}

Continuous optimization, stemming from the mathematical modeling of our world systems and theoretical physics approaches, universally exists not only in fundamental research but also in engineering including AI.
The classical gradient method, which is a standard algorithm for continuous optimization, often stagnates at local minima when the loss function is nonconvex.
To address this bottleneck, the authors of Ref.~\cite{2023-Leng-Hickan-Li-Wu} recently proposed a quantum algorithm called Quantum Hamiltonian Descent (QHD).
In QHD, the parameters to be optimized are reinterpreted as the position of the particle moving in a potential corresponding to the loss function as in classical optimization algorithms~\cite{2013-Itami,2015-Su-Boyd-Candes,2015-Krichene-Bayen-Bartlett,2015-Wibisono-Wilson,2016-Wibisono-Wilson-Jordan,2016-Wilson-Recht-Jordan,2021-Sanz-et-al,2021-Wilson-Recht-Jordan}, and their updates are described by the time evolution of a quantum mechanical system.
Due to the quantum effects such as tunneling effects and the energy dissipation via friction, the particle wave function can escape from local minima with the aid of quantum tunneling and be eventually localized at an optimal point that minimizes the loss function.
The schematic picture is shown in the left panel of Fig.~\ref{fig:summary}.
In this sense, QHD uses quantum dynamics to mitigate the stagnation problem in classical optimization.

QHD has recently attracted attentions as a new continuous-optimization algorithm with the capability of escaping from local solutions~\cite{2023-Leng-Hickan-Li-Wu,2023-Leng-Zhang-Wu,2023-Chen-et-al,2024-Leng-Li-Peng-Wu,2025-Kushnir-Leng-Peng-Fan-Wu,2025-Catli-Simon-Wiebe,2025-Chakrabarti-et-al,2025-Leng-et-al,Leng:2025msd,2025-Peng-et-al}.
From the viewpoint of quantum algorithms, QHD is a coherent quantum algorithm in the sense that it does not require intermediate measurements of the quantum state or updates conditioned on measurement outcomes.
Such quantum dynamics needs not be implemented solely by quantum circuits~\cite{2023-Leng-Hickan-Li-Wu,2023-Chen-et-al,2024-Leng-Li-Peng-Wu}.
In this regard, it stands apart from conventional quantum--classical hybrid algorithms~\cite{2019-Gilyen-et-al-grad}.

While QHD has an attractive aspect as a new continuous optimization algorithm that can potentially resolve the local-minimum stagnation problem, its formulation is essentially based on a Cartesian coordinate system where the Hamiltonian has a canonical kinetic term, and hence it lacks the freedom to incorporate prior knowledge about the parameter space.
This means that it is hard to reflect in the algorithm the geometric structure of the parameter space that naturally arises depending on the problem, or information about coordinate systems suitable for efficient optimization.

Motivated by this, we propose an extension of QHD, called Quantum Riemannian Hamiltonian Descent (QRHD), that allows one to incorporate such geometric information as prior knowledge.
The schematic picture is shown in the right panel of Fig.~\ref{fig:summary}.
In QRHD, we extend the kinetic term of QHD to a noncanonical form and introduce a new degree of freedom in the form of the metric tensor in the parameter space.
By choosing it appropriately, one can select general coordinate systems beyond Cartesian coordinates.
Although this extension is implemented as a modification of the kinetic term, quantum-mechanical properties such as tunneling are preserved, so the ability of QHD to escape from local minima will remain.
In addition, in QRHD, constraints imposed on the parameter space can be incorporated naturally through the metric, so constrained continuous optimization can be treated within the same framework.
By embedding the information of constraints directly into the metric, QRHD avoids the use of stiff penalty terms in the potential function, which would otherwise cause the spectral norm to blow up in the standard QHD framework.
QRHD provides a framework that extends QHD, while retaining its quantum-mechanical features, to a broader class of continuous optimization problems---namely, continuous optimization on Riemannian manifolds~\cite{1960-Rosen,1961-Rosen,1972-Luenberger,1982-Gabay,1994-Smith,1996-Amari,1998-Edelman-Arias-Smith,2016-Zhang-Suvrit,2018-Zhang-Suvrit,2019-Alimisis-et-al}.

The remainder of this paper is laid out as follows.
In Sec.~\ref{sec:QHD}, we give a brief review of QHD, which is a basic algorithm of our proposal.
QRHD, an extension of QHD to continuous optimization on Riemannian manifolds, is introduced in Sec.~\ref{sec:QRHD} and we discuss some quantum geometric corrections arising from the operator ordering.
In Sec.~\ref{sec:convergence}, we introduce the path integral formalisms for QHD and QRHD, and discuss the typical convergence time to the local minima by analyzing the quantum equations of motion.
We also show some numerical analyses of simple examples to demonstrate whether QRHD works as a quantum optimization algorithm.
We discuss the implementation of QRHD by a quantum circuit and estimate the complexity using the estimation of the convergence time in Sec.~\ref{sec:implementation}.
Sec.~\ref{sec:conclusions} is devoted to our conclusions.

\paragraph{Notation.}

In this paper, we work in natural units with $\hbar =1$ unless otherwise stated.
We use the Einstein notation or Einstein summation convention.
According to this convention, when an index variable appears twice in a single term, it implies summation of that term over all the indices and drops $\sum$:
\begin{align}
    \sum_{i} A_i B^i \to A_i B^i\,,
    \qquad 
    \sum_{i,j} g_{ij} A^i B^j \to g_{ij} A^i B^j\,.
\end{align}

\section{Quantum Hamiltonian Descent}
\label{sec:QHD}

Let us give a brief review of the quantum algorithm called \emph{Quantum Hamiltonian Descent (QHD)}~\cite{2023-Leng-Hickan-Li-Wu}, which gives a basic concept of our study.
QHD identifies the parameters to be optimized with the position of a particle moving in a potential, which corresponds to the loss function.
The optimization is realized by quantum mechanical time evolution.
In the following parts of this section, we begin with the Lagrangian and its classical dynamics, then move to the quantum dynamics for parameter optimization via QHD.

We consider the problem of finding the minimum point of a continuous and differentiable loss function $V(\bm{x})$ in the $N$-dimensional space, which is denoted by $\bm{x}_\ast$.
The coordinate of the particle is given by $\bm{x} = (x^1,\ldots,x^N)^\tt$.
For this optimization problem, the single point particle dynamics in the $N$-dimensional space feeling the potential $V(\bm{x})$ with a velocity viscosity is often used and it is described by the following (classical) Lagrangian 
\begin{align}
	L(\bm{x},\dot{\bm{x}}, t) = a(t) \biggl(
		\frac{m}{2} \dot{\bm{x}}^2 - \eta(t) V(\bm{x})
	\biggr)\,,
	\label{eq:QHD-Lagrangian}
\end{align}
where dot denotes the derivative with respect to time $t$.
We write the sum of squares of particle velocities as $\dot{\bm{x}}^2 = (\dot{x}^i)^2$.
$m$ is the particle mass and assume to be positive $m>0$.
$V(\bm{x})$ is real-valued function corresponding to the potential energy of the dynamics of the point particle.
The time-dependent coefficients $a(t)$ and $\eta(t)$ are real functions of time $t$, and $a(t)$ produces the dissipation of this system as shown below.
The time dependence of these quantities is dropped, and we write $\bm{x}(t) = \bm{x}$, $a(t) = a$, and $\eta(t) = \eta$ for simplicity unless otherwise stated.
The variation of \eqref{eq:QHD-Lagrangian} with respect to $\bm{x}$ gives the equation of motion (EOM) as 
\begin{align}
	m \ddot{\bm{x}} + m \frac{\dot{a}}{a} \dot{\bm{x}} + \eta \grad V(\bm{x}) =0\,,
	\label{eq:QHD-classial-eom}
\end{align}
which is so-called Euler--Lagrange equation.
$\grad$ denotes the gradient vector, namely $\grad = (\del_i,\ldots,\del_N)^\tt$ with $\del_i = \del / \del x^i$.
The time-dependence of the coefficient of the kinetic term produces a friction such as $\frac{\dot{a}}{a}\dot{\bm{x}}$ in EOM.

From this equation, we discuss the interpretation of $a(t)$ and $\eta(t)$ in optimization.
The time-dependence of $a$ leads to the energy dissipation with a friction term as mentioned above.
This dissipative term plays an important role so that we have a convergence process in optimization.
If we take $a(t) = \ee^{2 \gamma t}$ with a positive real value $\gamma$, Eq.~\eqref{eq:QHD-classial-eom} coincides with the continuous limit of the parameter update rule of the accelerated gradient method known as Polyak's heavy ball method, and it is known that the convergence speed to stationary points of the gradient function is characterized by $\gamma$ \cite{2021-Sanz-et-al}. 
For other choices of $a$, see, e.g., Refs.~\cite{2013-Itami,2015-Su-Boyd-Candes,2015-Krichene-Bayen-Bartlett,2015-Wibisono-Wilson,2016-Wibisono-Wilson-Jordan,2016-Wilson-Recht-Jordan,2021-Sanz-et-al,2021-Wilson-Recht-Jordan}.
Eq.~\eqref{eq:QHD-classial-eom} also tells that the parameters $\bm{x}$ are updated by the potential gradient, and $\eta$ being the coefficient of $\grad V$ is regarded as a learning parameter.

QHD~\cite{2023-Leng-Hickan-Li-Wu} extends the above process by incorporating quantum effects.
Since quantum mechanical features such as a tunneling effect would be included in the time evolution, QHD is expected to be an efficient quantum approach for optimization using the dynamics.

In order to consider the quantum dynamics of the system, we introduce the canonical formalism of the system \eqref{eq:QHD-Lagrangian}.\footnote{
In general, when the Lagrangian depends explicitly on time $t$, the system does not admit a canonical form; however, when the energy dissipation arises only from a friction term proportional to the velocity $\dot{\bm{x}}$, a canonical description is possible.
}
The conjugate momentum $\bm{p} = (p_1,\ldots,p_N)^\tt$ is defined by 
\begin{align}
	\bm{p} = \frac{\del L}{\del \dot{\bm{x}}} = a(t) m \dot{\bm{x}}\,,
\end{align}
and the Hamiltonian becomes 
\begin{align}
	H(t) = p_i \dot{x}^i - L(\bm{x}, \dot{\bm{x}}, t)
	= \frac{1}{a(t)} \left( \frac{\bm{p}^2}{2m} \right) + a(t) \eta(t) V(\bm{x})\,,
	\label{eq:QHD-Hamiltonian}
\end{align}
where $\bm{p}^2 = p_i^2$.
In order to move to quantum theory, let us introduce position states $\ket{\bm{x}}$ and their dual $\bra{\bm{x}}$ satisfying\footnote{
It is assumed that the coordinates $\bm{x}$ is appropriately discretized when we discuss quantum circuit implementation.
}
\begin{align}
	\hat{\bm{x}} \ket{\bm{x}} = \bm{x} \ket{\bm{x}}\,,
	\qquad 
	\int_{{\mbb{R}^N}} \dd^N {\bm{x}}\,
	\ket{{\bm{x}}}\!\!\bra{{\bm{x}}}=I
	\,,\qquad
	\braket{{\bm{x}}|{{\bm{x}}'}}={\delta^N}({\bm{x}}-{\bm{x}}')
	\,.
\end{align}
The quantum state $\ket{\psi(t)}$ describing the single particle follows the Schr\"odinger equation 
\begin{align}
	\iu \frac{\dd}{\dd t} \ket{\psi(t)} = \hat{H}(t) \ket{\psi(t)}\,.
	\label{eq:schrodinger-S}
\end{align}
$\hat{H}(t)$ is the Hamiltonian operator obtained from the classical Hamiltonian~\eqref{eq:QHD-Hamiltonian}
\begin{align}
	\hat{H}(t) = \frac{1}{a(t)} \left( \frac{\hat{{\bm{p}}}^2}{2m} \right) + a(t) \eta(t) V(\hat{\bm{x}})\,.
	\label{eq:QHD-quantum}
\end{align}
The coordinate operator $\hat{\bm{x}}$ and its conjugate momentum operator $\hat{\bm{p}}$ satisfy the canonical commutation relations 
\begin{align}
	[\hat{x}^i, \hat{p}_j] = \iu \delta^i_j\,.
\end{align}
Requiring the Hermicity of the momentum operator implies that $\hat{\bm{p}}$ in the position basis can be represented by the differential operator 
\begin{align}
	\hat{p}_i = - \iu \frac{\del}{\del x^i} = - \iu \del_i\,.
	\label{eq:p-QHD}
\end{align}
Then, the Hamiltonian operator \eqref{eq:QHD-quantum} in the position basis is given by 
\begin{align}
	\hat{H}(t) = \frac{1}{a(t)} \left( \frac{- \Delta}{2 m} \right) + a(t) \eta(t) V(\hat{\bm{x}})\,.
	\label{eq:QHD}
\end{align}
Here, $\Delta$ is the Laplace operator defined by 
\begin{align}
	\Delta = \del_i^2\,.
\end{align}
As a quantum algorithm, one implements the Hamiltonian simulation of Eq.~\eqref{eq:QHD} and  estimates the expectation value of the position of the output state $\ket{\psi(t)}$ as $\braket{\bm{x}(t)} = \braket{\psi(t)| \hat{\bm{x}} |\psi(t)}$.
With suitable choices of $a$ and $\eta$ for sufficiently long $t$, $\braket{\bm{x}(t)}$ should approach the desired value $\bm{x}_\ast$.

Several extensions of QHD have been proposed.
These include models that add thermal-noise-like effects to $\grad V({\bm x})$ and incorporate stochastic parameter updates \cite{2025-Peng-et-al,2025-Escalante}, as well as formulations that include contributions from a vector potential \cite{Leng:2025msd}.
These are frameworks that use quantum dynamics for parameter optimization as like QHD and they suggest performance advantages over classical continuous-optimization algorithms.

QHD and its extensions provide a new continuous optimization framework which can potentially resolve the local-minimum stagnation problem using a quantum feature.
However, it is not straightforward to adapt QHD to continuous optimization on Riemannian manifolds because the geometric structure of the parameter space is not reflected.
In this work, we propose an extension of QHD, that is applicable to continuous optimization on Riemannian manifolds.

\section{Quantum Riemannian Hamiltonian Descent}
\label{sec:QRHD}

In this section, we introduce our proposal of a quantum algorithm applicable to continuous optimization on Riemannian manifolds, which we call \emph{Quantum Riemannian Hamiltonian Descent (QRHD).}
This can be regarded as a natural extension of QHD by considering single point particle dynamics in curved space.

We extend the Lagrangian~\eqref{eq:QHD-Lagrangian} including effects of the background geometry and consider the dynamics of a single point particle in the $N$-dimensional curved space by
\begin{align}
	L(\bm{x}, \dot{\bm{x}}, t) = a(t) \biggl(
		\frac{m}{2} g_{ij}(\bm{x}) \dot{x}^i \dot{x}^j - \eta(t) V(\bm{x})
	\biggr) \,,
	\label{eq:QRHD-Lagrangian}
\end{align}
where $g_{ij}({\bm{x}})$ is a differentiable function of $\bm{x}$ and assumed to be a symmetric and positive definite matrix.
This is regarded as the metric of the $N$-dimensional space, which is assumed to be an $N$-dimensional Riemannian manifold $\mc{M}_N$ in this work.
The inverse matrix of $g_{ij}$ is denoted by $g^{ij}$ and satisfies 
\begin{align}
	g_{ij}({\bm{x}})g^{jk}({\bm{x}})=\delta^k_i\,.
\end{align}
QHD corresponds to the special choice $g_{ij}(\bm{x}) = \delta_{ij}$, which describes a particle mechanics in the flat space.\footnote{
Physical models with such a noncanonical kinetic term are know as nonlinear sigma models and actively discussed in wide contexts.
}
From \eqref{eq:QRHD-Lagrangian}, the classical EOM becomes 
\begin{align}
	m \l(
		\ddot{x}^i + \Gamma^i_{jk}(\bm{x}) \dot{x}^j \dot{x}^k
	\r) + m \frac{\dot{a}}{a} \dot{x}^i + \eta 
    (\grad_g V(\bm{x}))^i = 0\,,
	\label{eq:QRHD-classical-eom}
\end{align}
where $\Gamma^i_{jk}(\bm{x})$ is the Levi-Civita connection given by the derivative of $g_{ij}(\bm{x})$ as 
\begin{align}
	\Gamma^{i}_{jk}(\bm{x}) =  \frac{1}{2}g^{il}(\bm{x})\l(
		\del_j g_{lk} + \del_k g_{jl} - \del_l g_{jk}
	\r)\,.
\end{align}
The first two terms in Eq.~\eqref{eq:QRHD-classical-eom} correspond to the acceleration vector multiplied by the mass on the manifold.
When $\dot{a}=0$ and $V=0$, this acceleration vector characterizes the geodesic of this geometry.\footnote{
In this sense, one may also interpret the present algorithm as coherently performing an operation corresponding to the ``retraction'' used in Riemannian optimization algorithms on classical computers.
}
$(\grad_g V)^i$ is defined by 
\begin{align}
	(\grad_g V)^i \coloneqq g^{ij}({\bm{x}})\partial_j V({\bm{x}})\,,
\end{align}
which is the natural gradient of $V(\bm{x})$ in the Riemannian manifold $\mc{M}_N$.
Then, Eq.~\eqref{eq:QRHD-classical-eom} is regarded as the EOM of a particle moving on a Riemannian manifold with the natural gradient and the energy dissipation.
Choosing $g_{ij}$ appropriately can perform parameter updates while incorporating prior information about the geometric structure of the parameter space.

Let us move on to the quantization of the dynamics described by \eqref{eq:QRHD-Lagrangian}.
We start from the canonical formalism of the classical dynamics, in which the conjugate momentum is introduced by  
\begin{align}
	p_i = \frac{\del L}{\del \dot{x}^i} = a(t) m g_{ij}(\bm{x}) \dot{x}^j\,,
\end{align}
and the Hamiltonian is 
\begin{align}
	H(t) = p_i \dot{x}^i - L(\bm{x}, \dot{\bm{x}}, t)
	= \frac{1}{a(t)} \l(
		\frac{1}{2m} g^{ij}(\bm{x}) p_i p_j
	\r) + a(t) \eta(t) V(\bm{x})\,.
	\label{eq:QRHD-Hamiltonian}
\end{align}
Compared with the flat case \eqref{eq:QHD-Hamiltonian}, the inverse metric is found to appears in the kinetic term.
The existence of the function of $\bm{x}$ in the kinetic term causes the operator ordering problem in the quantization process, and careful treatment is required.

In this paper, we adopt the quantization procedure similar to QHD by taking into the account the appearance of $g^{ij}(\bm{x})$ and operator ordering.
Let us introduce the position states $\ket{\bm{x}}$ and their duals $\bra{\bm{x}}$ which satisfy the similar equations to QHD as 
\begin{align}
	\hat{\bm{x}} \ket{\bm{x}} = \bm{x} \ket{\bm{x}}\,,
	\quad 
	\int_{\mathcal{M}_N} \dd^N\bm{x} \sqrt{g({\bm{x}})}\ket{\bm{x}}\!\!\bra{\bm{x}} =I\,,
	\quad 
	\braket{\bm{x}|\bm{x}'} = g({\bm{x}})^{-\frac{1}{4}} \delta^N(\bm{x}-\bm{x}')
	g(\bm{x}')^{-\frac{1}{4}}\,,
	\label{eq:QRHD-complete-normalization}
\end{align}
where $g(\bm{x})$ denotes the determinant of the metric $g(\bm{x}) = \det g_{ij}(\bm{x})$.
$\int_\mc{M}$ denotes the integration in the parameter space $\mc{M}_N$.
It is noted that the integration measure $\dd^N\bm{x} \sqrt{g(\bm{x})}$ is the invariant volume element under point transformations (general coordinate transformations).
The factors $g^{-1/4}(\bm{x})$ in the normalization condition of Eq.~\eqref{eq:QRHD-complete-normalization} originate from the symmetric representation of the following relation:
\begin{align}
	1 = \int_{\mc{M}_N} \dd^N \bm{x} \delta^N(\bm{x} - \bm{x}') = \int_{\mc{M}_N} \dd^N \bm{x} \sqrt{g(\bm{x})} g^{-\frac{1}{4}}(\bm{x}) \delta^N(\bm{x} - \bm{x}') g^{- \frac{1}{4}}(\bm{x}')
	= \int_{\mc{M}_N} \dd^N\bm{x} \sqrt{g(\bm{x})} \braket{\bm{x}|\bm{x}'}\,.
\end{align}
The conjugate momentum operator $\hat{\bm{p}} = (\hat{p}_1,\ldots, \hat{p}_N)^\tt$ is also modified by the geometric properties.
The canonical commutation relations
\begin{align}
	[\hat{x}^i, \hat{p}_j] = \iu \delta^i_j\,,
\end{align}
implies $\hat{p}_i = - \iu \del_i + \iu f_i(\bm{x})$ using a real function $f_i(\bm{x})$.
The Hermicity of the momentum operator on $\mc{M}$ fixes the functional form as $f_i(\bm{x}) = - (1 /4 ) g(\bm{x})^{-1} \del_i g(\bm{x})$.
Then, the momentum operator in the position basis is given by \cite{1972-Omote-Sato}
\begin{align}
	\hat{p}_i = -\iu \del_i - \frac{\iu}{4} g(\bm{x})^{-1} \del_i g(\bm{x})
	= - \iu \left(
		\del_i + \frac{1}{2} \Gamma_i(\bm{x})
	\right)
	= g(\bm{x})^{-1/4} (- \iu \del_i ) g(\bm{x})^{1/4}\,.
	\label{eq:p-operator}
\end{align}
Here, $\Gamma_i(\bm{x}) \coloneqq \Gamma^j_{ij}(\bm{x}) = \frac{1}{2} g^{jk}(\bm{x})\del_i g_{jk}(\bm{x})$ is introduced.

Now we consider the quantum state of this system $\ket{\psi(t)}$ and its time evolution equation by the Schr\"odinger equation 
\begin{align}
	\iu \frac{\dd}{\dd t} \ket{\psi(t)} = \hat{H}(t) \ket{\psi(t)}\,.
    \label{eq:QRHD-Schroedinger}
\end{align}
The next question is what form does the Hamiltonian operator $\hat{H}(t)$ have.
This can be derived by the properties of the position basis wave function $\psi(\bm{x},t) = \braket{\bm{x}|\psi(t)}$.
By the completeness condition in \eqref{eq:QRHD-complete-normalization}, the inner product of two quantum states $\ket{\psi_1(t)}$ and $\ket{\psi_2(t)}$ becomes 
\begin{align}
	\braket{\psi_1(t)|\psi_2(t)} = \int_{\mc{M}_N} \dd^N\bm{x} \sqrt{g(\bm{x})} \psi^*_1(\bm{x},t) \psi_2(\bm{x},t)\,.
\end{align}
The inner product is a scalar and invariant under the coordinate transformation and we mention that the integration measure $\dd^N\bm{x} \sqrt{g(\bm{x})}$ is also invariant.
Then, we find that the wave function $\psi(\bm{x},t)$ behaves as a scalar function under the transformation and the Hamiltonian operator $\hat{H}(t)$ generating the time evolution is also invariant.
From these, the Hamiltonian operator of the curved space quantum system takes the following form~\cite{1928-Podolsky,1972-Omote-Sato}:
\begin{align}
	\hat{H}(t)=\frac{1}{a(t)}\l(\frac{-\Delta_g}{2m}\r) + a(t) \eta(t)V({\hat{\bm{x}}})\,.
	\label{eq:QRHD}
\end{align}
$\Delta_g$ is the covariant differential operator on the manifold $\mc{M}$ defined by\footnote{
If one wishes to add a vector field $A_i({\bm{x}})$ corresponding to a gauge field as in Ref.~\cite{Leng:2025msd}, one can replace the differential operator $\partial_i$ in Eq.~\eqref{eq:deltag} with $\partial_i- \iu e A_i({\bm{x}})$ ~\cite{1988-DOlivo-Torres} which is called the covariant derivative.
Here, $e$ is a constant called a gauge coupling.
}
\begin{align}
	\Delta_g = \frac{1}{\sqrt{g(\bm{x})}} \del_i \sqrt{g(\bm{x})} g^{ij}(\bm{x}) \del_j
	= g^{ij}(\bm{x}) \l(
		\del_i \del_j - \Gamma^k_{ij} \del_k
	\r)\,,
	\label{eq:deltag}
\end{align}
which is also called the Laplace--Beltrami operator~\cite{1997-Rosenberg}.

The metric also controls the propagation and spreading of the wave packet. 
In a local semiclassical description, the kinetic energy is
\begin{align}
    E_{\mathrm{kin}}(\bm{x},\bm{p}) \approx \frac{g^{ij}({\bm{x}})p_ip_j}{2ma(t)}\,,
\end{align}
and the corresponding group velocity is 
\begin{align}
    \dot{x}^{i} \approx \frac{1}{ma(t)}g^{ij}({\bm{x}})p_j\,.
\end{align}
The inverse metric therefore determines the directional response to momentum. 
In the local harmonic approximation near a minimum, choosing the metric to match the positive-definite Hessian equalizes the oscillator frequencies and allows a common friction coefficient to act similarly on all modes.
This provides a physical interpretation of how an appropriate metric can compensate for potential anisotropy and reduce persistent directional differences in wave-packet localization, as illustrated in Sec.~\ref{sec:numeric-flat}.

For the latter use to connect the operator formulation with the path integral formalism, we consider the rewriting of the Hamiltonian operator~\eqref{eq:QRHD}.
Using the momentum operator~\eqref{eq:p-operator}, $\hat{H}(t)$ is expressed as 
\begin{align}
	\hat{H}(t) = \frac{1}{a(t)} \left(
		\frac{1}{2m} g^{-1/4} (\hat{\bm{x}}) \hat{p}_i g^{ij}(\hat{\bm{x}}) \sqrt{g(\hat{\bm{x}})} \hat{p}_j g^{-1/4}(\hat{\bm{x}}) 
	\right) + a(t) \eta(t) V(\hat{\bm{x}})\,.
\end{align}
Furthermore, this can be rewritten as follows \cite{1975-Mizrahi,1988-DOlivo-Torres}:
\begin{align}
	\hat{H}(t) = \frac{1}{a(t)} \l(
		\frac{1}{2m} \l[
			g^{ij}(\hat{\bm{x}}) \hat{p}_i \hat{p}_j
		\r]_{\mr{W}} + \Delta V(\hat{\bm{x}})
	\r) + a(t) \eta(t) V(\hat{\bm{x}})\,.
	\label{eq:Weyl-ordering-Hamiltonian}
\end{align}
$[\hat{\bullet}]_{\mr{W}}$ denotes the Weyl ordering of the operator products $\hat{\bullet}$, which is often used when discussing the relation between the operator formalism and the path integral formalism.
$\Delta V({\bm{x}})$ is defined by \cite{1975-Mizrahi,Gervais:1976ws,Sakita:1985exh}
\begin{align}
    \Delta V(\bm{x}) = \frac{1}{8 m} \biggl(
        - \mc{R}(\bm{x}) + g^{ij}(\bm{x}) \Gamma^k_{il}(\bm{x}) \Gamma^l_{jk}(\bm{x})
    \biggr)\,.
    \label{eq:Delta-V}
\end{align}
Here $\mathcal{R}$ is the Ricci scalar defined by
\begin{align}
    \mc{R} = g^{ij} \Bigl(
        \del_k \Gamma^k_{ij} -\del_i \Gamma^k_{kj} + \Gamma^k_{km} \Gamma^m_{ij} - \Gamma^k_{im} \Gamma^m_{kj}
    \Bigr)\,.
\end{align}
It is noted that a geometric quantum correction $\Delta V(\bm{x})$ appears in addition to the naive promotion of the classical variables in Eq.~\eqref{eq:QRHD-Hamiltonian} to the operators with Weyl ordering.
$\Delta V$ is physically necessary to restore the general covariance of the full Hamiltonian because the Weyl ordered $[g^{ij} \hat{p}_i \hat{p}_j]_{\mr{W}}$ is not a general coordinate scalar.
In addition, this correction $\Delta V$ comes from the ordering in the kinetic term and the explicit time dependence is $1/a(t)$, which is same with the $\bm{p}^2$ term.

As an algorithm, we implement the Hamiltonian simulation for the time-dependent Hamiltonian~\eqref{eq:QRHD} and estimate the expectation value of the position with the output state $\ket{\psi(t)}$ in the same manner with QHD.
Suitable choices of $a$ and $\eta$ and sufficiently long $t$ would make $\braket{\bm{x}(t)} = \braket{\psi(t)| \hat{\bm{x}} | \psi(t)}$ approach the desired value $\bm{x}_\ast$ being the optimized point of the function $V(\bm{x})$.

As seen in this section, the key feature of the present algorithm is the additional degree of freedom $g_{ij}(\bm{x})$ which is trivial in QHD, through which one can incorporate the geometric information of the parameter space as prior information for optimization.
On the other hand, when we consider the quantum mechanical system extended by a nontrivial $g_{ij}({\bm{x}})$, quantum corrections arise, which are associated with the geometry of the parameter space such as Eq.~\eqref{eq:Delta-V}.
This can modify the quantum motion of the point particle.
In the next section, we analyze how such geometric corrections affect the dynamical features for optimization such as the convergence time.

Finally, we comment on the relation between QRHD and imaginary-time evolution.
Treating $\bm{x}$ as parameters (amplitudes) specifying a pure state, and equipping the parameter manifold with the structure of complex projective space endowed with the Fubini--Study metric, we find that, upon choosing the loss function to be normalized energy expectation value, the update rule \eqref{eq:QRHD-classical-eom} includes imaginary-time evolution \cite{2019-Yamamoto,2020-Stokes-et-al,2025-Suzuki-et-al}.
In this sense, with an appropriate choice of metric and cost function, QRHD can also be viewed as a (pure-state) ground state search algorithm.

\section{Convergence time}
\label{sec:convergence}

As discussed in the previous sections, Q(R)HD performs parameter optimization by simulating the quantum dissipative dynamics on a computer.
In order to obtain an optimal solution or evaluate a complexity, it is useful to evaluate the typical time scale of the quantum dynamics.
In this section, we derive the time-evolution equation of the quantum mechanical expectation value $\braket{\bm{x}}$ estimates the time scale required to obtain an optimal solution by analyzing this equation.
As seen in the following, this equation is regarded as the quantum version of EOM.

\subsection{QHD}
\label{sec:convergence-QHD}

We begin with the case of QHD.
In QHD, the time-evolution equation which $\braket{{\bm{x}}}$ obeys is, similarly to the classical equation of motion \eqref{eq:QHD-classial-eom},
\begin{align}
    m \braket{\ddot{\bm{x}}} + m \frac{\dot{a}}{a} \braket{\dot{\bm{x}}} + \eta \braket{\grad V(\bm{x})} = 0\,,
    \label{eq:QHD-EOM}
\end{align}
as shown in Ref.~\cite{2023-Leng-Zhang-Wu}.
This is derived from the Heisenberg equation with the Hamiltonian operator the operator Eq.~\eqref{eq:QHD-quantum} and the canonical commutation relation, but below we also derive this equation in the path integral formalism for analysis of the convergence time in QRHD.

Then, let us introduce the path integral formalism for QHD.
For path integrals, see, e.g., Refs.~\cite{Feynman:2010,Peskin:1995ev,MacKenzie:1999pu}.
The partition function for QHD introduced in Sec.~\ref{sec:QHD} is given by 
\begin{align}
    Z = \int \mc{D} \bm{x} \mc{D} \bm{p} \exp \iu \int \dd t \bigl( \bm{p} \cdot \dot{\bm{x}} - H(\bm{x}, \bm{p}) \bigr)\,,
\end{align}
where $H$ is Eq.~\eqref{eq:QHD-Hamiltonian}.
This is the path integral on phase space and the path integral measure $\mc{D}\bm{x} \mc{D}\bm{p}$ arises from inserting completeness relations in rewriting the transition amplitude.
Performing the momentum integral $\mc{D}\bm{p}$ gives the path integral in configuration space 
\begin{align}
	Z = \int \mc{D}\bm{x} \exp (\iu S[\bm{x}])\,,
	\qquad 
	S[\bm{x}] = \int \dd t L(\bm{x}, \dot{\bm{x}},t) = \int \dd t a(t) \biggl(
        \frac{m}{2} \dot{\bm{x}}^2 - \eta(t) V(\bm{x})
	\biggr)\,,
	\label{eq:partition-function-QHD}
\end{align}
where the path integral measure is defined by
\begin{align}
    \mc{D} \bm{x} \coloneqq \lim_{K\to \infty} \prod_{k=1}^{K}\left(\frac{m a(t_k)}{2 \pi \iu \Delta t} \right)^{N/2}\dd^N \bm{x}(t_k) \,,
\end{align}
with $K$ the number of time slices; in the above expression, $t_k$ denotes the time at the $k$-th step.
We take the time step size to be common between slices, $\Delta t = t_{k+1} - t_k$.

In the path integral formalism, the expectation value of an observable $\mc{O}(\hat{\bm{x}})$ is expressed as 
\begin{align}
	\braket{\mc{O}(\bm{x})} = \frac{1}{Z} \int \mc{D} \bm{x} \mc{O}(\bm{x}) \exp (\iu S[\bm{x}])\,.
\end{align}
The quantum EOM of $\braket{\mc{O}(\bm{x})}$ is derived from the partial integrability of the path integral
\begin{align}
	0 = \int \mc{D} \bm{x} \frac{\delta}{\delta x^i(t)} \biggl(
        \mc{O}(\bm{x}) \exp (\iu S[\bm{x}])
    \biggr)\,,
\end{align}
which yields 
\begin{align}
	 \Braket{\frac{\delta \mc{O}}{\delta x^i}} + \iu \Braket{\mc{O} \frac{\delta S}{\delta x^i}} =0 \,,
    \label{eq:SD-equation}
\end{align}
where $\delta / \delta x^i$ denotes the functional derivative with respect to the coordinate variable $x^i$.
Eq.~\eqref{eq:SD-equation} is known as the Schwinger--Dyson (SD) equation.
Setting $\mc{O} = 1$, we obtain the following quantum EOM which is the equation of the expectation value:
\begin{align}
	\Braket{\frac{\delta S}{\delta x^i}} = 0\,,
\end{align}
and the explicit form is found as 
\begin{align}
	\Braket{
    		a \delta_{ij}\biggl(
    			m \ddot{x}^j + m \frac{\dot{a}}{a} \dot{x}^j + \eta [\grad V(\bm{x})]^j
    		\biggr)
	} = 0\,.
\end{align}
Assuming $a(t) \neq 0$, this is the same with Eq.~\eqref{eq:QHD-EOM}.

\paragraph{Estimation of convergence time in QHD.}

Having derived the time-evolution equation for the parameters in QHD, we now estimate the time scale $t_\ast$ required for QHD to find an optimal solution.
$t_\ast$ depends on the choices of $a(t)$ and $\eta(t)$, and the properties of $V({\bm{x}})$.
In what follows, we specifically consider the case $a(t)=\ee^{2\gamma t}$ and $\eta(t)=\eta = \mr{const.} \geq 0$.
In this case, the time-evolution equation for the parameter $\braket{{\bm{x}}}$ approximately becomes
\begin{align}
	\Braket{\ddot{x}^i} + 2\gamma \braket{\dot{x}^i} + \frac{\eta}{m} \delta^{ik} \del_k V(\braket{\bm{x}}) = 0\,.
	\label{eq:eom-QHD-convex-approx}
\end{align}
Then $V(\bm{x})$ determines the convergence property.
Let us assume that $V(\bm{x})$ is differentiable and $\bm{x}_\ast$ is a local minimum $\grad V(\bm{x}_\ast) = 0$ such that the Hessian matrix at the point is positive semidefinite.
Expanding $V(\bm{x})$ around $\bm{x} = \bm{x}_\ast$~\cite{2025-Catli-Simon-Wiebe}, we obtain
\begin{align}
	V(\bm{x}) = V(\bm{x}_\ast) + \frac{1}{2} (\bm{x} - \bm{x}_\ast)^\tt \Hess V_\ast (\bm{x} - \bm{x}_\ast) + \mc{O} \bigl(|\bm{x} - \bm{x}_\ast|^3\bigr)\,,
	\label{eq:V_expand}
\end{align}
where the Hessian matrix of the potential $V(\bm{x})$ is introduced by 
\begin{align}
	(\Hess V)_{ij} \coloneqq \frac{\del^2 V(\bm{x})}{\del x^i \del x^j}\,,
\end{align}
and $\Hess V_\ast$ denotes the Hessian evaluated at the local minimum $\bm{x}_\ast$.
If we considering up to the quadratic terms in the Lagrangian or Hamiltonian (operator), the EOM reduces to
\begin{align}
	\braket{\ddot{x}^i} + 2 \gamma \braket{\dot{x}^i} + \frac{\eta}{m} \delta^{ij} (\Hess V_\ast)_{jk} (\braket{x^k} - x^k_\ast) =0\,.
	\label{eq:eom-QHD-convex}
\end{align}
describes the dynamics.
When $V(\bm{x})$ has a convex quadratic form, Eq.~\eqref{eq:eom-QHD-convex-approx} exactly becomes this equation.
Eq.~\eqref{eq:eom-QHD-convex} is the EOM for a damped oscillator in the $N$-dimensional flat space, which means that parameter $\bm{x}$ evolves toward $\bm{x} = \bm{x}_\ast$ and the convergence behavior is characterized by $\gamma$.
When $\gamma$ is smaller (larger) than $\sqrt{\frac{\eta}{m} \lambda_{\mr{min}}(\Hess V_\ast)}$, the system shows underdamped while converging (overdamped).
When $\gamma = \sqrt{\frac{\eta}{m} \lambda_{\mr{min}}(\Hess V_\ast)}$, it is critically damped.
Here, we introduce the notation of $\lambda_{\mr{min}}(A)$ to denote the minimum eigenvalue of the matrix $A$.
In this case, the convergence is typically evaluated as 
\begin{align}
	t_\ast \gtrsim \frac{1}{\sqrt{\frac{\eta}{m} \lambda_{\mr{min}}(\Hess V_\ast)}} \l[
		- W_{-1} \l(-\frac{\epsilon_\ast}{\ee} \r) - 1
	\r]\,,
	\label{eq:t_star_bound_QHD}
\end{align}
and the equality is achieved when
\begin{align}
	\gamma \sim \sqrt{{\frac{\eta}{m} \lambda_{\mr{min}}(\Hess V_\ast)}}\,.
\end{align}
Here, $t_\ast$ is introduced as the time when the condition $|\braket{{\bm{x}}(t_\ast)}-{\bm{x}}_\ast|\leq \epsilon_\ast|\braket{{\bm{x}}(0)}-{\bm{x}}_\ast|$ is first satisfied for a given small value $\epsilon_\ast$.
$W_{-1}(x)$ is the Lambert $W$ function satisfying $W(z) \ee^{W(z)} = z$ of the branch with $W_{-1} <-1$ defined on $-\ee^{-1} \leq z <0$.
This result is consistent with the estimate in Ref.~\cite{2025-Catli-Simon-Wiebe}.

\subsection{QRHD}
\label{sec:QRHD-pathintegral}

Let us move on to QRHD.
Following the previous discussion for QHD, we start our study from the derivation of the partition function in order to evaluate the expectation value of an observable in the path integral formalism for QRHD.
The partition function for a point particle in curved space on phase space is written as~\cite{Sato:1976hy,1988-DOlivo-Torres,Bastianelli:2017wsy}:
\begin{align}
    Z = \int \mc{D} \bm{x} \mc{D} \bm{p} \exp \iu \int \dd t \Bigl(
        p_i \dot{x}^i - \tilde{H}(\bm{x},\bm{p})
    \Bigr)\, ,
    \label{eq:phase-space-pathintegral-QRHD}
\end{align}
where the Hamiltonian $\tilde{H}$ is given by
\begin{align}
    \tilde{H}(\bm{x}, \bm{p}) = a(t) \biggl(\frac{1}{2 m } a^{-2}(t) g^{ij}(\bm{x}) p_i p_j + \eta(t) V(\bm{x}) + a^{-2}(t) \Delta V(\bm{x}) \biggr)\,,
    \label{eq:Htilde-pathintegral}
\end{align}
and $\Delta V$ is the correction term~\eqref{eq:Delta-V} arising from Weyl ordering introduced in Eqs.~\eqref{eq:Weyl-ordering-Hamiltonian}.%
It is noted that the power of $a(t)$ multiplying $\Delta V(\bm{x})$ is the same with that of the kinetic term in the Hamiltonian.
Performing the momentum integral $\mc{D}\bm{p}$ yields the configuration space path integral representation 
\begin{align}
	Z = \int_{\mc{M}_N} [\mc{D}\bm{x}] \exp(\iu \mc{S}[\bm{x}])\,,
	\label{eq:path-integral-QRHD}
\end{align}
where the action is 
\begin{align}
	\mc{S}[\bm{x}] = \int \dd t a(t) \biggl(
		\frac{m}{2} g_{ij}(\bm{x}) \dot{x}^i \dot{x}^j - \eta(t) V(\bm{x}) - a^{-2}(t) \Delta \mc{V}(\bm{x})
	\biggr)\, ,
	\label{eq:effective-action-QRHD}
\end{align}
and the additional correction to the potential is introduced by \cite{1988-DOlivo-Torres}
\begin{align}
	\Delta \mc{V}(\bm{x}) = \Delta V(\bm{x}) + \Delta V'(\bm{x})\,,
	\qquad 
	\Delta V'(\bm{x}) = \frac{1}{8m} g^{ij}(\bm{x}) \del_i \Gamma_j(\bm{x})\,.
    \label{eq:Delta_mathcalV}
\end{align}
It is noted that the individual correction terms appearing in this representation are regularization- and ordering-dependent and need not be generally covariant separately.
Together with the Weyl symbol of the kinetic operator and the time-slicing in the discretization, it ensures that the complete short-time evolution operator reproduces the generally covariant Hamiltonian operator~\eqref{eq:QRHD}.
An explicit derivation of this correspondence from the short-time kernel is given in App.~\ref{sec:relation-to-PI}.
The path integral measure is the covariant integration measure defined by 
\begin{align}
    [\mc{D}\bm{x}] = \lim_{K\to \infty} \prod_{k=1}^K \sqrt{g(\bm{x}(t_k))} \left( \frac{m a(t_k)}{2 \pi \iu \Delta t}\right)^{N/2} \dd^N \bm{x}(t_k)
    = \mc{D} \bm{x} \sqrt{\Det g}\,,
    \label{eq:path-integral-measure-QRHD}
\end{align}
where $\Det g$ denotes the functional determinant associated with the metric tensor $g_{ij}(\bm{x})$.
In Eqs.~\eqref{eq:phase-space-pathintegral-QRHD}--\eqref{eq:path-integral-measure-QRHD}, the phase-space path integral is defined by the midpoint time-slicing prescription associated with Weyl ordering.
The expectation value of an observable $\mc{O}(\hat{\bm{x}})$ in QRHD can be written as 
\begin{align}
    \braket{\mc{O}(\bm{x})} = \frac{1}{Z} \int_{\mc{M}_N} [\mc{D} \bm{x}] \mc{O}(\bm{x}) \exp(\iu \mc{S}[\bm{x}])\,.
\end{align}

To consistently track the metric-measure contribution and the associated contact terms before taking the continuum limit, we first derive the SD equation for QRHD at finite time slicing.
Let us write the coordinate at an interior time slice $t_n$ as $x_n^i = x^i(t_n)$, which notation is used in the notion of the path integral measure \eqref{eq:path-integral-measure-QRHD}.
The discretized path integral takes the schematic form
\begin{align}
    Z_K = \int_{\mc{M}_N} [\mc{D}\bm{x}]_K
    \exp \left( \iu \mc{S}_K[\bm{x}] \right)\,,
    \qquad 
    [\mc{D}\bm{x}]_K = \prod_{k=1}^{K} 
    \sqrt{g(\bm{x}_k)} \left(\frac{m a(t_k)}{2 \pi \iu \Delta t}\right)^{N/2} \dd^N \bm{x}_k\,,
\end{align}
where $\mc{S}_K[x]$ is the discretized effective action corresponding to \eqref{eq:effective-action-QRHD} which contains the midpoint-discretized kinetic term and the ordering counterterms required to reproduce the Hamiltonian \eqref{eq:QRHD}.
For an interior time slice, the partial integrability produces the following EOM for observable $\mc{O}(\bm{x})$:
\begin{align}
    0 &= \int_{\mc{M}_N} [\mc{D}\bm{x}]_K \frac{1}{\sqrt{g(\bm{x}_n})} \frac{\del}{\del x^i_n} \left[ \sqrt{g(\bm{x}_n)} \mc{O}(\bm{x}) \exp(\iu \mc{S}_K[\bm{x}]) \right]
    \\
    &= \Braket{\iu \mc{O}(\bm{x}) \frac{\del \mc{S}_K}{\del x^i_n} + \frac{\del \mc{O}(\bm{x})}{\del x^i_n} + \mc{O}(\bm{x}) \Gamma_i(\bm{x}_n)}_K\,,
\end{align}
where
\begin{align}
    \Gamma_i(\bm{x}) = \frac{1}{2}g^{jk}(\bm{x})\partial_i g_{jk}(\bm{x}) = \del_i\log\sqrt{g(\bm{x})}\,,
\end{align}
and $\braket{\bullet}_K$ denotes the expectation value of the path integral with $\exp(\iu \mc{S}_K)$ and $[\mc{D}\bm{x}]_K$.\footnote{In formal continuum notation, the functional derivative of the covariant measure produces a coincident-time contact term,
\begin{align}
    \frac{\delta}{\delta x^i(t)}
    \log\left[
        \prod_{t'}\sqrt{g(x(t'))}
    \right]
    =
    \delta(0)\,
    \Gamma_i(x(t))\,.
\end{align}
Here $\delta(0)$ is a regulator-dependent shorthand for the coincident-time singularity, and in a time-sliced regularization it is schematically of order $1/\Delta t$. 
This term is not an independent physical force. It combines with the corresponding contact terms from
the short-time kinetic kernel and with the ordering counterterms when the continuum limit is taken.}

We note that the term proportional to $\Gamma_i$ in the time-sliced identity should not be interpreted as an independent imaginary force or as a physical potential proportional to $-\iu\log\sqrt{g}$. In the continuum limit, this term combines with contact terms generated by the short-time kinetic kernel and with the ordering counterterms contained
in $\mc{S}_K[\bm{x}]$. 
With the same midpoint prescription used to construct the path integral, these contributions reconstruct the real and generally covariant equation associated with the Laplace--Beltrami operator in the system Hamiltonian. 
Thus, although $\sqrt{g}$ remains essential in the
functional measure, no separate imaginary potential
$-\I \log\sqrt{g}$ appears in the physical equation of motion.

The absence of an independent imaginary potential can also be verified or motivated in the wavefunction formalism.
The QRHD Schr\"odinger equation gives the following continuity equation of the probability
\begin{align}
    \del_t \rho + \frac{1}{\sqrt{g}} \del_i (\sqrt{g}j^i) = 2 a \eta \Im V \rho,
    \qquad 
    \rho =|\psi|^2\,,
    \qquad 
    j^i = \frac{1}{2\iu ma} g^{ij} (\psi^\ast \del_j \psi - \psi \del_j \psi^\ast)\,.
    \label{eq:prob-conti-eq}
\end{align}
For a real scalar potential, the right-hand side of the current divergence vanishes and the total probability is conserved, provided that the probability flux through the boundary vanishes. 
A genuine imaginary scalar potential would instead act as a local source or sink of probability. 
We derive this equation~\eqref{eq:prob-conti-eq} in App.~\ref{sec:conservation}.

Consequently, after consistently accounting for the time-slicing contact terms and ordering counterterms, the continuum SD equation can be expressed as
\begin{align}
    \Braket{\left[ \ddot{x}^i + \Gamma^i_{jk} (\bm{x}) \dot{x}^j \dot{x}^k\right]_{\mr{TS}} + \frac{\dot{a}}{a} \dot{x}^i + \frac{\eta}{m} (\grad_g \mc{V}(\bm{x}))^i } = 0\,,
    \label{eq:covariant-quantum-eom}
\end{align}
with the scalar potential including geometric corrections 
\begin{align}
    \mc{V}(\bm{x}) = V(\bm{x}) + \frac{1}{\eta a^2} (\Delta V(\bm{x}) + \Delta V'(\bm{x}))\,.
    \label{eq:mathcalV-eff}
\end{align}
Here $[\cdots]_{\mathrm{TS}}$ denotes the composite operator defined by the same time-slicing prescription as the discretized path
integral.
In particular, it includes the contact and ordering terms required for covariance.
In the following, we drop this notation for simplicity and consider the following quantum EOM\footnote{Strictly speaking, a physical equal-time expectation value
$\braket{\psi(t)|\hat{\mc{O}}|\psi(t)}$ is represented by a closed-time-path or Schwinger--Keldysh path integral~\cite{Schwinger:1960qe,Keldysh:1964ud} with the two branches matched at the final time.
In the main text, the branch labels are suppressed because the local time-sliced SD equation has the same form on each branch.
The resulting closed-time-path identity is equivalent to the Heisenberg expectation-value equation.}
\begin{align}
    \Braket{\ddot{x}^i + \Gamma^i_{jk} (\bm{x}) \dot{x}^j \dot{x}^k + \frac{\dot{a}}{a} \dot{x}^i + \frac{\eta}{m} (\grad_g \mc{V}(\bm{x}))^i} = 0\,.
    \label{eq:QRHD-SD-equation}
\end{align}
The counterterms in $\mc{V}$ are representation-dependent terms and are not separately covariant scalar potentials. 
Eqs.~\eqref{eq:covariant-quantum-eom}--\eqref{eq:QRHD-SD-equation} are to be interpreted with the same time-slicing prescription, including the associated contact terms, which implies that the covariance pertains to the complete operator identity rather than to each term considered separately.

\paragraph{$a(t)$ and semiclassical approximation.}

Before the discussion of the convergence time in QRHD, we give comments on the relation between the time-dependent coefficient of the kinetic term $a(t)$ and the semiclassical approximation.
$a(t)$ is treated as an overall factor in the action of Q(R)HD, which is written as\footnote{
Here we write the form of the action schematically.
$\mc{V}(\bm{x})$ is assumed to contain the quantum corrections from the geometry like Eq.~\eqref{eq:Delta_mathcalV} if needed.
} 
\begin{align}
	S[\bm{x}] = \int \dd t a(t) \biggl(
		\frac{m}{2} g_{ij} \dot{x}^i \dot{x}^j - \eta(t) \mc{V}(\bm{x})
	\biggr)\,.
\end{align}
Then the partition function becomes with $\hbar$
\begin{align}
	Z = \int \mc{D} \bm{x} \exp \frac{\iu}{\hbar} S[\bm{x}]
	= \int \mc{D} \bm{x} \exp \iu \int \dd t \frac{a(t)}{\hbar} \biggl(
        \frac{m}{2}
		g_{ij} \dot{x}^i \dot{x}^j - \eta(t) \mc{V}(\bm{x})
	\biggr)
	\,,
\end{align}
which shows that $1/\hbar$ and $a(t)$ appear with the same scaling, and $a(t) \to \infty$ corresponds to taking the semiclassical limit $\hbar \to 0$ through the time scheduling of this coefficient $a(t)$.
This correspondence is also seen by writing the Hamiltonian operator in Q(R)HD as 
\begin{align}
	\hat{H}(t) = \frac{1}{a(t)}  \biggl( 
		- \frac{\Delta_g}{2m}
	\biggr) + a(t) \eta(t) V(\hat{\bm{x}}) 
	= a(t) \Biggl[
		\frac{1}{a(t)^2} \biggl(
			\frac{-\Delta_g}{2m}
		\biggr) + \eta(t) V(\hat{\bm{x}})
	\Biggr]\,,
\end{align}
which is regarded as $\hat{H} / \hbar$ if we restore $\hbar$.
From this correspondence, we can understand the $a(t)$ dependence in Eq.~\eqref{eq:mathcalV-eff}.
It is known that the quantum corrections of $\Delta V$ and $\Delta V'$ are $\hbar^2$ corrections and these scalings are same with those of $1/a$ in Eq.~\eqref{eq:mathcalV-eff}.

This relation between $a(t)$ and $1/\hbar$ tells us some intuitive understandings of quantum effects in QHD and QRHD.
In the SD equation, the potential in the natural gradient is replaced with $V_\eff(\bm{x})$, which corrections arise from quantum effects in path integral measure and operator ordering of the kinetic term.
In terms of the powers of $a(t)$, these are suppressed as shown in \eqref{eq:mathcalV-eff}, which powers are same with those of $1/\hbar$ in a standard path integral calculation.
Then, we find that quantum effects are suppressed at late time in QHD and QRHD due to the behavior of $a$.
In this sense, the Q(R)HD algorithm is designed so that quantum effects are significant at early time, while at late time the contributions from the classical potential $V(\bm{x})$ become dominant.
We will demonstrate this behavior in the numerical simulations presented in the follows.

\paragraph{Estimation of convergence time in QRHD.}

We estimate the time $t_\ast$ required for QRHD to find an optimal solution from the time-evolution equation for the parameters \eqref{eq:QRHD-SD-equation}.
As in Sec.~\ref{sec:convergence-QHD}, we consider the case $a(t) = \ee^{2 \gamma t}$ and $\eta (t) = \eta = \mr{const.} \geq 0$.
In this case, the time-evolution equation becomes 
\begin{align}
	\Braket{ \ddot{x}^i + \Gamma^i_{jk}(\bm{x}) \dot{x}^j \dot{x}^k + 2 \gamma \dot{x}^i
	+ \frac{\eta}{m} g^{ij}(\bm{x}) \del_j V_{\mr{eff}}(\bm{x})} = 0\,.
	\label{eq:eom-QRHD-convex-approx}
\end{align}
The effective potential $V_\eff(\bm{x})$ contains the quantum corrections to the classical potential $V(\bm{x})$ as Eq.~\eqref{eq:mathcalV-eff}.
However, as discussed above, these quantum corrections are proportional to the higher powers of suppressed factor $1/a(t) = \ee^{-2 \gamma t}$, and their corrections are expected to be small near the optimal point, which means that the convergence behavior in late time would be mainly controlled by the classical potential $V(\bm{x})$.
Then, we drop these quantum corrections and approximate $V_\eff(\bm{x}) \approx V(\bm{x})$ in the following analysis.
The effects of the quantum corrections on the QRHD dynamics are checked in the numerical calculations in Sec.~\ref{sec:numerical-simulation} and App.~\ref{app:convergence-time}.

As in the similar manner to the case of QHD, we expand the potential $V(\bm{x})$ around the local optimal point $\bm{x} = \bm{x}_\ast$, which satisfies that $\del_i V (\bm{x}_\ast) = 0$ and the Hessian matrix is positive definite.
Then, the expansion of the classical potential is given by 
\begin{align}
	V(\bm{x}) = V(\bm{x}_\ast) + \frac{1}{2} (\bm{x} - \bm{x}_\ast)^\mr{t}\, \Hess_g V_\ast\, (\bm{x} - \bm{x}_\ast) + \mc{O}(|\bm{x} - \bm{x}_\ast|_g^3)\,,
\end{align}
where $|\bullet|_g$ denotes the norm on the Riemann manifold induced by the metric $g_{ij}$ such as $|\bm{x}|^2_g = g_{ij}(\bm{x}) x^i x^j$.
$\Hess_g V$ is the Hessian matrix on the Riemann manifold, which components are given by 
\begin{align}
	(\Hess_g V)_{ij} \coloneqq \frac{\del^2 V(\bm{x})}{\del x^i \del x^j} - \Gamma^k_{ij} \frac{\del V(\bm{x})}{\del x^k}\,,
\end{align}
which is a covariant form of the second order derivatives.
$\Hess_g V_\ast$ is the Hessian matrix evaluated at $\bm{x} = \bm{x}_\ast$.
Considering up to the quadratic terms in the Lagrangian or Hamiltonian (operator), the EOM reduces to 
\begin{align}
	\braket{\ddot{x}^i} + \braket{\Gamma^i_{jk}(\bm{x}) \dot{x}^j \dot{x}^k} + 2 \gamma \braket{\dot{x}^i} + \frac{\eta}{m} \braket{g^{ij}(\bm{x}) (\Hess_g V_\ast)_{jk} (x^k - x^k_\ast)} = 0\,.
	\label{eq:eom-QRHD-convex-pre}
\end{align}

This EOM is still nonlinear because it contains $\Gamma^i_{jk}(\bm{x})$ in the acceleration term and $g^{ij}(\bm{x})$ in the natural gradient.
To simplify the this equation, we put a physically natural assumption that $\bm{x} - \bm{x}_\ast$ is approximated as a locally flat coordinate system.
This is systematically treated by introducing a normal coordinate for $\bm{x} - \bm{x}_\ast$,\footnote{The applications of the normal coordinate to the theoretical physics have been discussed \cite{DeWitt:1964mxt,DeWitt:1967ub,Gilkey:1975iq,Alvarez-Gaume:1981exa,Mukhi:1985vy}.} which leads to 
\begin{align}
	\Gamma^i_{jk}(\bm{x}_\ast) = 0\,,
\end{align}
and the expansions of $\Gamma^i_{jk}(\bm{x})$ and $g^{ij}(\bm{x})$ are given by 
\begin{align}
	\Gamma^i_{jk}(\bm{x}) = \mc{O}(|\bm{x} - \bm{x}_\ast|_g)\,,
	\qquad 
	g^{ij}(\bm{x}) = g^{ij}(\bm{x}_\ast) + \mc{O}(|\bm{x}- \bm{x}_\ast|^2_g)\,.
\end{align}
Under these assumptions and taking just leading term into the account, we obtain
\begin{align}
	\braket{\ddot{x}^i} + 2 \gamma \braket{\dot{x}^i} + \frac{\eta}{m} g^{ij}(\bm{x}_\ast) (\Hess_g V_\ast)_{jk} (\braket{x}^k - x_\ast^k) =0\,,
	\label{eq:eom-QRHD-convex}
\end{align}
which is also regarded as the EOM of a damped oscillator like Eq.~\eqref{eq:eom-QHD-convex} in QHD.
Here, we neglect the quantum correlations among the parameters and assume $\braket{x^i x^j} = \braket{x^i} \braket{x^j}$.
The validity of this approximation will be examined in the following numerical simulations.
Therefore, following the same analysis with that deriving \eqref{eq:t_star_bound_QHD} in QHD, we obtain a lower bound on the convergence time $t_\ast$ as 
\begin{align}
	t_\ast \gtrsim \frac{1}{\sqrt{ \frac{\eta}{m} \lambda_{\mr{min}}(g^{-1} \Hess_g V_\ast)}} \left[
		- W_{-1} \left( -\frac{\epsilon_\ast}{\ee} \right)-1
	\right]\,,
	\label{eq:t_star_bound}
\end{align}
and the equality is achieved when 
\begin{align}
	\gamma \sim \sqrt{\frac{\eta}{m}\lambda_{\mr{min}}
(g^{-1}\Hess_g V_\ast)}\,.
\end{align}
The convergence time $t_\ast$ is introduced as the time when the condition $|\braket{{\bm{x}}(t_\ast)}-{\bm{x}}_\ast|_{g_\ast} \leq \epsilon_\ast|\braket{{\bm{x}}(0)}-{\bm{x}}_\ast|_{g_\ast}$ is first satisfied for a given small value $\epsilon_\ast$ as in the QHD case.
The components of $g^{-1} \Hess_g V$ are given by 
\begin{align}
	(g^{-1} \Hess_g V)^i{}_j = g^{ik} (\Hess_g V)_{kj}\,,
\end{align}
and it is known that the eigenvalues of $g^{-1}\Hess_g V_\ast$ do not depend on the detailed choice of the coordinate system at the stationary point $\del_i V=0$.
Thus, under the approximations that quantum correlations among parameters and quantum corrections to the potential $V({\bm{x}})$ are neglected, we estimate the lower bound \eqref{eq:t_star_bound} on the local convergence time $t_\ast$ of QRHD to an optimal point.

\paragraph{Comments on quantum corrections and quantum correlations.}

Some quantum corrections appear in the quantum EOMs but we expect these effects to be small due to the behavior of $a(t)$.
In order to check this assumpution, we numerically solve the differential equation \eqref{eq:eom-v} for several examples and confirm that the estimated lower bound on $t_\ast$ satisfies Eq.~\eqref{eq:t_star_bound} in App.~\ref{app:convergence-time}.
This is consistent with our observation that the quantum corrections to the potential contribute with a suppression factor $1/a$, so while they may affect the early-stage dynamics, their influence on the final convergence time can be regarded as small.

Before ending this subsection, we give a comment on the effect of quantum correlations among parameters.
In QRHD the parameter evolution involves nontrivial quantum correlations that depend on the geometry of the parameter space and on the choice of coordinates through the nonlinear terms in the equation.
To analyze this effect, one needs to solve the Schr\"odinger equation with the Hamiltonian operator \eqref{eq:QRHD}, which requires an analysis beyond the above approximation.
In order to see whether QRHD works well as a quantum optimization algorithm, we numerically solve the QRHD Schr\"odinger equation for simple examples and check the behaviors of the solutions in the next subsection.

\subsection{Numerical simulation of quantum dynamics}
\label{sec:numerical-simulation}

In this subsection, we numerically solve the time evolution of the wave function in QRHD for two examples,\footnote{ 
    The methodology for solving the Schr\"odinger equation with the time-dependent Hamiltonian operator~\eqref{eq:QRHD} is summarized in App.~\ref{app:simulation}.
    }
and confirm that QRHD works as an optimization algorithm.
For these examples, we consider the cases where the parameter space geometries are flat or curved.

\subsubsection{Flat case}
\label{sec:numeric-flat}

The first case is the flat parameter space.
We take $V(\bm{x})$ to be a quadratic convex function and choose $g_{ij}(\bm{x})$ to be the Hessian matrix of $V(\bm{x})$, which is globally constant matrix.
It is found that $\braket{{\bm{x}}}$ follows an equation similar to the classical equation of motion, up to a shear transformation of the potential.

\begin{figure}[t]
  \centering
  \includegraphics[width=16cm]{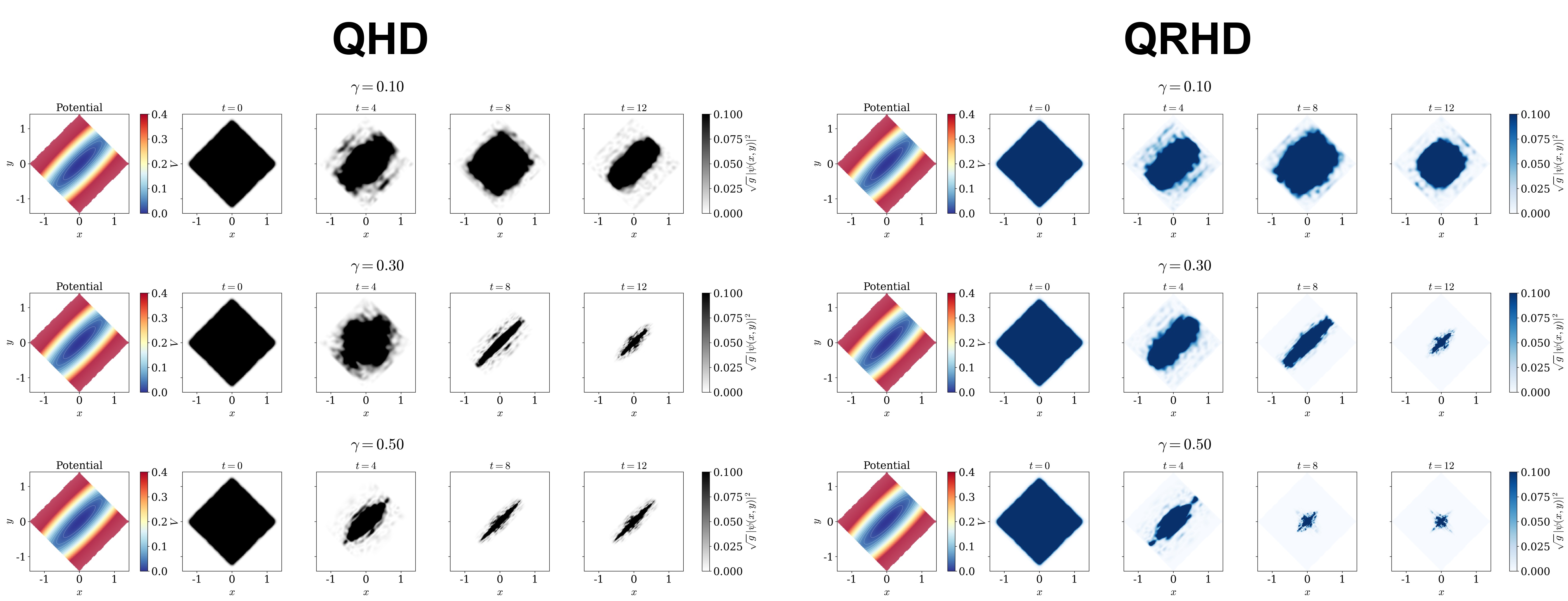}
  \caption{
   Time evolution of the wave functions in the flat space case ($N=2$).
   The left panels show results of the standard QHD ($g_{ij} = \delta_{ij}$), and the right panels show those of the QRHD ($g_{ij} = (A_1)_{ij}$).
   The friction coefficient varies as $\gamma = 0.1, 0.3,$ and $0.5$ from top to bottom.
   In both cases, insufficient friction ($\gamma=0.1$) causes under-damped oscillations.
   Notably, QRHD contracts the wave function along both principal directions of $A_1$, whereas QHD leaves it elongated along the soft direction $x^1 + x^2$.
  }
  \label{fig:flat-quad-wavefunc}
\end{figure}

\begin{figure}[t]
  \centering
  \includegraphics[width=16cm]{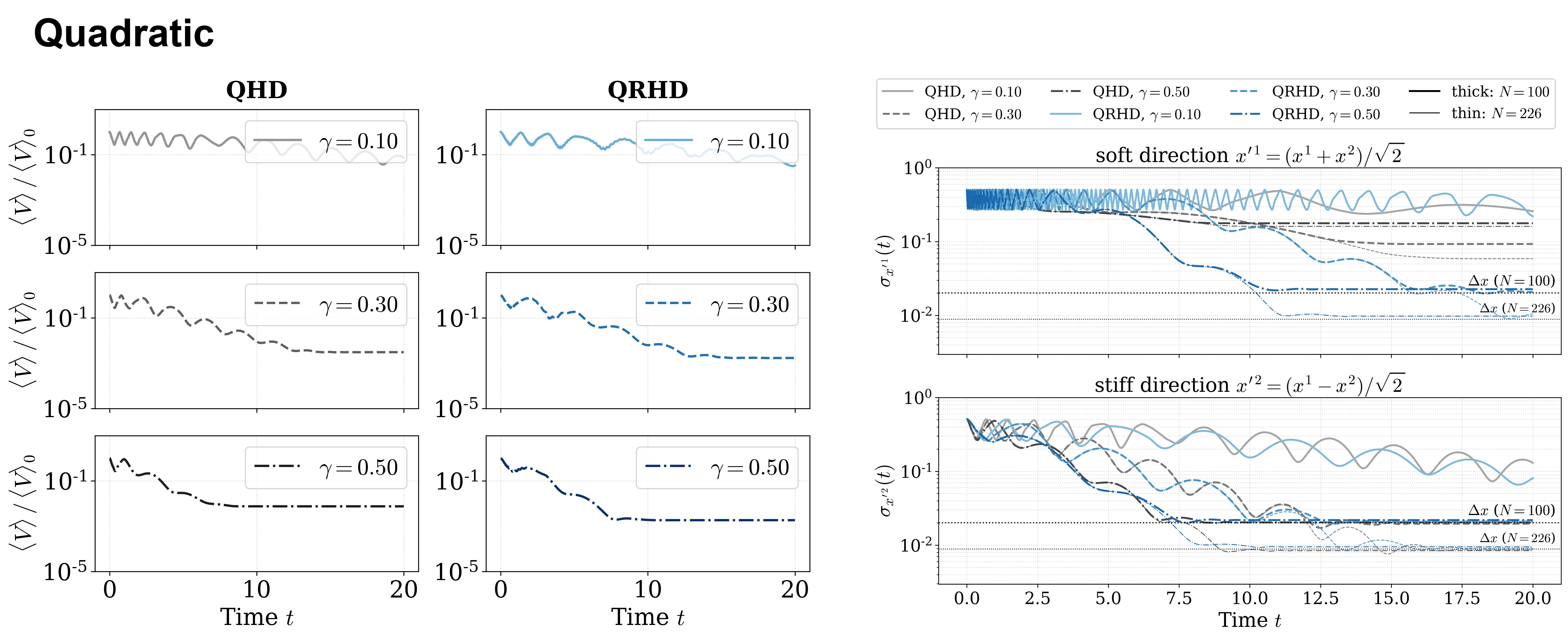}
  \caption{
   Time evolution of the cost function and of the widths of the wave function.
   The left panels plot the expectation value $\braket{V(t)}/\braket{V(0)}$ for QHD (black) and QRHD (blue), and the right panels plot the widths $\sigma_{x'^1}$ and $\sigma_{x'^2}$ \eqref{eq:def-width} along the soft and stiff principal axes of $A_1$.
   QRHD contracts the wave function along both axes at the same rate, whereas in QHD the width along the soft direction decreases more slowly.
   The late-time plateaus of the widths and of $\braket{V}$ depend on the grid and are not used in the comparison as discussed in App.~\ref{app:accuracy}.
  }
  \label{fig:flat-quad-costfunc}
\end{figure}

For $N=2$, we consider the following potential
\begin{align}
    V(\bm{x}) = \frac{m}{2} \bm{x}^\tt A_1 \bm{x}\,,
    \qquad
    A_1 =
    \begin{pmatrix}
    1 & -0.9 \\
    -0.9 & 1
    \end{pmatrix}\,,
    \label{eq:potential-flatcase}
\end{align}
and we set $m =\eta=1$ and $a(t)=\ee^{2\gamma t}$.
The numerical results of the time evolution in this system are displayed in Figs.~\ref{fig:flat-quad-wavefunc} and \ref{fig:flat-quad-costfunc}.
Fig.~\ref{fig:flat-quad-wavefunc} shows the time evolution of the wave functions for the QHD case with $g_{ij}(\bm{x}) = \delta_{ij}$ in the left panel and the QRHD case with $g_{ij}(\bm{x}) = (A_1)_{ij}$ in the right panel under various friction coefficients, $\gamma = 0.1, 0.3,$ and $0.5$ from top to bottom.
In both cases, the convergence behavior highly depends on the value of $\gamma$.
When $\gamma$ is too small, the insufficient friction prevents the wave function from settling down into the minimum.
Focusing on the convergence capability, one can clearly observe that the wave function in QRHD contracts toward the global minimum along both principal directions of $A_1$, whereas that in QHD remains elongated along the soft direction $x^1 + x^2$ (eigenvalue $0.1$ of $A_1$).

This difference is further elucidated by the behavior of the cost function shown in Fig.~\ref{fig:flat-quad-costfunc}, which plots the expectation value of the potential $V(\bm{x})$ and the widths of the wave function for both algorithms.
Here, the widths are evaluated as
\begin{align}
    \sigma_{x'^a}^2 = \braket{(x'^a)^2} - \braket{x'^a}^2\,,
    \qquad
    x'^1 = \frac{x^1 + x^2}{\sqrt{2}}\,,
    \quad
    x'^2 = \frac{x^1 - x^2}{\sqrt{2}}\,.
    \label{eq:def-width}
\end{align}
As shown in the right panels of Fig.~\ref{fig:flat-quad-costfunc}, in QRHD both widths decrease at the same rate, whereas in QHD the width along the soft direction decreases much more slowly.
This behavior reflects the fact that the metric $g = A_1$ equalizes the frequencies of the two normal modes, so that a single friction coefficient damps both, whereas in QHD the soft mode is damped slowly for the values of $\gamma$ that damp the stiff mode (Sec.~\ref{sec:QRHD}).
In other words, the potential is shear-transformed and the condition number of the effective Hessian matrix becomes smaller in QRHD.
Since $g = A_1$ is constant, this effect is attributable to preconditioning rather than to intrinsic curvature.
We discuss this point later in Sec.~\ref{sec:complexity-flat}.

\begin{figure}[t]
  \centering
  \includegraphics[width=16cm]{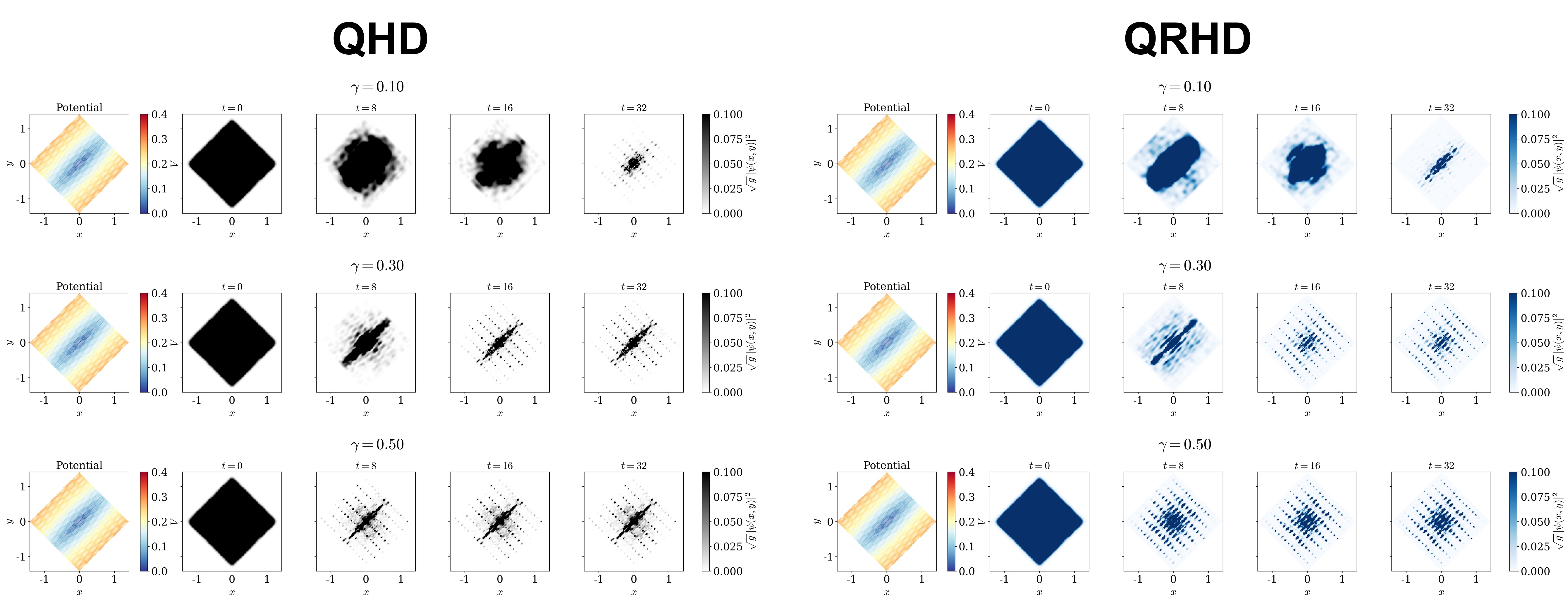}
  \caption{
   Time evolution of the wave functions in the non-convex space case with the Ackley function.
   The left panels show the results of standard QHD, and the right panels show those of QRHD.
   The friction coefficient varies as $\gamma = 0.1, 0.3,$ and $0.5$ from top to bottom.
  }
  \label{fig:flat-ackley-wavefunc}
\end{figure}

\begin{figure}[t]
  \centering
  \includegraphics[width=16cm]{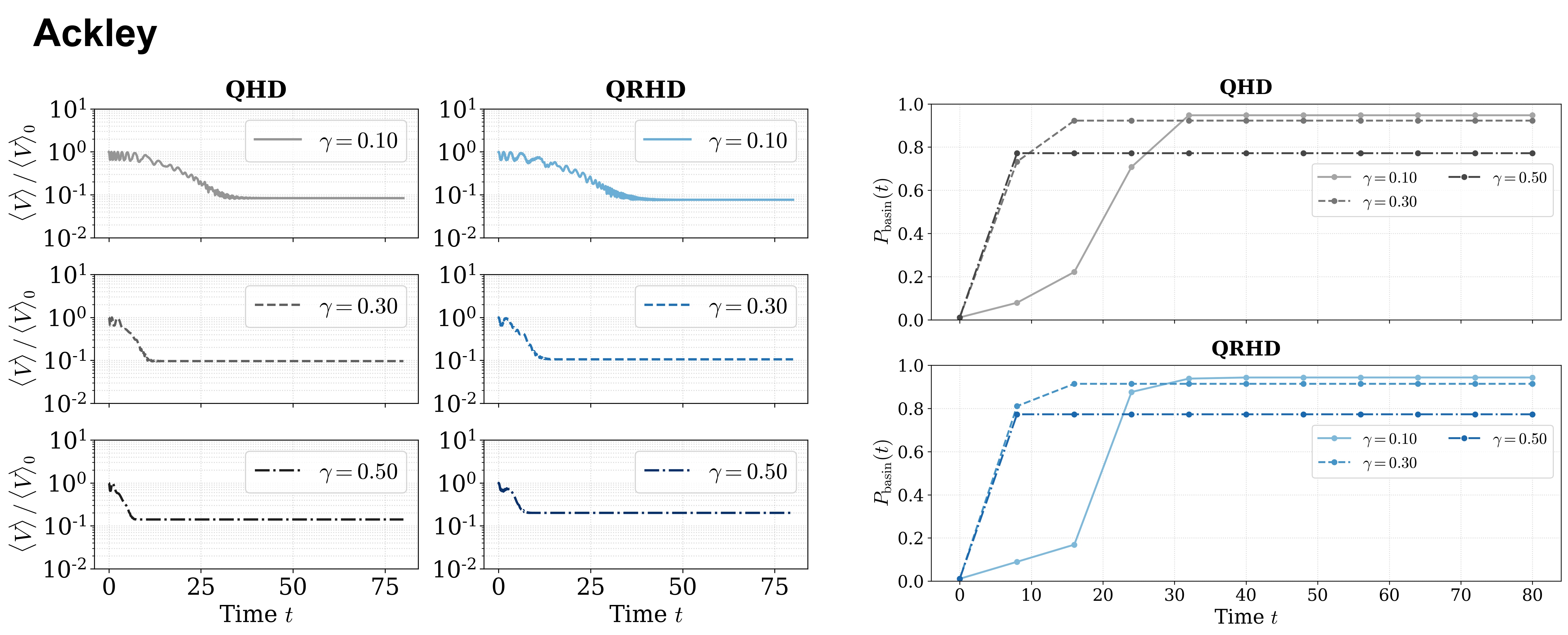}
  \caption{
   Time evolution of the cost function and of the basin probability (Ackley function).
   The left panels plot the expectation value $\braket{V(t)}/\braket{V(0)}$ for QHD (black) and QRHD (blue), and the right panels plot the probability $P_{\rm basin}(t)$ of finding the system defined by \eqref{eq:def-pbasin} in the basin of the global minimum for QHD (top) and QRHD (bottom).
   The two methods give almost the same $P_{\rm basin}$ for every $\gamma$.
  }
  \label{fig:flat-ackley-costfunc}
\end{figure}

In order to verify the efficacy of QRHD, we perform a similar analysis on a non-convex potential problem. 
Specifically, we consider a two-dimensional Ackley function that is shear-transformed via the matrix $A_1^{1/2}$ given by Eq.~\eqref{eq:potential-ackley} in App.~\ref{app:setup}, and the corresponding numerical results are displayed in Figs.~\ref{fig:flat-ackley-wavefunc} and \ref{fig:flat-ackley-costfunc}.
Fig.~\ref{fig:flat-ackley-costfunc} shows the evolution of the cost-function expectation value and the basin probability $P_{\mr{basin}}(t)$ defined in Eq.~\eqref{eq:def-pbasin}. 
As discussed in App.~\ref{app:accuracy}, the late-time value of $\braket{V}$ is sensitive to the grid resolution because of the narrow local wells. 
We therefore also use $P_{\mr{basin}}$ to compare the convergence behavior of QHD and QRHD.
As in the previous convex case, the convergence behavior depends on the choice of the friction coefficient $\gamma$, with $\gamma=0.1$ giving the most favorable finite-time convergence among the values tested in this non-convex landscape.
When $\gamma$ exceeds this value, the convergence capability significantly degrades.
The deterioration at larger $\gamma$ is consistent with stronger damping suppressing quantum transport across the potential landscape, including tunneling contributions. This behavior is observed in both QHD and QRHD.
This is a characteristic feature common to both the QHD and QRHD frameworks.
Regarding the convergence efficiency, QRHD performs on a par with standard QHD: the values of $P_{\rm basin}$ of the two methods agree to within $0.03$ for all $\gamma$ shown in App.~\ref{app:accuracy}, and $\braket{V}$ decreases at a similar rate.
This result implies that even for non-convex optimization problems, QRHD retains the ability of QHD to escape from local minima by tunneling, while the metric by itself does not enhance it.
We finally remark that the favorable values of $\gamma$ observed in these examples are not universal. These are empirically favorable values for the examples studied. 

\subsubsection{Curved case}
\label{sec:numeric-curved}

The second case is that the parameter space is not flat and has nontrivial geometry.
In addition to the optimization of the {differentiable} function in $N$-dimensional flat space, we impose the constraint $\bm{x}^2 = R^2$ on the parameters, where $R$ is a positive real number.
This constraint makes the parameter space be the $(N-1)$-dimensional sphere of the radius $R$.
{For the case with $V({\bm{x}})$ being a quadratic potential, the corresponding constrained problem is so-called Rayleigh quotient minimization problem.}
QRHD provides a quantum approach to solve this kind of problem dynamically by setting $g_{ij}(\bm{x})$ to be the metric on the sphere.
We should note that $g_{ij}(\bm{x})$ depends on the choice of coordinates and a selection of an \emph{appropriate coordinate system} can affect the performance of the present algorithm.

\begin{figure}[t]
  \centering
  \includegraphics[width=15cm]{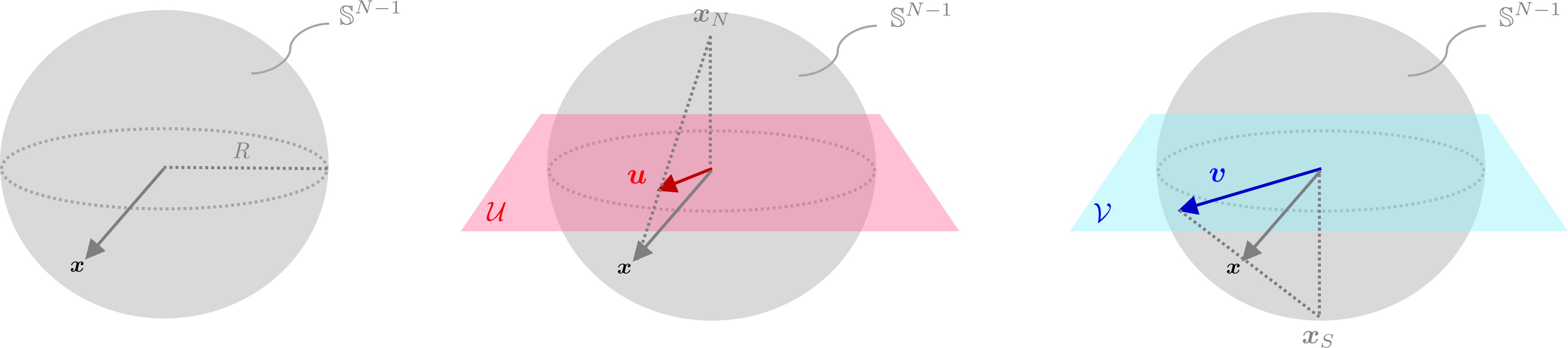}
  \caption{
	A schematic illustration of a conformally flat coordinate systems on the sphere.
	(Left) The $N$-dimensional vector $\bm{x}$ constrained to line on the sphere $\mbb{S}^{N-1}$ by $\bm{x}^2 = R^2$.
	(Middle) The correspondence between $\bm{x}$ and the stereographic projection $\bm{u}$ from the north pole $\bm{x}_N$.
	$\bm{u}$ denotes the $(N-1)$-dimensional coordinates.
	(Right) The correspondence between $\bm{x}$ and the stereographic projection $\bm{v}$ from the south pole $\bm{x}_S$.
	$\bm{v}$ denotes the $(N-1)$-dimensional coordinates.
  }
  \label{fig:coordinate-sphere}
\end{figure}

A standard choice for the sphere would be spherical coordinates, but it contains points where a component of metric vanishes, which can lead to the divergences of operators such as the Laplace--Beltrami operator.
In order to avoid the divergent issues, we consider coordinate systems which does not include coordinate singularities.
These singularities are properties of the coordinate system, not of the sphere, whose intrinsic curvature is constant and finite everywhere; the stereographic charts used below are singular only at the projection pole, which lies outside the computational domain.
In this work, we use two conformally flat coordinate systems of dimension one less than the ambient space.
A schematic illustration is shown in Fig.~\ref{fig:coordinate-sphere}, which shows that any point on the sphere is specified by the stereographic projections $\bm{u}$ or $\bm{v}$ from the north pole $\bm{x}_N$ and the south pole $\bm{x}_S$, respectively.
If we take the north and south poles in $N$-th direction, the coordinates are related by 
\begin{align}
	u^i = \frac{x^i}{1 - x^N/R}\,, 
	\qquad 
	v^i = \frac{x^i}{1 + x^N/R}\,,
    \label{eq:uv-stereographic}
\end{align}
for $i = 1,\ldots, N-1$ and their metric tensors have the following conformally flat forms as
\begin{align}
	g_{ij}(\bm{u}) = \biggl( \frac{2}{1 + \bm{u}^2/R^2} \biggr)^2 \delta_{ij}\,,
	\qquad 
	g_{ij}(\bm{v}) = \biggl( \frac{2}{1 + \bm{v}^2/R^2} \biggr)^2 \delta_{ij}\,,
    \label{eq:uv-metric-onsphere}
\end{align}
where $\bm{u}^2 = (u^i)^2$ and $\bm{v}^2 = (v^i)^2$.
The domains $(\mc{U}, \mc{V})$ of the conformally flat coordinates $({\bm{u}},{\bm{v}})$ on $\mathbb{R}^{N-1}$ depend on the choice of the atlas on $\mathbb{S}^{N-1}$, which involves a certain degree of arbitrariness.
As will be demonstrated later, this choice directly affects the computational complexity of the resulting algorithm.
Some formulae and calculations for the conformally flat coordinates of $\mbb{S}^{N-1}$ are summarized in App.~\ref{sec:coordinate-SN-1}.
To apply QRHD to this problem, we can run the quantum time evolution in parallel in the two coordinate systems (${\bm{u}},{\bm{v}}$) and compare the output results of the cost-function values via classical postprocessing.

\begin{figure}[t]
  \centering
  \includegraphics[width=16cm]{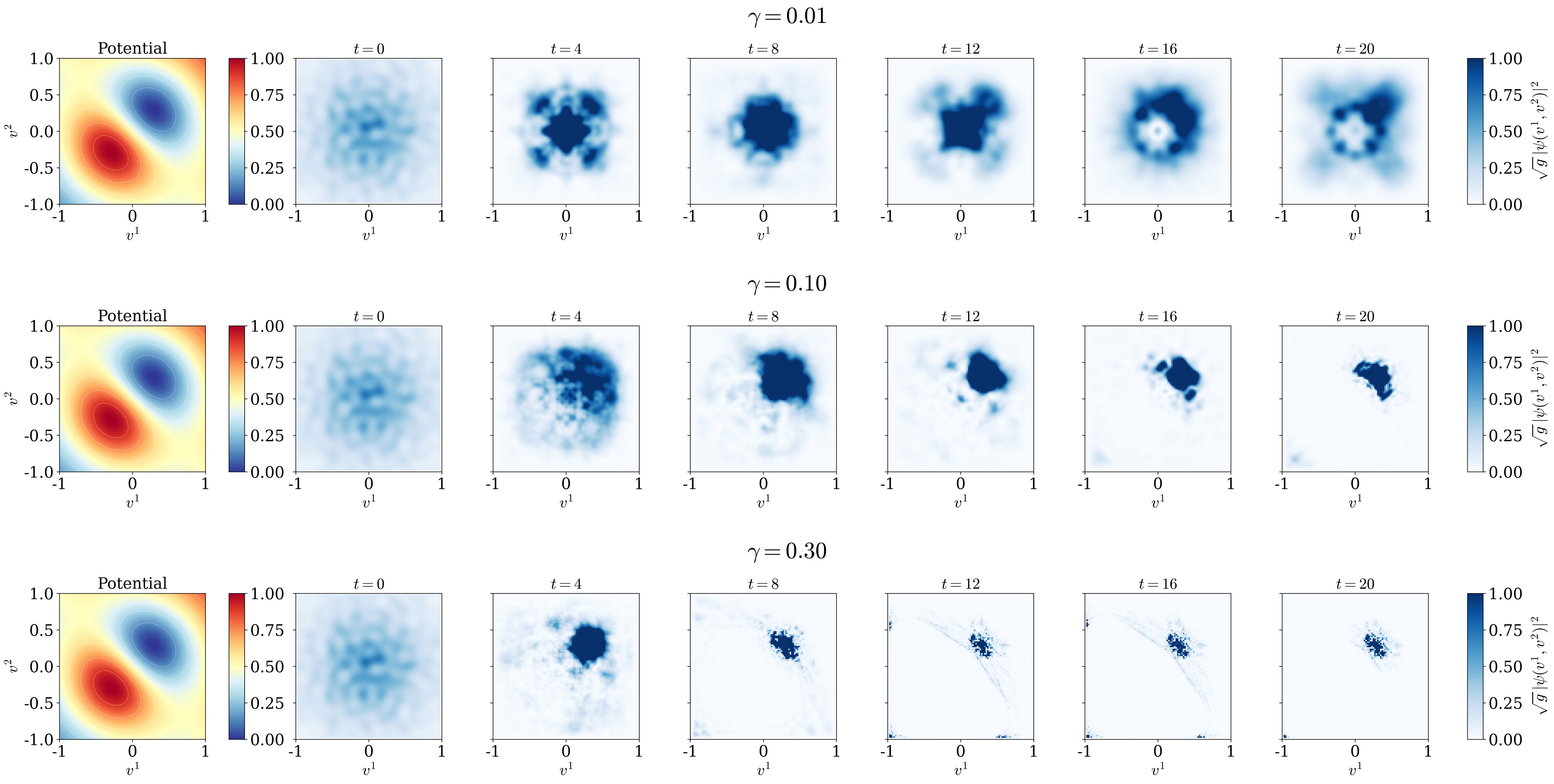}
  \\
  \vspace{1cm}
  \includegraphics[width=16cm]{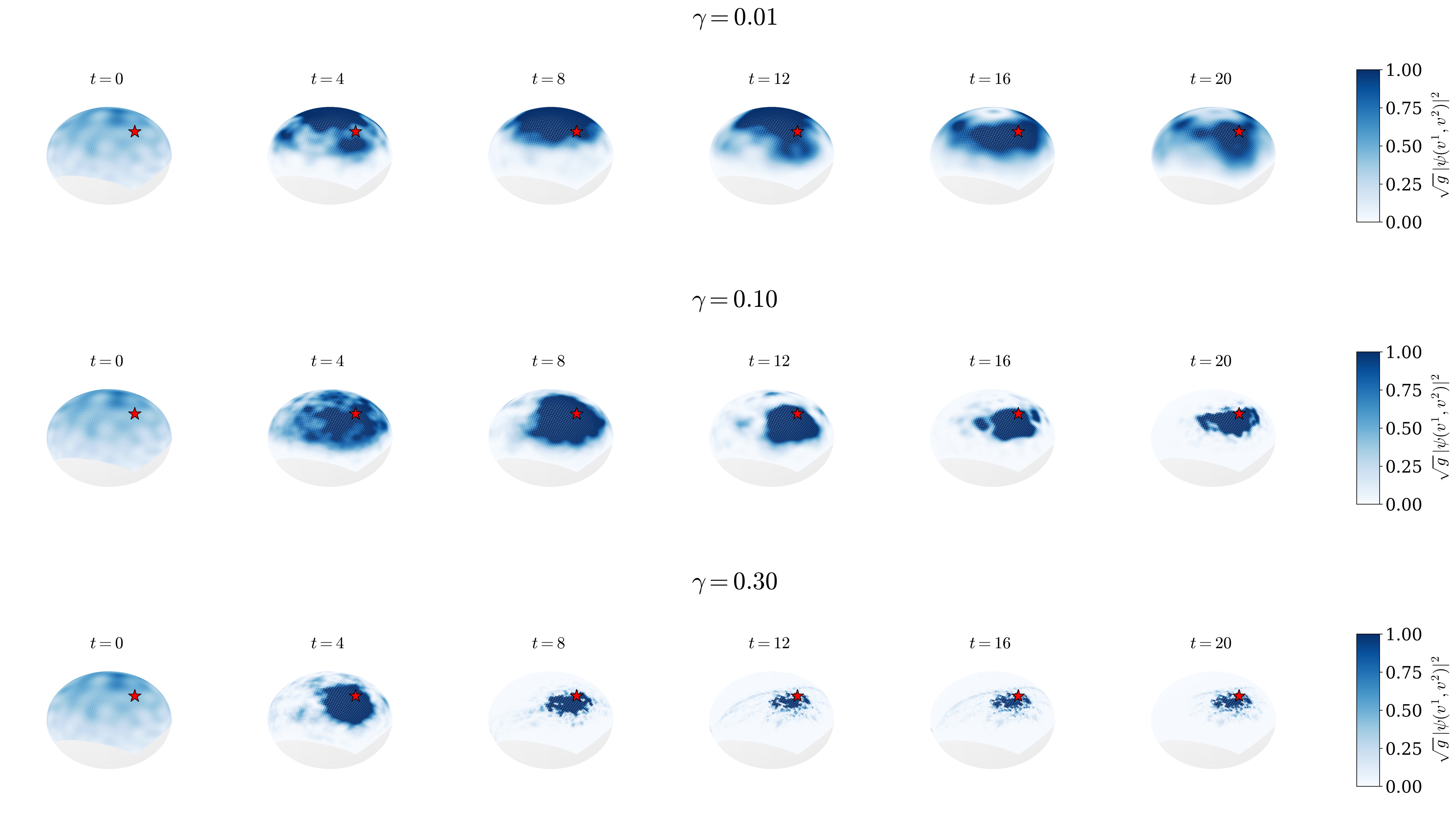}
  \caption{
  	Time evolution of the wave functions on the curved parameter space ($N=3$).
    The top panels show the density $\sqrt{g}\,|\psi(\bm{v}, t)|^2$ in the conformally-flat coordinates ($\bm{v}$), and the bottom panels show the same density mapped onto the sphere embedded in the flat space through $\bm{x} = \bm{x}(\bm{v})$ in Eq.~\eqref{eq:uv-stereographic}.
    The red star denotes the global minimum on the sphere.
    From top to bottom, the friction coefficient is varied as $\gamma = 0.01, 0.1,$ and $0.3$.
    We find that insufficient friction ($\gamma = 0.01$) fails to drive convergence, whereas an appropriately tuned friction ($\gamma = 0.3$) leads the wave function to efficiently settle into the global minimum.
    }
  \label{fig:curved-wavefunc}
\end{figure}

\begin{figure}[t]
  \centering
  \includegraphics[width=16cm]{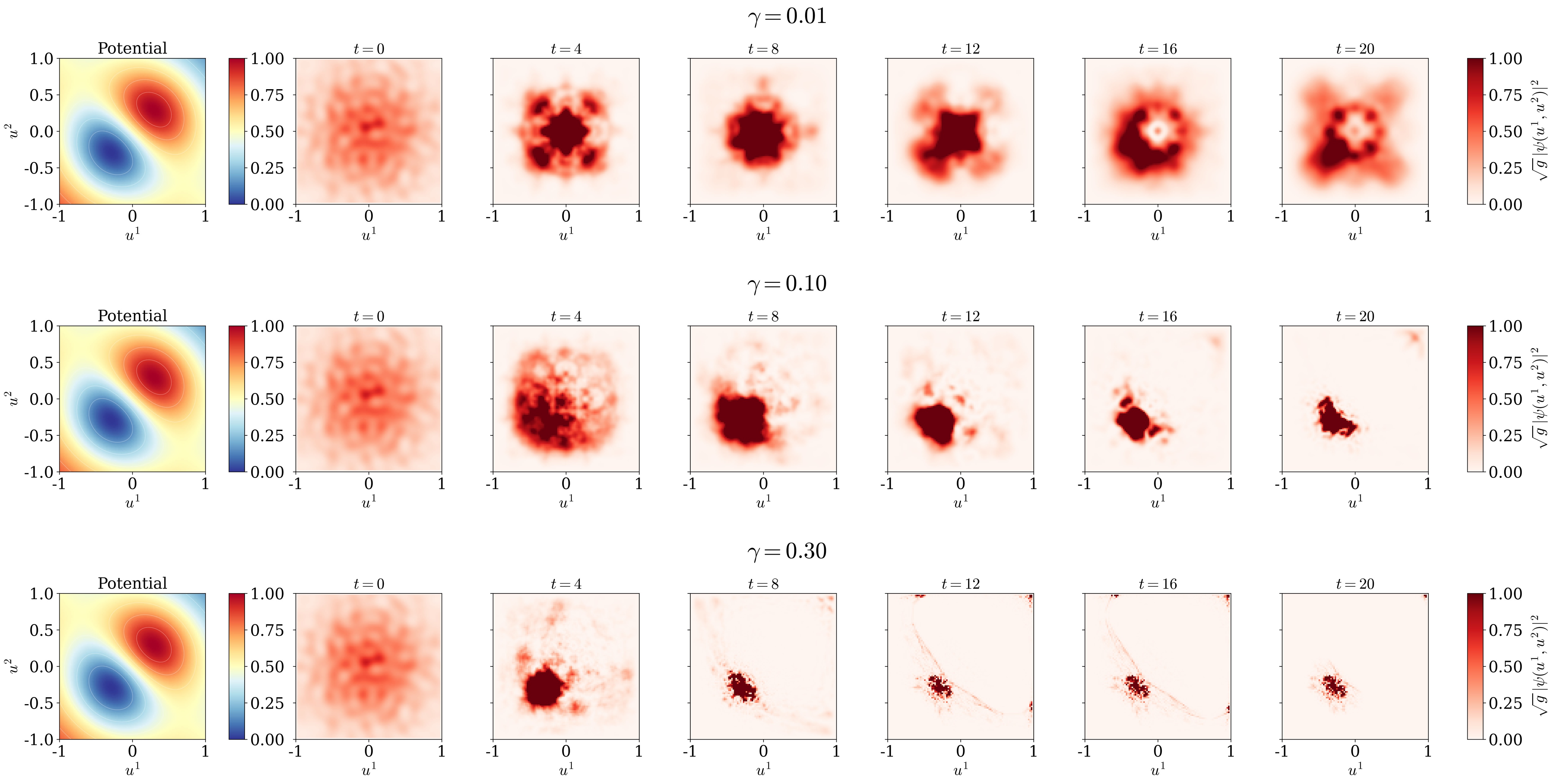}
  \\
  \vspace{1cm}
  \includegraphics[width=16cm]{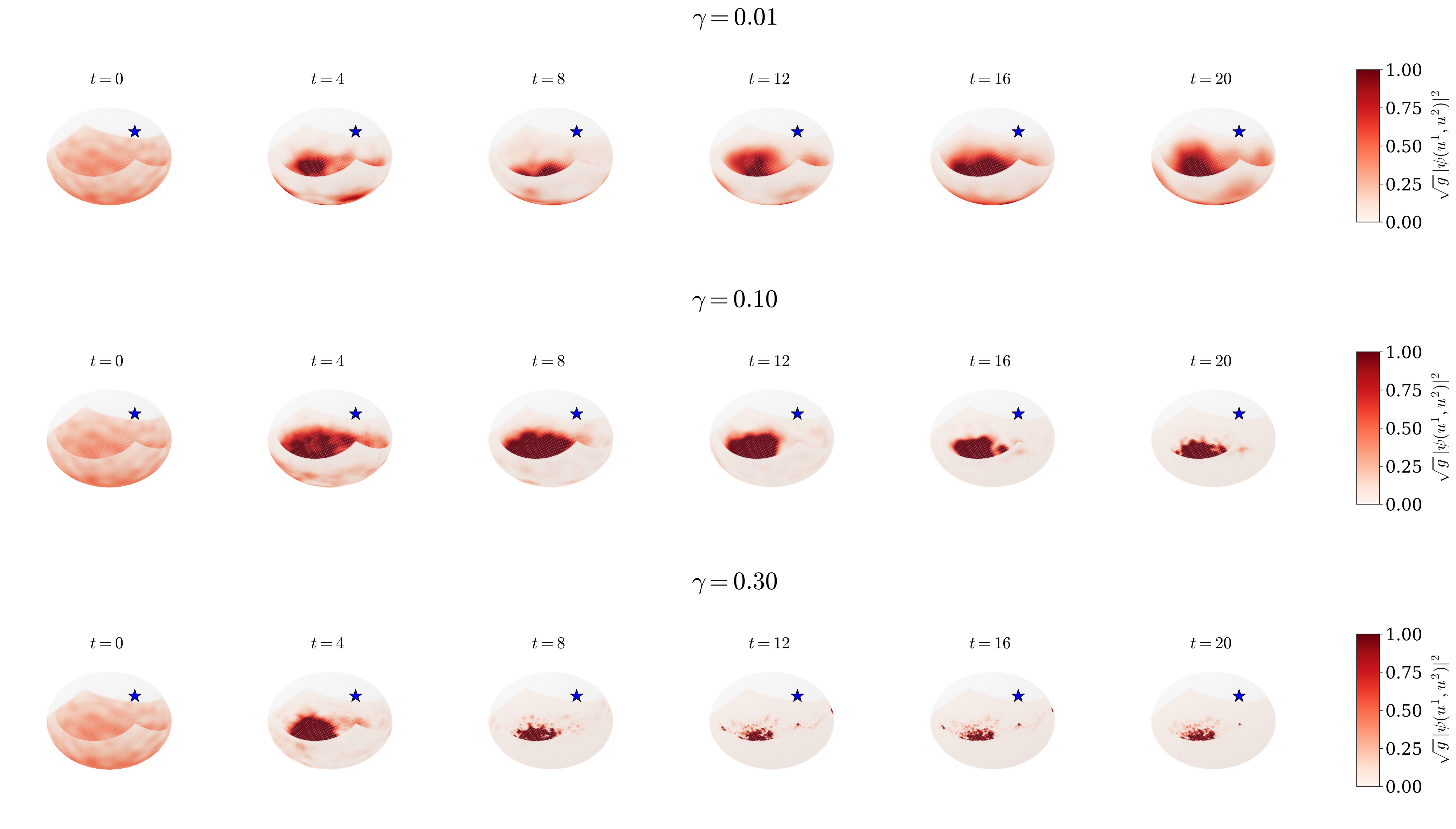}
  \caption{
  	Time evolution of the wave functions on the curved parameter space ($N=3$) in the other chart.
    The top panels show the density in the conformally-flat coordinates ($\bm{u}$), and the bottom panels show the same density mapped onto the sphere embedded in the flat space, as in Fig.~\ref{fig:curved-wavefunc}.
    The blue star denotes the global minimum $\bm{x}_\ast$ of Fig.~\ref{fig:curved-wavefunc}.
    In this chart the wave function converges to the antipodal point $-\bm{x}_\ast$, which is also a global minimum since $V(-\bm{x}) = V(\bm{x})$.
    From top to bottom, the friction coefficient is varied as $\gamma = 0.01, 0.1,$ and $0.3$. 
    }
  \label{fig:curved-wavefunc-2}
\end{figure}

\begin{figure}[t]
    \centering
    \includegraphics[width=16cm]{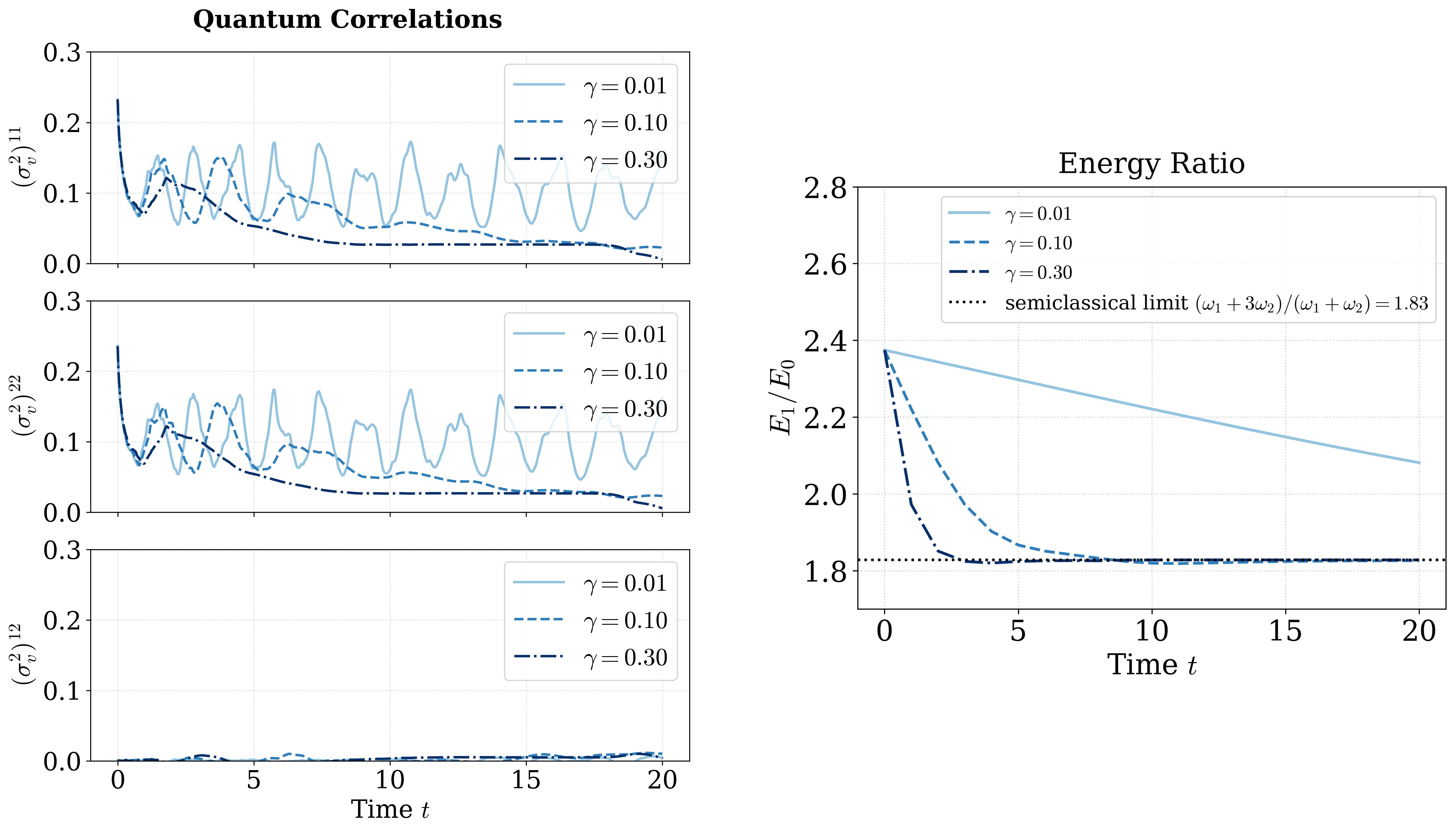}
    \caption{
        Time evolution of the quantum correlations and energy spectrum for QRHD on sphere. 
        The lighter-blue solid line, blue dashed line, and darker-blue dot-dashed line are results for $\gamma = 0.01, 0.10,$ and $0.30$, respectively. 
        The left panel plots the variance $\sigma^2_{ii}(t)$ and covariance $\sigma^2_{12}(t)$ of the parameters $v^i$, which shows that for $\gamma = 0.1$ and $0.3$ quantum fluctuations are well-suppressed in the late-time regime due to the friction term.
        The right panel shows the dimensionless instantaneous spectral ratio $E_1(t)/E_0(t)$ of the two lowest eigenvalues of the Hamiltonian on the computational domain of the chart $\bm{v}$.
        The dotted line is the value $(\omega_1 + 3\omega_2)/(\omega_1 + \omega_2) = 1.83$ of the harmonic approximation around $\bm{x}_\ast$, with the normal-mode frequencies $\omega_1 = \sqrt{2}$ and $\omega_2 = 1$.
        This ratio characterizes the low-energy spectral structure of the Hamiltonian.
        }
    \label{fig:curved-quantum}
\end{figure}

We consider the $N=3$ case and the following potential
\begin{align}
V(\bm{x}) = \frac{m}{2} \bm{x}^\tt A_2 \bm{x},\qquad
A_2 = \begin{pmatrix}
1 & 0 & - \frac{1}{\sqrt{2}} \\
0 & 1 & - \frac{1}{\sqrt{2}} \\
- \frac{1}{\sqrt{2}} & -\frac{1}{\sqrt{2}} & 1
\end{pmatrix}, 
\label{eq:text-matrix-A}
\end{align}
and we set $m=R=\eta=1$ and $a(t)=\ee^{2\gamma t}$.
It is known that the global minimum for the potential~\eqref{eq:text-matrix-A} is given by the eigenvector of $A_2$ associated with the minimum eigenvalue, $\bm{x}_\ast = (1/2, 1/2, 1/\sqrt{2})^\tt$.
For this setup, we numerically solve the Schr\"odinger equation with the time-dependent Hamiltonian operator. 
The results are displayed in Figs.~\ref{fig:curved-wavefunc}, \ref{fig:curved-wavefunc-2}, and \ref{fig:curved-quantum}.
Figs.~\ref{fig:curved-wavefunc} and \ref{fig:curved-wavefunc-2} show the time evolution of the wave function on the two-dimensional conformally-flat coordinates of $\bm{v}$ and $\bm{u}$ in the top panels, respectively.  
The corresponding configurations on the two-dimensional sphere are in the bottom panels of the both figures.
In these calculations, various friction coefficients are chosen as $\gamma = 0.01, 0.1,$ and $0.3$ from top to bottom.
Here, the domain is taken to be $v^i \in [-R,R] = [-1,1]$.

The convergence behavior depends sensitively on the friction coefficient $\gamma$, as in the flat-space example of Sec.~\ref{sec:numeric-flat}.
Within the range of parameters that can be reliably simulated with the present discretization, $\gamma=0.3$ exhibits the clearest finite-time convergence toward the minimum. 
We therefore use $\gamma=0.3$ as a representative successful example rather than claiming that it is the globally optimal value of the friction coefficient.\footnote{
Simulations at larger values of $\gamma$ become increasingly demanding because the norm of the potential term grows as $\|a(t)\eta\hat V\|\propto \ee^{2\gamma t}$. 
Consequently, resolving the corresponding time evolution over the same time interval requires substantially finer temporal and spatial discretizations.
A systematic numerical determination of the optimal $\gamma$ beyond the presently accessible range is therefore left for future work.}

Next, we consider the numerical results in Fig.~\ref{fig:curved-quantum} to investigate the underlying physical mechanisms in this optimization. 
The left panel of Fig.~\ref{fig:curved-quantum} plots the time evolution of the quantum correlations among the coordinate variables $v^i$,
\begin{align}
	(\sigma_v^2)^{ij} = \braket{v^i v^j} - \braket{v^i} \braket{v^j}\,.
\end{align}
The results show that the covariance $\sigma^2_{12}$ remains consistently small and that, for $\gamma = 0.1$ and $0.3$, the variances $\sigma^2_{ii}$ are also well-suppressed except during the very early stage of the optimization.
This suppression supports our previous theoretical analysis that quantum effects are efficiently mitigated by the friction term.

We characterize the evolution of the low-energy spectral structure by plotting the ratio of the instantaneous eigenvalues of the Hamiltonian.
Following Ref.~\cite{2023-Leng-Hickan-Li-Wu}, the right panel of Fig.~\ref{fig:curved-quantum} shows the ratio between the first excited state and the ground state, $E_1/E_0$, which is used in Ref.~\cite{2023-Leng-Hickan-Li-Wu} as a qualitative indicator of the dynamical phases of QHD.
The ratio is a property of the instantaneous Hamiltonian $\hat{H}(t)$, which depends on $t$ only through $a(t) = \ee^{2\gamma t}$, and not of the state.
For $\gamma=0.1$ and $0.3$, the ratio approaches the asymptotic value of approximately $1.83$ by $t \approx8$ and $t \approx 3$, respectively.
This value, indicated by the dotted line in Fig.~\ref{fig:curved-quantum}, agrees with the harmonic approximation around the stationary point $\bm{x}_\ast$, which gives 
\begin{align}
    \frac{E_1}{E_0} = \frac{\omega_1 + 3 \omega_2}{\omega_1+ \omega_2}\,,
    \qquad 
    \omega_1 = \sqrt{2}\,,
    \qquad 
    \omega_2 = 1\,.
\end{align}
These $\omega$'s are the normal-mode frequencies, i.e. the square roots of the nonzero eigenvalues of $A_2$.
For $\gamma = 0.01$ the ratio decreases only slowly and has not settled at $t = 20$.
Taken together, the suppression of coordinate fluctuations and the associated evolution of the low-energy spectral structure are consistent with a progressive semiclassical localization of the optimization dynamics.
Such a crossover toward a semiclassical regime is a common feature of both standard QHD and the proposed QRHD framework, although it should not be interpreted as decoherence, since the quantum evolution remains unitary.

\clearpage

\section{Implementation by quantum circuit and its complexity}
\label{sec:implementation}

Let us consider the implementation of QRHD by a quantum circuit and evaluate the circuit complexity having in mind Hamiltonian simulation based on the interaction picture~\cite{2018-Dyson-IntPic}.
Assuming phase-oracle access to the cost function or potential function $V(\bm{x})$ and sparse-access oracles to the kinetic energy, the query complexity of the above Hamiltonian simulation is estimated as $\mathcal{O}(\alpha_H\beta_H t^2_{\ast})$ \cite{2025-Catli-Simon-Wiebe}.
$t_\ast$ is the time scale required for optimization via the quantum dynamics discussed in Sec.~\ref{sec:convergence}.
For details of this estimation, see App.~\ref{app:QHD-query}.
$\alpha_H$ and $\beta_H$ are 
\begin{align}
	\alpha_H \coloneqq \max_t \biggl[
		\frac{1}{a(t)} \frac{\| \Delta_g \|}{2m}
	\biggr]\,,
	\qquad 
	\beta_H \coloneqq \max_t \biggl[
		a(t) \eta(t) \| V(\bm{x}) \|
	\biggr]\,,
\end{align}
where $\|\bullet\|$ denotes the spectral norm.
$\alpha_H$ depends on the metric tensor in QRHD.
$\beta_H$ is proportional to the maximum value of the cost function $V_{\mr{max}} \coloneqq \max_t V(\bm{x})$.
It is noted that $\alpha_H$ and $t_\ast$ depend on the choice of coordinates in QRHD.
In the following parts, we examine the difference between QHD and QRHD through some examples.

\subsection{Flat case}
\label{sec:complexity-flat}

We first consider the problem discussed in Sec.~\ref{sec:numeric-flat}, namely, quadratic convex optimization.
The cost function is Eq.~\eqref{eq:potential-flatcase} and we solve this in two ways, which are QHD with $g_{ij} = \delta_{ij}$ and QRHD with $g_{ij} = (A_1)_{ij}$.
Here, we compare $\alpha_H$ and $t_\ast$ derived by these two approaches.

Let $\alpha_{H,\mr{Q(R)HD}}$ denote $\alpha_H$ in Q(R)HD and the ratio is evaluated as
\begin{align}
	\frac{\alpha_{H,\mr{QRHD}}}{\alpha_{H,\mr{QHD}}} = \frac{\|\Delta_g\|}{\| \Delta \|} 
    \leq \|A_1^{-1}\| = \frac{1}{\lambda_{\mr{min}}(A_1)}\,,
    \label{eq:ratio-of-alpha_H}
\end{align}
where $\lambda_{\mr{min}}(A_1)$ denotes the minimum eigenvalue of the matrix $A_1$ introduced in Sec.~\ref{sec:convergence-QHD}.
Speaking of the convergence time $t_\ast$, it is estimated from the quantum EOM.
In the QHD case, the equation becomes 
\begin{align}
	\Braket{\ddot{\bm{x}} + 2 \gamma \dot{\bm{x}} + \eta A_1 \bm{x}} = 0\,,
\end{align}
and the typical lower bound on $t_\ast$ is estimated by the same analysis in Sec.~\ref{sec:numeric-flat} as 
\begin{align}
	t_{\ast, \mr{QHD}} \gtrsim \frac{1}{\sqrt{\eta \lambda_{\mr{min}}(A_1)}}\,,
\end{align}
because $t_\ast$ should be larger than the longest time scale of the dynamics, which is associated with the smallest eigenvalue of $A_1$ in this forced oscillation case.
The equality is satisfied when $\gamma \approx \sqrt{\eta \lambda_{\mr{min}}(A_1)}$.
On the other hand, if we take $g_{ij} = (A_1)_{ij}$, the EOM becomes 
\begin{align}
	A_1 \Braket{\ddot{\bm{x}} +2 \gamma \dot{\bm{x}} + \eta \bm{x}} =0\,,
\end{align}
and we estimate the convergence time as 
\begin{align}
	t_{\ast,\mr{QRHD}} \gtrsim \frac{1}{\sqrt{\eta}}\,.
\end{align}
The equality is achieved when $\gamma \approx \sqrt{\eta}$.
The lower bound does not depend on the eigenvalue of $A_1$ because it appears as the overall factor in the EOM.
The ratio of $t_\ast$ between two approaches is 
\begin{align}
	\frac{\min t_{\ast,\mr{QRHD}}}{\min t_{\ast,\mr{QHD}}} \approx \sqrt{\lambda_{\mr{min}}(A_1)}\,,
\end{align}
This reduction in the characteristic convergence time is a preconditioning effect. Indeed, setting $\bm{y}=A_1^{1/2}\bm{x}$ gives $\Delta_g=\Delta_y$ and $V=m|y|^2/2$. 
With the initial state and computational domain transformed consistently, QRHD with $g=A_1$ is therefore equivalent to QHD on an isotropic quadratic objective in the $y$ coordinates.
The constant metric has vanishing intrinsic curvature.
This is consistent with the numerical results in Sec.~\ref{sec:numeric-flat}.

Finally, let us compare the query comlexities of QHD and QRHD for this problem.
From the above calculations and $\beta_{H,\mr{QRHD}}/ \beta_{H,\mr{QHD}} \approx \mc{O}(1)$, we find that the ratio of query complexities remains order unity,
\begin{align}
    \frac{n_{\mr{query,QRHD}}}{n_{\mr{query,QHD}}} \approx \mc{O}(1)\,,
\end{align}
where we use $n_{\mr{query}}=\mathcal{O}(\alpha_H \beta_H t^2_\ast)$. See App.~\ref{app:QHD-query} for the derivation.
This indicates that for this problem, there is no substantial difference in query complexity between QHD and QRHD.
This fact originates from the use of Hamiltonian simulation as the basis of the quantum algorithm, which means that the complexity depends on the product of the Hamiltonian norm and $t_\ast$.
As seen in Sec.~\ref{sec:numeric-flat}, QRHD significantly reduces the convergence time, but the norm of the kinetic term $\alpha_{H}$ increases due to the metric tensor, and these contributions cancel each other.
This seems to remind us of the conjecture \cite{2015-Brown-Roberts-Susskind-Swingle-Zhao} that the ``complexity'' of a quantum system is determined by the action integral of the system.

\subsection{Curved case}

The second example is the problem discussed in Sec.~\ref{sec:numeric-curved}.
The cost function is Eq.~\eqref{eq:text-matrix-A} with the constraint $\bm{x}^2 = R^2$.
We solve this problem using the conformally flat coordinate systems $({\bm{u}},{\bm{v}})$, which metrics are given by Eq.~\eqref{eq:uv-metric-onsphere}.
From this, we write explicitly the Laplace--Beltrami operator as 
\begin{align}
	\Delta_g = \biggl( \frac{\bm{u}^2 + R^2}{2 R^2} \biggr)^2 \biggl[
		\Delta - 2 (N-3) \frac{u^i}{\bm{u}^2 + R^2} \frac{\del}{\del u^i}
	\biggr]\,.
\end{align}
The same holds for $\bm{v}$.
We introduce the conformal coordinates so that $\bm{u}^2$ and $\bm{v}^2$ are at most of order $R^2$ (and if the factor $N-3$ is not so large), then we find that $\|\Delta_g\| \sim \|\Delta\|$, which implies the not so large finite $\alpha_H$.
However if we change the regions of $\mc{U}$ and $\mc{V}$, $\alpha_H$ can increase due to the relatively large $\bm{u}^2$ and $\bm{v}^2$.
In this sense, the implementation cost of QRHD depends on the choice of the atlas for the manifold of the problem we consider. 
Generally, to prevent the explosion of the Hamiltonian norm, one should select a coordinate chart where both the metric tensor $g_{ij}(\bm{x})$ and its inverse $g^{ij}(\bm{x})$ remain non-singular and well-behaved throughout the computational domain. 
Such a choice of atlas characteristically encodes prior knowledge regarding the target constraint into the bounded spectral radius of the operator, thereby structurally mitigating the risk of numerical divergence without relying on artificial penalty terms.

In the present problem, the loss functions in the conformal coordinates $V(\bm{x}(\bm{u}))$ and $V(\bm{x}(\bm{v}))$ are nonlinear functions of $\bm{u}$ and $\bm{v}$.
It is generally hard to estimate the convergence time to the global minimum.
As a simple estimation, we instead evaluate the convergence time to a local minimum as Eq.~\eqref{eq:t_star_bound} with some approximations in Sec.~\ref{sec:QRHD-pathintegral}.
In the present case, it is noted that the eigenvalues $g^{-1} \Hess_g V$ at the stationary point are $m(\lambda(A_2) - \lambda_{\mr{min}}(A_2)$, where $\lambda(A_2)$ represents a eigenvalue of the matrix $A_2$.
Then, we obtain 
\begin{align}
	t_\ast \gtrsim \frac{1}{\sqrt{\eta ( \lambda_{\mr{nmin}}(A_2) - \lambda_{\mr{min}}(A_2))}} \biggl[
		-W_{-1}\left(-\frac{\epsilon_\ast}{\ee}\right) - 1
	\biggr]\,,
	\label{eq:t_star_sphere}
\end{align}
and the lower bound is achieved when $\gamma \approx \sqrt{\eta(\lambda_{\mr{nmin}}(A_2) - \lambda_{\mr{min}}(A_2))}$.
$\lambda_{\mr{nmin}}(A_2)$ is the next smallest eigenvalue of $A_2$.
That is, by choosing $\gamma \approx \sqrt{\eta (\lambda_{\mr{nmin}}(A_2)-\lambda_{\mr{min}}(A_2)}$, one finds that the algorithm would converge within a time which does not depend on the parameter dimension.

\section{Conclusions}
\label{sec:conclusions}

In this paper, we have proposed QRHD, which generalizes QHD for continuous optimization on Riemannian manifolds.
In QRHD, the kinetic term is extended to a noncanonical form in order to incorporate the geometric information of the parameter space into the quantum dynamics through the metric tensor $g_{ij}(\bm{x})$.
The introduction of the metric enables one to incorporate prior knowledge about coordinate choices and geometric constraints into the dynamics such as a norm constraint $\bm{x}^2 = R^2$ in Rayleigh quotient.

The QRHD Hamiltonian operator we propose contains the Laplace--Beltrami operator $\Delta_g$ in the kinetic term.
We also introduce the path integral formalism describing Q(R)HD to analyze the quantum dynamics of Q(R)HD and estimate the convergence time.
It is found that quantum corrections from the geometric properties appear in the effective potential, but they are suppressed by the higher powers of $a(t)$ compared to the classical contribution.
This implies the convergence behavior in the late time is mainly controlled by the classical potential and quantum corrections can be dominant in early stage of the dynamics although the convergence behavior of QRHD can essentially depend on the geometry of the parameter space.
In addition, as numerical demonstrations, we show follows:
\begin{enumerate}[label=(\roman*)]

	\item 
    For a quadratic potential on a flat space, QRHD with the metric chosen as the Hessian localizes the wave function around the minimum faster than QHD.
    In QHD the convergence is limited by the soft direction, along which the wave function remains elongated, whereas in QRHD the wave function contracts along all directions on a single time scale.
    This is because the metric equalizes the frequencies of all modes, so that a single friction coefficient damps the stiff and soft directions alike.
    For a non-convex landscape with multiple local minima (the Ackley function), QRHD locates the global minimum as reliably as QHD.
    The ability to escape from local minima by tunneling is retained under the metric.

	\item By using the spherical metric, we treat optimization with a norm constraint within the same framework and confirm that QRHD functions as an optimization algorithm even on curved parameter spaces. 
    This geometric approach allows QRHD to naturally eliminate the need for stiff penalty terms in the potential function, thereby avoiding the potential risk of a blowing-up spectral norm that could typically arise when incorporating such penalties into the standard QHD framework.

\end{enumerate}
These results are numerical demonstrations in low dimensions.
They do not by themselves establish a computational advantage of QRHD over classical optimization methods, which is beyond the scope of this work.
We also discuss a quantum circuit implementation based on time-dependent Hamiltonian simulation and evaluate the query complexity of QRHD.

These results suggest that QRHD provides not only a practical extension of quantum optimization algorithms but also a conceptual bridge between geometry and quantum dynamics. 
In particular, the freedom to incorporate geometric structures which reminiscent of those appearing in gravitational or gauge theoretic contexts may play an important role both in designing efficient optimization algorithms and in deepening our theoretical understanding of quantum dynamics on structured parameter spaces. 
In this sense, QRHD offers a framework that unifies geometry with continuous optimization driven by quantum dynamics, and is expected to stimulate further developments both in theory and applications.

\section*{Acknowledgments}
\noindent
The authors thank Naoki Yamamoto for the grateful comments and discussions.
This work is supported by MEXT Quantum Leap Flagship Program Grant No. JPMXS0120319794.
Y.A. acknowledges support by JST Grant No. JPMJPF2221.

\appendix

\section{Quantum circuit for time-dependent Hamiltonian simulation}
\label{app:QHD-query}

Q(R)HD is a quantum algorithm in which the time evolution of a quantum state works as the optimization process.
When we implement this, the main interest is the time-evolution unitary with the Hamiltonian operator describing the optimization dynamics
\begin{align}
	\hat{U} = \msc{T} \exp \left(
		- \iu \int_0^t \dd t' \hat{H}(t')
	\right)\,,
	\label{eq:hatU}
\end{align}
where $\msc{T}$ denotes the time ordering operator.
The quantum circuit of this unitary is basic on implementing Q(R)HD and getting the expectation value of the position operator with the output quantum state.

\begin{figure}[t]
  \centering
  \includegraphics[width=15cm]{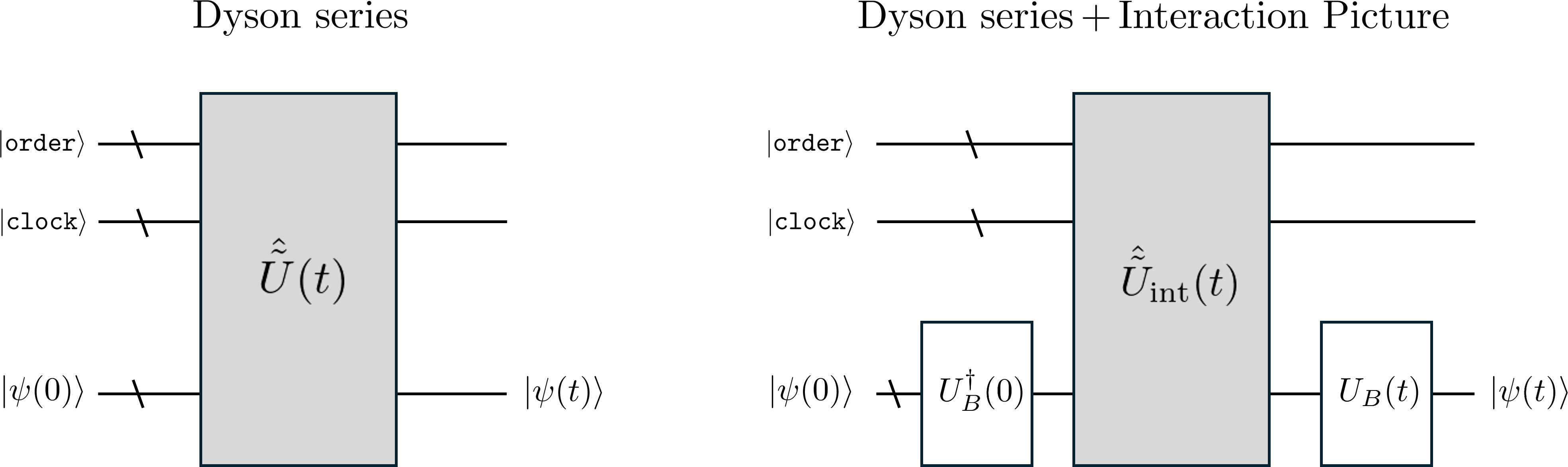}
  \caption{
  Schematic quantum circuit diagrams for time-dependent Hamiltonian simulation.
  The left panel is based on the Dyson series expansion \cite{2018-Dyson}, and the right panel further adopts the interaction picture \cite{2018-Dyson-IntPic}.
  In both panels, $\ket{\mt{clock}}$ is a register that manages time ordering, and $\ket{\mt{order}}$ is a register that manages the order in the Dyson series expansion.
  }
  \label{fig:circuit}
\end{figure}

It is known that the unitary \eqref{eq:hatU} can be implemented by quantum algorithms based on the Dyson series expansion~\cite{2018-Dyson}.
This approach approximates Eq.~\eqref{eq:hatU} by an enlarged system unitary $\hat{\tilde{U}}(t)$, where $\hat{\tilde{U}}(t)$ is constructed from a block encoding (BE) of $\hat{H}(t)$.
A schematic circuit diagram is shown in the left panel of Fig.~\ref{fig:circuit}, which corresponds to Fig.~1 of Ref.~\cite{2018-Dyson}.
Since this quantum algorithm is based on a BE of $\hat{H}$, its complexity is characterized by the maximum spectral norm of $\hat{H}(t)$, $\|\hat{H}\|_{\mr{max}} = \max_t \|\hat{H}\|_{\mr{max}}$.
However, in Q(R)HD, it is needed to increase the norm of the potential term at late time for the convergence of the dynamics, which leads to the large $\|\hat{H}\|_{\mr{max}}$ and the circuit complexity.

A possible way to avoid the operator norm increase of the potential term is Hamiltonian simulation based on the interaction picture~\cite{2018-Dyson-IntPic}, in which the contribution of the large operator norm is removed by the introduction of the interaction picture associated with the problematic operator.
Let us split the Hamiltonian operator as 
\begin{align}
	\hat{H}(t) = \hat{A}_H(t) + \hat{B}_H(t)\,,
	\qquad 
	\hat{A}_H = \frac{1}{a(t)} \biggl( \frac{-\Delta_{(g)}}{2m} \biggr)\,,
	\qquad 
	\hat{B}_H = a(t) \eta(t) V(\hat{\bm{x}})\,,
	\label{eq:AH-QHD}
\end{align}
and consider the interaction picture where $\hat{B}_H$ is treated as the non-perturbative term.
The interaction picture state is defined by 
\begin{align}
    \ket{\psi_I(t)} = \hat{U}_B^\dagger \ket{\psi(t)}\,,
	\qquad 
	\hat{U}_B(t) = \msc{T} \exp \biggl( - \iu \int_0^t \dd t' \hat{B}_H(t') \biggr)\,.
\end{align}
The time evolution of $\ket{\psi_I(t)}$ is given by 
\begin{align}
	\iu \frac{\dd}{\dd t} \ket{\psi_I(t)} = \hat{H}_{\mr{int}}(t) \ket{\psi_I(t)}\,,
	\qquad 
	\hat{H}_{\mr{int}} \coloneqq \hat{U}^\dagger_B(t) \hat{A}_H(t) \hat{U}_B(t)\,,
	\label{eq:schrodinger-I}
\end{align}
which is the Schr\"{o}dinger equation for the interaction picture $\ket{\psi_I(t)}$ with the Hamiltonian operator $\hat{H}_{\mr{int}}(t)$.
The time evolution operator in the interaction picture is given by 
\begin{align}
    \hat{U}_{\mr{int}} = \msc{T} \exp \left(
        -\iu \int_0^t \dd t' \hat{H}_{\mr{int}}(t')
    \right)\,,
\end{align} 
and we write an enlarged system unitary as $\hat{\tilde{U}}_{\mr{int}}$.
The quantum circuit of this approach is shown in the right panel of Fig.~\ref{fig:circuit}.
The circuit complexity characterized by the time evolution generator norm, which is $\|H_{\mr{int}}\| = \|A_H\|$ because $\hat{U}_B(t)$ is unitary.
This implies that the circuit complexity does not show the unbounded growth via due to $a(t)$.
In the following analysis, we assume the implementation based on this interaction picture.

Let us move on to analyzing the complexity of a quantum circuit implementing QHD.
The complexity of $\hat{\tilde{U}}_{\mr{int}}(t)$ is expressed as 
\begin{align}
	(\text{complexity of a BE of $\hat{H}_{\mr{int}}$}) \times \mc{O} \left(
		\alpha_H T \frac{\log (\alpha_H T/ \epsilon)}{\log \log(\alpha_H T / \epsilon)}
 	\right)\,,
 	\label{eq:complexity-Uint}
\end{align}
where we define \cite{2018-Dyson}
\begin{align}
	\alpha_H = \max_t \| A_H(t)\|\,.
\end{align}
$T$ is the simulation time we typically take the convergence time $T = t_\ast$, and $\epsilon$ is the desired error in the time evolution.
The second factor in Eq.~\eqref{eq:complexity-Uint} arises because $\hat{\tilde{U}}_{\mr{int}}(t)$ is constructed based on a truncated Dyson series expansion~\cite{2018-Dyson}.
$\alpha_H T$ represents the number of time segments and $\log(\alpha_H T/\epsilon)/ \log\log(\alpha_H T / \epsilon)$ corresponds to the truncation order of the Dyson series.
The fact that this factor depends only on $\alpha_H$ is due to taking the interaction picture.
On the other hand, the construction of the BE of $\hat{H}_{\mr{int}}$ \footnote{A BE of $\hat{H}_{\mr{int}}$ is denoted by $\HAMT$ in Refs.~\cite{2018-Dyson-IntPic,2025-Catli-Simon-Wiebe}.} consists of implementing the BE of $\hat{A}_H$ and the unitary $\hat{U}_B$, both which are controlled on ancilla qubits.
Let $n_A$ and $n_{U_B}$ denote the numbers of oracle queries required to construct these, respectively.
Then, the number of oracle queries to construct $\hat{\tilde{U}}_{\mr{int}}(t)$ is written as 
\begin{align}
	\mc{O} \left(
		(n_A + n_{U_B}) \alpha_H T \frac{\log(\alpha_H T/\epsilon)}{\log\log(\alpha_H T/\epsilon)}
	\right)\,.
\end{align}

In QHD, we estimate the expectation value of the particle position $\braket{\bm{x}(T)} = \braket{\psi(T)| \hat{\bm{x}} | \psi(T)}$, and require the probability that the absolute deviation between the estimator and the true mean exceeds $\epsilon$ is at most $\delta$ implies that $\mathcal{O}(\log(1/\delta))$ samples are needed via a concentration inequality.
This leads to the $\mc{O}(\log(1/\delta))$ times implementation of Hamiltonian simulation.

Combining these evaluations, the number of queries required to implement QHD denoted by $n_{\mr{query}}$ is given by 
\begin{align}
	n_{\mr{query}} &= n_{\mr{query},A} + n_{\mr{query}, U_B}\,,
	\label{eq:nquery}
	\\
	n_{\mr{query}, A} &= \mc{O} \left(
		n_A \alpha_H T \frac{\log(\alpha_H T/\epsilon)}{\log\log(\alpha_H T /\epsilon)} \log(1/\delta)
	\right)\, ,
	\label{eq:nquery-2}
	\\
	n_{\mr{query}, U_B}&= \mc{O} \left(
		n_{U_B} \alpha_H T \frac{\log(\alpha_H T/\epsilon)}{\log\log(\alpha_H T /\epsilon)} \log(1/\delta)
	\right)\,.
	\label{eq:nquery-3}
\end{align}

\paragraph{Evaluation of $n_{U_B}$.}

Let us estimate the number of queries for the ancilla-controlled $\hat{U}_B$.
Since $\hat{U}_B$ depends on the cost function $V(\bm{x})$, implementing ancilla-controlled $\hat{U}_B$ requires a mechanism to access the structure of $V(\bm{x})$.\footnote{
In Ref.~\cite{2025-Sato-et-al}, a quantum circuit functioning as a PDE solver serves as an oracle for the cost function $V({\bm{x}})$.
}
Following Ref.~\cite{2025-Catli-Simon-Wiebe}, we assume the phase oracle $O_p$ for the cost function 
\begin{align}
	O_p \ket{\bm{x}} = \ee^{\iu F}\ket{\bm{x}},
	\qquad 
	\hat{F} = \sum_{\bm{x}} \frac{V(\bm{x})}{2 V_{\mr{max}}} \ketbra{\bm{x}}{\bm{x}}\,.
\end{align}
We define $n_{U_B}$ as the number of accesses to $O_p$ and estimate as 
\begin{align}
    n_{U_B} = \mc{O} \left(
        \left[ \int_0^T \dd t \| B_H(t)\| \right] \log^2(1/\epsilon)
    \right)\,.
	\label{eq:nUB}
\end{align}
The factor $\log^2(1/\epsilon)$ comes because the query complexity of the BE of $\hat{F}$ scales as $\log(1/\epsilon)$ and the number of ancilla qubits scales as $\log(1/\epsilon)$.

For the case of $a= \ee^{2 \gamma t}$ and $\eta=1$, the integral in Eq.~\eqref{eq:nUB} is evaluated as 
\begin{align}
	\int_0^T \dd t B_H(t) = \int_0^T \dd t \ee^{2 \gamma t} V \leq \frac{1}{2\gamma} \biggl( \ee^{2 \gamma T} - 1 \biggr) V_{\mr{max}}
	\approx \frac{1}{2\gamma} \ee^{2 \gamma T} V_{\mr{max}} = \frac{\beta_H}{2\gamma}\,,
	\label{eq:VmaxfromBH}
\end{align}
where $\beta_H$ is the maximal operator norm of the potential term defined by
\begin{align}
	\beta_H = \max_t \| \hat{B}_H(t) \| = \max_t\| a(t) \eta(t) V(\hat{\bm{x}})\| = \ee^{2\gamma T} V_{\mr{max}}\,.
	\label{eq:beta-def}
\end{align}
Then, we find for $a(t)= \ee^{2\gamma t}$ and $\eta(t)=1$
\begin{align}
	n_{\mr{query},U_B} = \mc{O} \left(
		\frac{\beta_H}{\gamma} \alpha_H T \log^2 (1/\epsilon) \frac{\log(\alpha_H T/\epsilon)}{\log \log(\alpha_H T/\epsilon)} \log(1/\delta)
	\right)\,,
	\label{eq:nquery-B}
\end{align}
which corresponds to Eq.~(322) in Ref.~\cite{2025-Catli-Simon-Wiebe}.

\paragraph{Evaluation of $n_A$.}

Let us consider the estimation of the number of queries to construct the ancilla-controlled BE of $\hat{A}_H$.\footnote{
For concrete examples of BEs of the Laplacian, see also Refs.~\cite{2025-Kharazi-et-al,2025-Sturm-Schillo}.
}
In order to construct this BE, we need a mechanism to access the structure of the matrix representation of the kinetic term $\hat{A}_H$.
Noting that the matrix representation of $\hat{A}_H$, i.e., that of the Laplace operator $\Delta$, is typically a sparse matrix,\footnote{To specify the explicit structure, one must choose a discretization of the space ${\bm{x}}$, a finite-difference scheme, and boundary conditions.} we may regard the matrix representation $\hat{A}_H$ as sparse.
We therefore assume standard sparse-matrix access oracles for $\hat{A}_H$:
\begin{align}
	\msc{O}_r \ket{i}\ket{j} = \ket{i} \ket{r_{ij}}\,,
	\qquad 
	\msc{O}_c \ket{i} \ket{j}= \ket{c_{ij}} \ket{j}\,,
	\qquad 
	\msc{O}_a \ket{i}\ket{j} \ket{0}^{\otimes b} = \ket{i} \ket{j}\ket{a_{ij}}\,,
\end{align}
where $r_{ij}$ ($c_{ij}$) denotes the index of the $j$-th nonzero element in the $i$-th row (the $i$-th nonzero element in the $j$-th column) of $\hat{A}_H$, and $a_{ij}$ denotes the $b$-bit representation of the $(i,j)$ entry of $\hat{A}_H$.
We define $n_A$ as the number of accesses to $\msc{O}_{r,c,a}$ required to construct the controlled BE of $\hat{A}_H$.\footnote{
Such sparse-matrix access oracles were first introduced in Ref.~\cite{2003-Aharonov-TaShma}, and have since been widely used in the context of Hamiltonian simulation \cite{2007-Berry-et-al,2009-Berry-Childs,2016-Low-Chuang-QSP,2019-Childs-Kothari}.
Note, however, that constructing such black-box oracles is, in general, a highly nontrivial task \cite{2022-Camps-Lin-VanBeeumen-Yang}.}

From Lemma 49 of Ref.~\cite{2019-Gilyen-Su-Low-Wiebe}, we find
\begin{align}
	n_A = \mc{O} \left(
		\sqrt{\frac{s(\hat{A}_H)}{\alpha_H}} \log^2(1/\epsilon)
	\right)\,,
\end{align}
where $s(\hat{A}_H)$ denotes the sparsity of $\hat{A}_H$, and the factor $\log(1/\epsilon)$ corresponds to the number of ancilla qubits.
Then, it is found that the number of queries is given by 
\begin{align}
	n_{\mr{query},A} = \mc{O} \left(
		T \sqrt{s(\hat{A}_H) \alpha_H} \log^2(1/\epsilon) \frac{\log(\alpha_H T/\epsilon)}{\log\log(\alpha_H T /\epsilon)} \log (1/\delta)
	\right) \in \tilde{\mc{O}} \left( T \sqrt{s(\hat{A}_H) \alpha_H} \log(1/\delta) \right)\,.
	\label{eq:nquery-A}
\end{align}

\paragraph{Total query.}

From Eqs.~\eqref{eq:nquery}, \eqref{eq:nquery-2}, \eqref{eq:nquery-3}, \eqref{eq:nquery-B}, and \eqref{eq:nquery-A}, we obtain the query complexity of Q(R)HD.
\begin{align}
	n_{\mr{query}} = \mc{O} \left(
		\left( \sqrt{s(\hat{A}_H) \alpha_H} + \alpha_H V_{\mr{max}} \left[ \int_0^T \dd t a(t) \eta(t) \right] \right)
		T \log^2(1/\epsilon)  \frac{\log(\alpha_H T/\epsilon)}{\log \log(\alpha_H T/\epsilon)} \log(1/\delta) 
	\right)\,.
\end{align}
For $a(t) = \ee^{2\gamma t}$ and $\eta =1$, the integral from the contribution of $n_{U_B}$ is evaluated as $\ee^{2\gamma T}$.
The analysis result Eq.~(322) of Ref.~\cite{2025-Catli-Simon-Wiebe} agrees with ours when the contribution of $n_{\mr{query},A}$ is neglected.
Noting that if $T = t_\ast \approx 1/\gamma$, we have $n_{\mr{query},U_B} \approx \mc{O}(\alpha_H\beta_H t^2_\ast)$.

\section{Convention of geometry}
\label{sec:geometry}

In this appendix we summarize the conventions and definitions of geometric quantities used in this paper.

\subsection{Convention}

We consider a Riemannian manifold such that, in coordinates $\bm{x}$, the line element is given by
\begin{align}
    \dd s^2 = g_{ij}(\bm{x}) \dd x^i \dd x^j,
\end{align}
where $g_{ij}(\bm{x})$ is a real symmetric matrix.
We introduce its inverse matrix $g^{ij}$ by requiring
\begin{align}
    g^{ik} g_{kj} = \delta^i_j,
\end{align}
to hold.
Raising and lowering indices for vectors and tensors are defined using this metric tensor:
\begin{align}
    A_i = g_{ij} A^j\,,
    \qquad 
    A^i = g^{ij} A_j\,.
\end{align}

For this metric, the Levi--Civita connection is introduced by
\begin{align}
    \Gamma^i_{jk} = \frac{1}{2} g^{il}( \del_j g_{lk} + \del_k g_{jl} - \del_l g_{jk})\,,
\end{align}
and the curvature tensors are defined by
\begin{align}
    R_{ij}{}^k{}_l &= \del_i \Gamma^k_{jl} - \del_j \Gamma^k_{il} + \Gamma^k_{im} \Gamma^m_{jl} - \Gamma^k_{jm} \Gamma^m_{il}\,,
    \\
    R_{ij} &= R_{ki}{}^k{}_j = 
        \del_k \Gamma^k_{ij} -\del_i \Gamma^k_{kj} + \Gamma^k_{km} \Gamma^m_{ij} - \Gamma^k_{im} \Gamma^m_{kj}\,,
    \\
    \mc{R} &= g^{ij} R_{ij}\,.
\end{align}
In this convention, the Ricci scalar of the sphere becomes positive as seen below.

\subsection{Coordinates on $\mbb{S}^{N-1}$}
\label{sec:coordinate-SN-1}

We summarize the coordinates on the $N$-dimensional sphere $\mbb{S}^{N-1}$ used in this paper.
A sphere $\mbb{S}^{N-1}$ of radius $R$ is characterized as the subspace in $\mbb{R}^N$ satisfying
\begin{align}
    \bm{x}^2 = (x^i)^2 = R^2\,,
\end{align}
We introduce coordinates on the sphere by stereographic projection from the north pole or the south pole to the equatorial plane.
Taking the north and south poles at $x^N = \pm R$, the stereographic coordinates $(\bm{u},\bm{v})$ are related to the coordinates $\bm{x}$ in $\mbb{R}^N$ by
\begin{align}
    u^i = \frac{x^i}{1 - x^N/R}\,,
    \qquad 
    v^i = \frac{x^i}{1 + x^N/R}\,,
    \qquad (i=1,\ldots,N-1)
\end{align}
The inverse transformations are given by
\begin{align}
    \begin{cases}\displaystyle
        x^i = \frac{2u^i}{1 + \bm{u}^2/R^2}
        \\
        \\\displaystyle
        x^N = R \frac{\bm{u}^2 - R^2}{\bm{u}^2 + R^2}
    \end{cases}\,,
    \qquad 
    \begin{cases}\displaystyle
        x^i = \frac{2v^i}{1 + \bm{v}^2/R^2}
        \\
        \\\displaystyle
        x^N = R \frac{R^2 - \bm{v}^2}{R^2 + \bm{v}^2}
    \end{cases}\,,
\end{align}
where $\bm{u}^2 = (u^i)^2$ and $\bm{v}^2 = (v^i)^2$.
The metric takes the same form in either coordinate system:
\begin{align}
    \dd s^2 = \frac{4}{(1 + \bm{u}^2/R^2)^2} \delta_{ij} \dd u^i \dd u^j\,,
    \qquad 
    \dd s^2 = \frac{4}{(1 + \bm{v}^2/R^2)^2} \delta_{ij} \dd v^i \dd v^j\,.
\end{align}
That is,
\begin{align}
	g_{ij}({\bm{u}})
	=
	\frac{4}{(1 + \bm{u}^2/R^2)^2}
	\delta_{ij}
	\,,\qquad
	g_{ij}({\bm{v}})
	=
	\frac{4}{(1 + \bm{v}^2/R^2)^2}
	\delta_{ij}\,.
 	\label{eq:metric-app}
\end{align}
From this metric, for the coordinate system $\bm{u}$, we find
\begin{align}
    \Gamma^i_{jk}(\bm{u})
    &=\frac{1}{2}\Bigl[ \delta^i_k \del_j \xi(\bm{u}) + \delta^i_j \del_k \xi(\bm{u}) - \delta_{jk} \del^i \xi(\bm{u})\Bigr]
    \,,\\
    R_{ij}{}^k{}_l({\bm{u}})
    &= \frac{1}{2} \Bigl[
    \delta^k_j \del_i \del_l \xi(\bm{u}) - \delta_{jl} \del_i \del^k \xi(\bm{u}) - \delta^k_i \del_j \del_l \xi (\bm{u}) + \delta_{il} \del_j \del^k \xi (\bm{u})
    \Bigr]
    \nn\\
    & \quad
    + \frac{1}{4} \Bigl[
        2 \delta^k_i \del_j \xi (\bm{u})\del_l \xi(\bm{u}) + 2 \delta^k_{(j} \del_{l)} \xi(\bm{u}) \del_i \xi(\bm{u}) - 2 \delta_{i(j} \del_{l)} \xi (\bm{u}) \del^k\xi(\bm{u})
        - 2 \delta^k_j \del_i \xi(\bm{u}) \del_l \xi (\bm{u}) 
        \nn \\
        &\quad 
        - 2 \delta^k_{(i} \del_{l)} \xi (\bm{u}) \del_j \xi (\bm{u}) + 2 \delta_{j(i} \del_{l)} \xi (\bm{u}) \del^k \xi (\bm{u})
        -2 \delta^k_{[i} \delta_{j]l} (\del \xi (\bm{u}))^2
    \Bigr]
    \,,\\
    R_{ij}(\bm{u})
    &=\frac{1}{2} \Bigl[
        - (N-3) \del_i \del_j \xi(\bm{u}) - \delta_{ij} \del^2 \xi(\bm{u})
    \Bigr] + \frac{N-3}{4}\Bigl[
        \del_i \xi(\bm{u})\del_j \xi(\bm{u}) - \delta_{ij} (\del \xi(\bm{u}))^2
    \Bigr]
    \,,\\
    \mc{R}(\bm{u})
    &= \ee^{-\xi(u)} \biggl(- (N-2) \del^2 \xi(\bm{u}) - \frac{(N-2)(N-3)}{4}(\del \xi(\bm{u}))^2 \biggr)
    = \frac{(N-1)(N-2)}{R^2}\,.
\end{align}
In these formulae, we write the overall factor in the metric \eqref{eq:metric-app} as
\begin{align}
    \xi(\bm{u}) \coloneqq \log \biggl( \frac{2}{1 + \bm{u}^2/R^2} \biggr)^2\,,
    \qquad 
    \ee^{\xi(\bm{u})} = \frac{4}{(1 + \bm{u}^2/R^2)^2}\,,
\end{align}
and its derivatives are given by
\begin{align}
    \del_i \xi (\bm{u}) &= - \frac{4}{1 + \bm{u}^2/R^2} \frac{u_i}{R^2}\, ,
    \qquad 
    \del_i \del_j \xi(\bm{u}) = -\frac{4}{1 + \bm{u}^2/R^2} \frac{1}{R^2} \delta_{ij} + \frac{8}{(1 + \bm{u}^2/R^2)^2} \frac{u_i u_j}{R^4}\, ,
    \\
    \del^2 \xi(\bm{u}) &= \del_i \del_i \xi(\bm{u}) = - \frac{4}{1+\bm{u}^2/R^2}\frac{N-1}{R^2} + \frac{8u^2/R^4}{(1 + \bm{u}^2 /R^2)^2}\,,
    \\
    (\del \xi(\bm{u}))^2 &= (\del_i \xi(\bm{u}))^2\,.
\end{align}
It is noted that the contraction of the indices are done by $\delta_{ij}$ and $\delta^{ij}$ and $\sum_i \delta_{ii} = N-1$ here.
For the coordinate $\bm{v}$, the results are given by replacing $\bm{u}$ in the above equations with $\bm{v}$.

\section{Consistency checks of the QRHD quantum formulation}

In this appendix, we verify the consistency between the operator and path integral formulations of QRHD. 
We first show that the short-time kernel reproduces the Schr\"odinger evolution generated by the QRHD Hamiltonian operator with the Laplace--Beltrami operator in the kinetic term in App.~\ref{sec:relation-to-PI}. 
We then derive the covariant continuity equation and demonstrate probability conservation for real scalar potentials under appropriate boundary conditions in App.~\ref{sec:conservation}.

In this appendix, we mainly work in the configuration basis and Eq.~\eqref{eq:QRHD-Schroedinger} becomes
\begin{align}
    \iu \del_t \psi(\bm{x},t) = \hat{H} \psi(\bm{x},t)\,,
    \label{eq:QRHD-Schrodinger}
\end{align}
where we write the QRHD Hamiltonian as 
\begin{align}
    \hat{H} = - \frac{1}{2ma(t)} \Delta_g + a(t) \eta(t) V(\bm{x})
    = - \frac{1}{2M(t)} \Delta_g + U(\bm{x},t)\,.
    \label{eq:QRHD-Ham-2}
\end{align}
We simply write $M(t) = m a(t)$ and $U(\bm{x},t) = a(t) \eta(t) V(\bm{x})$.

\subsection{Relation between path integral and wavefunction formalisms}
\label{sec:relation-to-PI}

To establish the correspondence between the path integral and wavefunction formalisms, we start from the infinitesimal time evolution
\begin{align}
    \psi(\bm{x},t+ \Delta t) = (\mc{T}_{\Delta t} \psi)(\bm{x}) = \int_{\mc{M}} \dd^N\bm{y} \sqrt{g(\bm{y})} \mc{K}_{\Delta t}(\bm{x},\bm{y};t) \psi(\bm{y},t)\,,
    \label{eq:time-evolution-kernel}
\end{align}
with the kernel
\begin{align}
    \mc{K}_{\Delta t}(\bm{x}, \bm{y};t) = \frac{1}{g(\bm{x})^{1/4} g(\bm{y})^{1/4}} \int \frac{\dd^N\bm{p}}{(2\pi)^N} \exp \biggl[
        \iu p_i \Delta x^i - \iu \Delta t \tilde{H}\Bigl(\frac{\bm{x}+\bm{y}}{2}, \bm{p}, t \Bigr)
    \biggr]\,.
    \label{eq:kernel-mathcalK}
\end{align}
Here, $\Delta x^i = x^i - y^i$, and the Hamiltonian in this kernel is written as
\begin{align}
    \tilde{H}(\bm{x}, \bm{p},t) = \frac{1}{2M(t)} g^{ij}(\bm{x}) p_i p_j + U(\bm{x},t) + W(\bm{x},t)\,,
    \qquad
    W(\bm{x},t) = \frac{\Delta V(\bm{x})}{a(t)}\,,
\end{align}
which agrees with Eq.~\eqref{eq:Htilde-pathintegral}.
The explicit form of $\Delta V(\bm{x})$ is given by Eq.~\eqref{eq:Delta-V}.
Using this kernel, the transition amplitude becomes
\begin{align}
    \mc{K}(\bm{x}_f,t_f;\bm{x}_i,t_i) &= \lim_{K\to\infty} \int \prod_{n=1}^{K-1} \dd^N \bm{x}_n \sqrt{g(\bm{x}_n)} \prod_{n=0}^{K-1} \mc{K}_{\Delta t}(\bm{x}_{n+1}, \bm{x}_n;t_n)
    \\
    &=[g(\bm{x}_f) g(\bm{x}_i)]^{-1/4} \lim_{K\to \infty} \int \prod_{n=1}^{K-1} \dd^N\bm{x}_n \int \prod_{n=0}^{K-1} \frac{\dd^N \bm{p}_n}{(2\pi)^N}
    \nn \\
    &\quad \times 
    \exp\left\{
        \iu \sum_{n=0}^{K-1} \left[
            \bm{p}_n\cdot (\bm{x}_{n+1} - \bm{x}_n) - \Delta t \tilde{H}((\bm{x}_{n+1}+\bm{x}_n)/2,\bm{p}_n,t_n)
        \right]
    \right\}\,,
\end{align}
and this is the discretized description of the path integral in the phase space for QRHD introduced in Eq.~\eqref{eq:phase-space-pathintegral-QRHD}.

Now we evaluate the right-hand side of \eqref{eq:time-evolution-kernel}.
In order to perform the integrals, it is useful to rescale the wavefunction as
\begin{align}
    \chi(\bm{x},t) \coloneqq g(\bm{x})^{1/4} \psi(\bm{x},t)\,.
\end{align}
We then calculate $(\mc{T}_{\Delta t} \psi)(\bm{x},t)$ as
\begin{align}
    &(\mc{T}_{\Delta t}(t) \psi)(\bm{x},t) 
    = \int_{\mc{M}} \dd^N \bm{y} \sqrt{g(\bm{y})} \mc{K}_{\Delta t}(\bm{x},\bm{y};t) \psi(\bm{y},t)
    \nn \\
    &= \int_{\mc{M}} \dd^N\bm{y} \sqrt{g(\bm{y})} \frac{1}{[g(\bm{x}) g(\bm{y})]^{1/4}} \int \frac{\dd^N\bm{p}}{(2\pi)^N} \exp\left[
        \iu p_i \Delta x^i - \iu \Delta t \tilde{H}((\bm{x}+\bm{y})/2,\bm{p},t)
    \right] \frac{\chi(\bm{y}, t)}{g(\bm{y})^{1/4}}
    \\
    &\approx \frac{1}{g(\bm{x})^{1/4}} \int \frac{\dd^N \bm{y} \dd^N\bm{p}}{(2\pi)^N} \exp\left( \iu p_i \Delta x^i\right) \left(
        1 - \iu \Delta t \tilde{H}((\bm{x}+\bm{y})/2,\bm{p},t)
    \right) \chi(\bm{y},t)
    \\
    &= \frac{1}{g(\bm{x})^{1/4}} \int \dd^N\bm{y} \delta^N(\bm{x}-\bm{y}) \nn \\
    &\quad \times \left[
        1 -\iu \Delta t \left(
            - \frac{1}{2M} \left( g^{ij} \del_i \del_j + (\del_i g^{ij} )\del_j + \frac{1}{4} (\del_i \del_j g^{ij}) \right)
            + U + W
        \right)
    \right] \chi(\bm{y},t)
    \\
    &= \frac{1}{g(\bm{x})^{1/4}} \left[
        1 - \iu \Delta t \left(  
        - \frac{1}{2M} \left( \del_i g^{ij} \del_j  - \left( \frac{1}{2} \del_i (g^{ij} \Gamma_j) + \frac{1}{4} g^{ij} \Gamma_i \Gamma_j \right) \right) + U 
    \right)
    \right] \chi(\bm{x},t)
    \\
    &= \frac{1}{g(\bm{x})^{1/4}} \left[ 
        1 - \iu \Delta t \left(
            - \frac{1}{2M} g^{1/4} \Delta_g g^{-1/4} + U
        \right)
    \right] \chi(\bm{x},t)
    \\
    &= \psi(\bm{x},t) - \iu \Delta t \left( - \frac{1}{2M} \Delta_g + U \right) \psi(\bm{x},t)\,,
\end{align}
where the following equations are used in this derivation:
\begin{align}
    &\int \dd^N\bm{y}\int \frac{\dd^N\bm{p}}{(2\pi)^N} \exp \left( \iu p_i \Delta x^i \right) \frac{1}{2M} g^{ij}(\bm{z}) p_i p_j \chi(\bm{y})
    \\
    &=- \int \dd^N\bm{y} \int \frac{\dd^N\bm{p}}{(2\pi)^N} \frac{1}{2M} g^{ij}(\bm{z}) \left[\del_{y^i} \del_{y^j}\exp \left( \iu p_i \Delta x^i \right)\right] \chi(\bm{y})
    \\
    &= - \int \dd^N\bm{y} \frac{1}{2M} g^{ij}(\bm{z}) \Bigl(\del_{y^i} \del_{y^j} \delta^N(\bm{x}- \bm{y}) \Bigr) \chi(\bm{y})
    \\
    &=  - \int \dd^N \bm{y} \frac{1}{2M}\delta^N(\bm{x}-\bm{y})\del_{y^i} \del_{y^j} \Bigl(g^{ij}(\bm{z}) \chi(\bm{y}) \Bigr)
    \\
    &= - \int \dd^N \bm{y} \frac{1}{2M} \delta^N(\bm{x} -\bm{y})  \left[g^{ij}\del_i \del_j + (\del_i g^{ij}) \del_j + \frac{1}{4} (\del_i \del_j g^{ij}) \right] \chi(\bm{y})\,.
\end{align}
Here, the factor $1/4$ comes from the derivative of $\bm{z} = (\bm{x} + \bm{y})/2$ with respect to $\bm{y}$.
We then find that
\begin{align}
    (\mc{T}_{\Delta t}(t) \psi)(\bm{x}) = \psi(\bm{x},t) - \iu \Delta t \hat{H} \psi(\bm{x},t) + \mc{O}(\Delta t^2)\,,
    \qquad 
    \hat{H} = - \frac{1}{2M} \Delta_g + U\,.
\end{align}
This equation shows that the time evolution with the kernel \eqref{eq:kernel-mathcalK} produces the Schr\"odinger equation with the QRHD Hamiltonian operator \eqref{eq:QRHD} or \eqref{eq:QRHD-Ham-2}.
In this derivation, we use the following equations:
\begin{align}
    \Delta V = \frac{1}{8m} \left(- \mc{R} + g^{ij} \Gamma^k_{il} \Gamma^l_{jk} \right) = \frac{1}{8m} \left( \del_i \del_j g^{ij} + 2 \del_i (g^{ij} \Gamma_j) + g^{ij} \Gamma_i \Gamma_j \right)\,,
\end{align}
\begin{align}
    g^{1/4} \Delta_g g^{-1/4} = \del_i(g^{ij} \del_j) - \left(
             \frac{1}{2} \del_i (g^{ij} \Gamma_j) + \frac{1}{4} g^{ij} \Gamma_i \Gamma_j
        \right)\,.
\end{align}
It is noted that the non-invariant correction to the potential $\Delta V$ is needed to reproduce successfully the Laplace--Beltrami operator acting on the wavefunction.

\subsection{Continuity equation of probability for QRHD}
\label{sec:conservation}

An important consistency condition on the continuum QRHD dynamics is the conservation of total probability. 
Since the Hamiltonian \eqref{eq:QRHD-Ham-2} is intended to generate unitary time evolution on the Hilbert space equipped with the covariant inner product, the norm
\begin{align}
    \int_{\mc{M}} \dd^N \bm{x} \sqrt{g(\bm{x})}|\psi(\bm{x},t)|^2
\end{align}
is needed to be conserved. 
This also provides an independent check of the path-integral discussion in Sec.~\ref{sec:QRHD-pathintegral}, where the metric-dependent functional measure and coincident-time contact terms may superficially appear as imaginary contributions to the EOM. 
We therefore derive the covariant continuity equation directly from the Schr\"odinger equation and show that a genuine imaginary scalar potential would act as a source or sink of probability. 
Consequently, the physical scalar potential in QRHD is real, and the metric factor $\sqrt{g}$ in the path-integral measure does not generate an independent imaginary quantum potential.

The probability density and probability current are defined by
\begin{align}
    \rho(\bm{x},t) \coloneqq |\psi(\bm{x},t)|^2\,,
    \qquad 
    j^i(\bm{x},t) \coloneqq \frac{1}{2\iu m a} g^{ij}(\psi^\ast(\bm{x},t) \del_j \psi(\bm{x},t) - \psi(\bm{x},t) \del_j \psi^\ast(\bm{x},t))\,,
    \label{eq:def-of-current}
\end{align}
and the continuity equation or probability conservation law is given by\footnote{
We use the following equation:
\begin{align}
    \nabla_i j^i = \del_i j^i + \Gamma^i_{ij} j^j = \del_i j^i + \left(\frac{1}{\sqrt{g}}\del_i \sqrt{g} \right) j^i
    =\frac{1}{\sqrt{g}}(\del_i \sqrt{g}j^i)\,.
\end{align}
}
\begin{align}
    \del_t \rho +  \nabla_i j^i = \del_t \rho + \frac{1}{\sqrt{g}} \del_i (\sqrt{g} j^i) = 0\,.
    \label{eq:probability-conservation}
\end{align}
We check whether this equation is satisfied.
From the Schr\"odinger equation \eqref{eq:QRHD-Schrodinger} and its complex conjugate, the time derivative of the probability density is evaluated as 
\begin{align}
    \del_t \rho &= - \frac{1}{\sqrt{g}} \del_i \left[
        \sqrt{g} \frac{1}{2M \iu} g^{ij} \left(
            \psi^\ast \del_j \psi - \psi \del_j \psi^\ast
        \right)
    \right] - \iu (U - U^\ast) \rho
    \\
    &= - \nabla_i j^i + 2 \Im U \rho\,.
\end{align}
For a real scalar potential and vanishing boundary flux, we find that the total probability is thus conserved. 
In particular, the metric-dependent measure does not generate an independent imaginary potential in the QRHD Hamiltonian.

\paragraph{Field theoretic derivation.}

We give a brief comment on the derivation of \eqref{eq:def-of-current} from the field theoretic perspective.
The Lagrangian for a so-called Schr\"odinger field is
\begin{align}
    \mc{L} = \sqrt{g} \biggl[
        \iu \psi^\ast \del_t \psi - \frac{1}{2M} g^{ij} \del_i \psi^\ast \del_j \psi - U |\psi|^2
    \biggr]\,,
\end{align}
and the variation with respect to $\psi^\ast$ provides the Schr\"odinger equation \eqref{eq:QRHD-Schrodinger}.
It is noted that the overall $\sqrt{g}$ factor is needed to get the Laplace--Beltrami operator.
This Lagrangian is invariant under the following global phase rotation of $\psi$
\begin{align}
    \psi \mapsto \ee^{\iu \alpha} \psi\,,
    \qquad 
    \psi^\ast \mapsto \ee^{-\iu \alpha} \psi^\ast\,,
    \label{eq:phase-rotation}
\end{align}
and the corresponding conservation law is nothing but the probability conservation~\eqref{eq:probability-conservation}.
The conserved current is read from the coefficient when the transformation parameter in \eqref{eq:phase-rotation} is formally replaced with the arbitrary function $\alpha(\bm{x},t)$ as
\begin{align}
    \delta \int \dd t \dd^N\bm{x} \mc{L} &= \int \dd t \dd^N \bm{x} \sqrt{g}\left[
        - (\del_t \alpha) |\psi|^2 - (\del_i \alpha) \frac{1}{2 M\iu} g^{ij} (\psi^\ast \del_j \psi - \psi \del_j \psi^\ast)
    \right]
    \\
    &= \int \dd t \dd^N \bm{x} \sqrt{g} \left[
        - (\del_t \alpha) \rho - (\del_i \alpha) j^i
    \right]\,.
\end{align}

\section{Estimation of convergence time from (semi-)classical equations of motion}
\label{app:convergence-time}

In this appendix, we analyze the parameter evolution given by the equation \eqref{eq:eom-QRHD-convex-approx} for several concrete examples, and verify that the (local) convergence time $t_\ast$ to the optimal solution satisfies the inequality \eqref{eq:t_star_bound}.
we consider the problem treated in Sec.~\ref{sec:numeric-curved}, namely, the problem of finding $\bm{x}=\bm{x}_\ast$ that minimizes
$V(\bm{x})=\frac{m}{2}\bm{x}^\tt A \bm{x}$ for ${\bm{x}}=(x^1,\ldots,x^N)^\tt$ satisfying $\bm{x}^2=R^2$.
In this appendix, we take the matrix $A$ is diagonalized by the appropriate similarity transformation as $\diag(1^2, 2^2,\ldots, N^2)$, and assume that the $N$-th component of the optimal solution $\bm{x} = \bm{x}_\ast$ is positive. 
In addition, we set $m = R = \eta(t) = 1$ for simplicity.

We consider the above optimization problem in the conformally flat coordinate as Sec.~\ref{sec:numeric-curved}.
To analyze the equation, we put assumptions that quantum correlations among parameters are neglected and the expectation value satisfies $\braket{v^iv^j}=\braket{v^i}\braket{v^j}$.
In this set up, the time-evolution equation for $\braket{\bm{v}}$ can be written as 
\begin{align}
	\braket{\ddot{v}^i} + \Gamma^i_{kl}(\braket{\bm{v}}) \braket{\dot{v}^k}\braket{\dot{v}^l} + 2\gamma \braket{\dot{v}^i} +\frac{\eta}{m} g^{ij}(\braket{\bm{v}}) \partial_j \mathcal{V}(\braket{\bm{v}})=0\,.
    \label{eq:eom-v}
\end{align}

The behavior of the solution to the differential equation \eqref{eq:eom-v} depends on the matrix $A$ and the initial conditions $\braket{\bm{v}(0)}$ and $\braket{\dot{\bm{v}}(0)}$.
There are infinitely many choices of $A$ satisfying the conditions above. 

To eliminate any statistical bias and ensure that our benchmark removes directional bias relative to the chosen coordinate chart on the sphere without any directional preference (such as near the equator or poles), we strictly generate the matrix $A$ based on the Haar measure sampling.
Specifically, we draw an $N \times N$ matrix with independent standard normal entries and orthonormalize its rows by the Gram--Schmidt procedure, i.e.\ the standard QR-based construction of a Haar-random orthogonal matrix (the QR decomposition of a Gaussian matrix with the diagonal of $R$ taken positive).
The rows $\bm{x}_1, \ldots, \bm{x}_N$ are taken as the eigenvectors of $A$.
To respect the geometric constraint of our coordinate chart (i.e., restricting $\bm{x}_1$ to the northern hemisphere), we invert the signs of all components of the generated basis if the $N$-th component of $\bm{x}_1$ is negative.
Then, the matrix $A$ is constructed as
\begin{align}
    A = \omega^2_i \bm{x}_i \bm{x}^{\mr{t}}_i\,,
\end{align}
where the eigenvalues are configured as $\omega^2_i = i^2$ and $i \in \mbb{N}$.
In this analysis, we generate (for fixed dimension $N$) 100 random instances of $A$ satisfying these conditions and numerically solve the differential equation \eqref{eq:eom-v} for each instance. For the initial conditions, for each instance we randomly choose ${\bm{v}}(0)$ satisfying $|\braket{{\bm{v}}(0)}-{\bm{v}}_\ast |=0.1R$ and set $\dot{\bm{v}}(0)$ as $\braket{\dot{\bm{v}}(0)}={\bm{0}}$.

\begin{figure}[t]
  \centering
  \includegraphics[width=15cm]{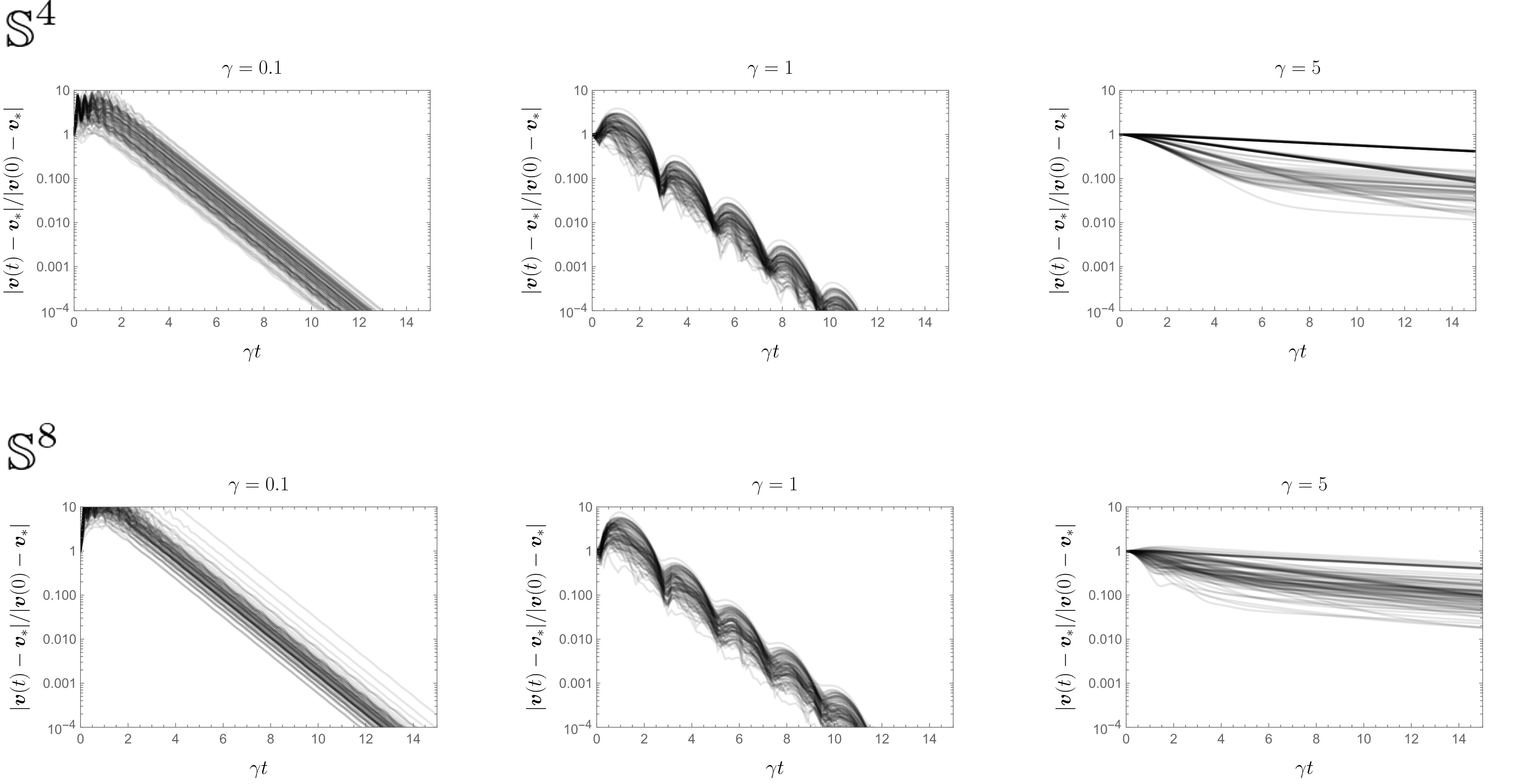}
  \caption{
  Behaviors of the solutions of the equation \eqref{eq:eom-v}.
  The upper panels show the results for $N=5$, and the lower panels show the results for $N=9$.
  }
  \label{fig:parameters}
\end{figure}

Fig.~\ref{fig:parameters} shows the numerical results of the solution of Eq.~\eqref{eq:eom-v}.
In both the upper and lower panels, the horizontal axis is $\gamma t$, and the vertical axis shows $|\braket{{\bm{v}}(t)}-{\bm{v}}_\ast|/|\braket{{\bm{v}}(0)}-{\bm{v}}_\ast|$.
From left to right, we change $\gamma$ as $\gamma=0.1,1,5$.
The upper panels results with $N=5$ and the lower panels $N=9$.
Each plot shows 100 black curves corresponding to different choices of the matrix $A$ and initial conditions.
In all cases, the parameter $\braket{{\bm{v}}(t)}$ converges to the optimal point ${\bm{v}}_\ast$ as time evolves, which means $|\braket{{\bm{v}}(t)}-{\bm{v}}_\ast|/|\braket{{\bm{v}}(0)}-{\bm{v}}_\ast|$ goes to 0.

\begin{figure}[t]
  \centering
  \includegraphics[width=15cm]{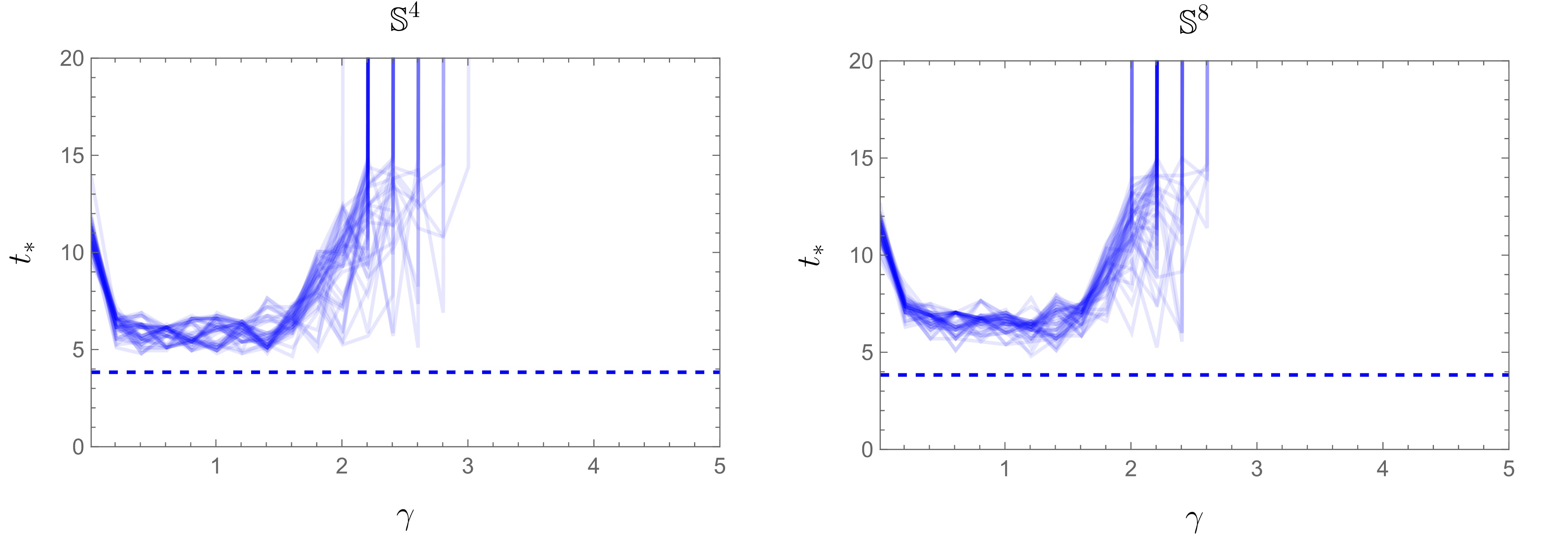}
  \caption{Relation between the convergence time $t_\ast$ and $\gamma$.
  The left panel is the results for $N=5$, and the right panel for $N=9$.
  The blue dotted curve indicates the lower bound in inequality \eqref{eq:t_star_sphere}.}
  \label{fig:t_star}
\end{figure}

Next, we estimate the relation between the convergence time $t_\ast$ and $\gamma$.
As a convergence criterion, we define $t_\ast$ as the time satisfying
$|\braket{{\bm{v}}(t)}-{\bm{v}}_\ast|/|\braket{{\bm{v}}(0)}-{\bm{v}}_\ast|\leq \epsilon_\ast=0.01$.
Figure~\ref{fig:t_star} shows the numerical results for the relation between $t_\ast$ and $\gamma$ for $N=5,9$.
Each plot shows 100 blue curves corresponding to different choices of the matrix $A$ and initial conditions.
The blue dotted curve indicates the lower bound in inequality \eqref{eq:t_star_bound} (specifically, the lower bound in Eq.~\eqref{eq:t_star_sphere}).
Thus, even though the differential equation \eqref{eq:eom-v} contains nonlinear terms such as $\Gamma^i_{jk}$ and $\mc{V}$, we find that the convergence time satisfies inequalities \eqref{eq:t_star_bound} and \eqref{eq:t_star_sphere}.

\section{Numerical simulation of Q(R)HD}
\label{app:simulation}

In this appendix, we summarize the numerical methodologies and specific discretization parameters employed to solve the Schr\"{o}dinger equation governed by the time-dependent Q(R)HD Hamiltonian operator. 
In the numerical simulations, we use spectral (sine-transform- and fast Fourier transform (FFT)-based) approaches, which are detailed below.

\subsection{Discretization parameters, boundary, and initial conditions}
\label{app:setup}

Before explaining the numerical schemes, we specify the simulation setup. The explicit discretization parameters and boundary conditions used across all numerical demonstrations are summarized in Tab.~\ref{tab:numerical-parameters}.
The accuracy checks are collected in App.~\ref{app:accuracy} and Tab.~\ref{tab:numerical-accuracy}.

\begin{table}[t]
    \centering
    \footnotesize
    \begin{tabular}{lccc}
        \hline\hline
        \textbf{Parameter} &
        \textbf{Flat: quadratic} & \textbf{Flat: Ackley} &
        \textbf{Curved (sphere)}
        \\ \hline
        Computational domain & \multicolumn{2}{c}{$x'^1, x'^2 \in [-1, 1]$ (principal axes of $A_1$)} & $v^1, v^2 \in [-1, 1]$ \\
        Grid points per axis $N$ & $100$ & $100$ & $100$ \\
        Grid spacing $\Delta x = 2/(N-1)$ & $0.0202$ & $0.0202$ & $0.0202$ \\
        Final time $T$ & $20$ & $80$ & $20$ \\
        Time step $\Delta t$ & $3.3 \times 10^{-5}$ & $2.7 \times 10^{-5}$ & $3.7 \times 10^{-5}$ / $1.1 \times 10^{-5}$ \\
        Number of steps $N_t$ & $6 \times 10^{5}$ & $3 \times 10^{6}$ & $5.4 \times 10^{5}$ / $1.9 \times 10^{6}$ \\
        Rule for $\Delta t$ & $C_1 \Delta t^2 \le 10^{-4}$ \eqref{eq:strang-bound} & $\Delta t/2$ check & $\lambda_{\max} \Delta t \le 1.7$ \eqref{eq:rk4-cfl} \\
        Initial state & \multicolumn{3}{c}{band-limited random field, Eq.~\eqref{eq:initial-state} (seed $0$, $k_c = 30$, $N_{\rm ref} = 100$)} \\
        Taper at the walls, Eq.~\eqref{eq:taper} & $\delta = 0.2$ & $\delta = 0.2$ & none \\
        Boundary & \multicolumn{2}{c}{exact Dirichlet walls (DST-I)} & absorbing outermost layer (FFT) \\
        \hline\hline
    \end{tabular}
    \caption{
        Summary of the numerical discretization parameters.
        In the flat case the computational domain is the square $[-1, 1]^2$ in the principal-axis frame of $A_1$, i.e.\ rotated by $45^\circ$ with respect to the original coordinates, and the wave function vanishes exactly on its boundary (App.~\ref{app:flat-solver}). On the sphere the outermost row and column of the periodic FFT grid are set to zero after every step (App.~\ref{app:sphere-solver}).
        The same $\Delta t$ is used for all $\gamma$ of a given potential in the flat case (App.~\ref{app:flat-solver}). For the sphere the two values refer to $\gamma \le 0.1$ and $\gamma = 0.3$, for which the stability condition is most restrictive at $t = 0$ and at $t = T$, respectively (App.~\ref{app:sphere-solver}).
        The grid-convergence checks use $N = 226$ ($\Delta x = 0.0089$) in the flat case and $N = 200$ ($\Delta x = 0.0101$) on the sphere, with the same $\Delta t$ and the same initial state, and the time-step checks use $\Delta t/2$ (App.~\ref{app:accuracy}).
    }
    \label{tab:numerical-parameters}
\end{table}

\paragraph{Potentials and basin probability.}

The quadratic potential is Eq.~\eqref{eq:potential-flatcase}.
The Ackley potential of Figs.~\ref{fig:flat-ackley-wavefunc} and \ref{fig:flat-ackley-costfunc} is the two-dimensional Ackley function in the sheared coordinates $\bm{y} = L A_1^{1/2} \bm{x}$,
\begin{align}
    V(\bm{x}) = \frac{1}{L_0} \biggl[ -20 \exp\Bigl( -0.2 \sqrt{\tfrac{1}{2} |\bm{y}|^2} \Bigr) - \exp\Bigl( \tfrac{1}{2} \bigl( \cos 2\pi y^1 + \cos 2\pi y^2 \bigr) \Bigr) + 20 + \ee \biggr]\,,
    \label{eq:potential-ackley}
\end{align}
with $L = 8$ and the normalization $L_0 = 65.536$, so that $\max V \approx 0.3$ on the computational domain. The global minimum is at $\bm{x} = 0$, and the local minima are separated by $1/L = 0.125$ in $\bm{y}/L$.
The basin probability of Fig.~\ref{fig:flat-ackley-costfunc} is the probability of finding the system in the unit cell of the local minima around $\bm{y} = 0$,
\begin{align}
    P_{\rm basin}(t) = \int_{|y^1|, |y^2| < 1/2} \dd^2 x\, \sqrt{g}\, |\psi(\bm{x}, t)|^2\,.
    \label{eq:def-pbasin}
\end{align}

\paragraph{Initial state.}
The initial wave function is a band-limited random field, to which the boundary window (flat case) or mask (sphere) described below is applied.
On a reference grid of $N_{\rm ref} \times N_{\rm ref}$ points ($N_{\rm ref} = 100$) we draw $\psi_{\rm ref}[i,j] = u_{ij} + \iu\, u'_{ij}$ with $u_{ij}, u'_{ij}$ independent and uniformly distributed in $[0,1)$ (NumPy \texttt{RandomState}, seed $0$), remove the Fourier components with $|\bm{k}| > k_c = 30$, and evaluate the resulting trigonometric polynomial on the computational grid,
\begin{align}
    \psi_0(\bm{x}) = \frac{1}{N_{\rm ref}^2} \sum_{|\bm{k}_{pq}| \le k_c} \tilde{\psi}_{\rm ref}[p,q]\, \ee^{\iu \bm{k}_{pq} \cdot (\bm{x} - \bm{x}_0)}\,,
    \qquad
    \bm{k}_{pq} = \frac{2\pi}{2 x_{\max}} (p, q)\,,
    \label{eq:initial-state}
\end{align}
where $\tilde{\psi}_{\rm ref}$ is the discrete Fourier transform of $\psi_{\rm ref}$, $\bm{x}_0 = (-1,-1)$ is the corner of the domain and $x_{\max} = 1$.
The cut-off $k_c = 30$ corresponds to a minimum wavelength $2\pi/k_c \approx 0.21$, i.e.\ about ten grid points, which sets the correlation length of the random field, so that $|\psi_0|^2$ is a random density that covers the whole domain (Figs.~\ref{fig:flat-quad-wavefunc}, \ref{fig:flat-ackley-wavefunc}, \ref{fig:curved-wavefunc}, and \ref{fig:curved-wavefunc-2}, $t=0$).
Because Eq.~\eqref{eq:initial-state} is a function rather than a grid vector, the same initial state is used for $N = 100$ and for the finer grids of the grid-convergence checks.
In the flat case the field is generated in the coordinates $(x'^1, x'^2)$ of the computational square introduced below and multiplied by a window that brings it smoothly to zero at the walls,
\begin{align}
    \psi_0(\bm{x}') \to \psi_0(\bm{x}')\, w(x'^1)\, w(x'^2)\,,
    \qquad
    w(u) =
    \begin{cases}
        1 & |u| \le 1 - \delta\,, \\[2pt]
        \sin^2\!\big[\pi (1 - |u|) / (2\delta)\big] & 1 - \delta < |u| < 1\,,
    \end{cases}
    \label{eq:taper}
\end{align}
with $\delta = 0.2$, so that the initial kinetic energy is independent of the grid.
On the sphere the outermost row and column of grid points are set to zero.
In all cases the state is normalized as
\begin{align*}
    \sum_{i,j} \sqrt{g(\bm{x}_{ij})}\, |\psi_0(\bm{x}_{ij})|^2\, \Delta x^2 = 1\,,
\end{align*}
and the same seed and cut-off are used in the flat and curved cases and for QHD and QRHD.

\paragraph{Boundary conditions and norm.}

\emph{Flat case.}
For the constant metric $g = A_1$ the kinetic operator $\hat{K}_0 = g^{ij} \hat{p}_i \hat{p}_j / (2m)$ contains the cross term $\hat{p}_1 \hat{p}_2$, so that Dirichlet walls parallel to the $x^1$ and $x^2$ axes cannot be imposed in a factorized basis.
We therefore work in the principal axes of $A_1$, $x'^1 = (x^1 + x^2)/\sqrt{2}$ and $x'^2 = (x^1 - x^2)/\sqrt{2}$, in which $g = \diag(\lambda_{\rm soft}, \lambda_{\rm stiff}) = \diag(0.1,\, 1.9)$ ($\lambda_{\rm soft, stiff} = 1 \mp 0.9$ are the eigenvalues of $A_1$) and $\hat{K}_0 = \hat{p}_{1'}^2 / (2m \lambda_{\rm soft}) + \hat{p}_{2'}^2 / (2m \lambda_{\rm stiff})$.
The computational domain is the square $x'^1, x'^2 \in [-1, 1]$, i.e.\ a square rotated by $45^\circ$ in the original coordinates (Figs.~\ref{fig:flat-quad-wavefunc} and \ref{fig:flat-ackley-wavefunc}), and the potential is the same function $V(x^1, x^2)$ evaluated at the rotated grid points. The same frame is used for QHD ($g_{ij} = \delta_{ij}$).
The wave function is represented on the $(N-2)^2$ interior grid points and expanded in the sine basis $\prod_i \sin[n_i \pi (x'^i + 1)/2]$, $n_i = 1, \ldots, N-2$, by the type-I discrete sine transform (DST-I), which satisfies $\psi = 0$ at the walls exactly. The kinetic propagator is diagonal in this basis (App.~\ref{app:flat-solver}).
The norm $\|\psi\|^2 = \sum \sqrt{g} |\psi|^2 \Delta x^2$ is conserved to round-off, $|1 - \|\psi\|^2| < 3 \times 10^{-10}$ over every flat run.

\emph{Sphere.}
The FFT-based solver of App.~\ref{app:sphere-solver} works on the periodic grid $v^i_j = -1 + j \Delta x$, $j = 0, \ldots, N-1$, $\Delta x = 2/(N-1)$.
The boundary condition is imposed as a mask: after every RK4 step and before the renormalization, the wave function is set to zero on the outermost row and column of the grid, so that $\psi = 0$ at $v^i = \pm 1$ after each step.
This one-cell layer acts as an absorbing wall for the components that reach the boundary, and the time step of Tab.~\ref{tab:numerical-parameters} is small enough that no component crosses the layer within one step.
The norm is recorded after this operation and before the subsequent renormalization to unity (``$\|\psi\|^2_{\rm pre}$''), so that $1 - \|\psi\|^2_{\rm pre}$ is the change of the norm in one step. It contains the part removed by the layer and the error of the RK4 step, which are not separated.
It is reported in App.~\ref{app:accuracy} (Fig.~\ref{fig:numerics-norm}).

\subsection{Flat parameter space solver}
\label{app:flat-solver}
Let us first consider the flat case where the metric tensor is constant.
In this case, we employ the pseudo-spectral method (or the split-step spectral method) introduced below, on the sine basis of App.~\ref{app:setup}.
The total Hamiltonian operator is split into the kinetic and potential operators, $\hat{H}(t) = \hat{K}(t) + \hat{V}(t)$. 
The time-evolution operator for a small time step $\Delta t$ is approximated via the second-order Suzuki-Trotter decomposition:
\begin{equation}
    \psi(\bm{x}, t + \Delta t) \approx \exp\left( - \iu \hat{V}(t) \frac{\Delta t}{2} \right) \exp\left( - \iu \hat{K}(t) \Delta t \right) \exp\left( - \iu \hat{V}(t) \frac{\Delta t}{2} \right) \psi(\bm{x}, t)\,.
\end{equation}
The potential operator $\exp(- \iu \hat{V}(t) \Delta t / 2)$ is diagonal in the coordinate representation and evaluated as a pointwise multiplication on the spatial grid $\bm{x}$. 
Conversely, the kinetic operator $\exp(-\iu \hat{K}(t) \Delta t)$ is diagonal in the sine basis of App.~\ref{app:setup}, which reproduces the Dirichlet walls exactly.
We transform the wave function between the coordinate representation and the sine basis using the forward and inverse 2D DST-I routines:
\begin{equation}
    \psi(\bm{x}', t+\Delta t) = \exp\left( - \iu V(t) \frac{\Delta t}{2} \right)
    \mathcal{S}^{-1} \left[ 
        \exp\left( - \iu \tilde{K}(\bm{n}, t) \Delta t \right) \mathcal{S}\left[
        \exp\left( -\iu V(t) \frac{\Delta t}{2} \right)
        \psi(\bm{x}', t)
        \right]
    \right],
\end{equation}
where $\mathcal{S}[\bullet]$ denotes the 2D DST-I on the interior grid points, $\mc{S}^{-1}[\bullet]$ is its inverse, and
$\tilde{K}(\bm{n}, t) = [\kappa_{n_1}^2/\lambda_{\rm soft} + \kappa_{n_2}^2/\lambda_{\rm stiff}] / [2m\, a(t)]$ with $\kappa_n = n \pi / 2$ is the kinetic operator in the sine basis ($n_i = 1, \ldots, N-2$). 
For QHD, it is found that $\lambda_{\rm soft} = \lambda_{\rm stiff} = 1$.

The time-dependent coefficients $a(t)$ and $1/a(t)$ are evaluated at the mid-point $t + \Delta t/2$ of each step, so that the scheme is second order also with respect to the explicit time dependence.
For $a(t) \gg 1$ the splitting error is dominated by the double commutator $[\hat{V}_{\rm eff}, [\hat{V}_{\rm eff}, \hat{K}]]$ of $\hat{V}_{\rm eff}(t) = a(t) \eta V(\hat{\bm{x}})$ and $\hat{K}(t) = \hat{K}_0/a(t)$, which is the multiplication operator $-a(t) \eta^2 |\nabla V|_g^2 / m$ with $|\nabla V|_g^2 = g^{ij}\, \del_i V\, \del_j V$. Replacing $|\nabla V|_g^2$ by its maximum over the domain and summing over the steps gives the estimate
\begin{align}
    \epsilon_{\rm Strang} \approx C_1 \Delta t^2\,,
    \qquad
    C_1 = \frac{\eta^2}{24} \frac{\max_{\bm{x}} |\nabla V|_g^2}{m} \int_0^T a(t)\, \dd t
        = \frac{\eta^2}{24} \frac{\max_{\bm{x}} |\nabla V|_g^2}{m} \frac{\ee^{2\gamma T} - 1}{2\gamma}\,.
    \label{eq:strang-bound}
\end{align}
We choose $\Delta t$ such that $C_1 \Delta t^2 \le 10^{-4}$ for the quadratic potential with $\gamma = 0.3$ and use the same $\Delta t$ for all $\gamma$ of this potential. 
For the Ackley potential, a smaller $\Delta t$ is used, which is shown in Tab.~\ref{tab:numerical-parameters}.
For larger $\gamma T$ the estimate \eqref{eq:strang-bound} grows as $\ee^{2\gamma T}$ and becomes uninformative, because it takes the maximum of $|\nabla V|_g^2$ over the whole domain, whereas the wave function is localized around the minima, where $\nabla V \approx 0$, by the time $a(t)$ becomes large.
The time-step accuracy of all runs is therefore established by the direct check of App.~\ref{app:accuracy}, in which the runs are repeated with $\Delta t/2$.

\subsection{Curved parameter space solver for spherical geometry}
\label{app:sphere-solver}

Let us next consider the curved case, focusing on the scenario where the parameter space is a sphere. 
Taking the conformally-flat coordinates ${\bm{v}}$ for the $N=3$ case considered in Sec.~\ref{sec:numeric-curved}, the kinetic term can be expressed as $\hat{K}=c(\bm{v})\hat{K}_0$, where $\hat{K}_0$ represents the kinetic term for the flat case. In this framework, the split-step pseudo-spectral method cannot be applied directly because the position-dependent geometric factor $c(\bm{v})$ does not commute with the Laplacian operator. Instead, we solve the Schr\"{o}dinger equation by combining FFT-based spectral differentiation with the explicit fourth-order Runge--Kutta (RK4) time-integration scheme. Let the time derivative of the wave function be defined by
\begin{equation}
    \frac{\partial \psi}{\partial t} = f(t, \psi) = -\iu \left[ -\frac{1}{2m a(t)} c(\bm{v}) \mathcal{F}^{-1}\left[ -|{\bm{k}}|^2 \mathcal{F}[\psi] \right] + \eta a(t) V(\bm{v}) \psi \right]\,.
\end{equation}
Here, the spatial derivatives are evaluated with high spectral accuracy by transforming the wave function into the momentum space via the FFT, multiplying by the momentum eigenvalues $-|{\bm{k}}|^2=-k^2_i$, and transforming back to the coordinate space. Once the spatial action of the full Hamiltonian $f(t, \psi)$ is obtained, the temporal evolution is advanced using the standard RK4 step:
\begin{align}
	\bm{r}_1 &= f(t, \psi_n)\,, \\
	\bm{r}_2 &= f\left(t + \frac{\Delta t}{2}, \psi_n + \frac{\Delta t}{2} \bm{r}_1\right)\,, \\
	\bm{r}_3 &= f\left(t + \frac{\Delta t}{2}, \psi_n + \frac{\Delta t}{2} \bm{r}_2\right)\,, \\
	\bm{r}_4 &= f(t + \Delta t, \psi_n + \Delta t \bm{r}_3)\,, \\
	\psi_{n+1} &= \psi_n + \frac{\Delta t}{6} (\bm{r}_1 + 2\bm{r}_2 + 2\bm{r}_3 + \bm{r}_4)\,.
\end{align}
The wave function is explicitly projectively re-normalized at the end of each macroscopic step using the invariant volume element weight $\sqrt{g(\bm{v})}$ to preserve the total probability.

To guarantee the numerical convergence and absolute stability of the RK4 integration scheme under the time-dependent Hamiltonian, the time step size $\Delta t$ is carefully determined. 
The procedure is as follows.
\begin{enumerate}[label=(\roman*)]
    \item We evaluate the flat Laplacian operator in the momentum basis, $\tilde{\Delta}_0 = -|\bm{k}|^2$ and write the maximum absolute value of its diagonal elements as $\tilde{\Lambda}_{0,\mr{max}}$.

    \item We obtain the matrix representation of the position-dependent geometric factor
    \begin{align}
        c(\bm{v}) = \left(\frac{|{\bm{v}}|^2 + R^2}{2R^2} \right)^2\,,
    \end{align}
    in the position basis, and define its maximum absolute value as $c_{\mr{max}}$.
    The upper bound of the kinetic operator magnitude is then computed as
    \begin{align}
        |K(t)|_{\mr{max}} = \frac{\exp(-2\gamma t)}{2m} c_{\mr{max}} \tilde{\Lambda}_{0,\mr{max}}\,.
    \end{align}

    \item We find the matrix representation of the potential function $V(\bm{v})$ in the position basis and determine its maximum absolute magnitude $|V|_{\mr{max}}$.
    The upper bound of the potential operator magnitude is then given by
    \begin{align}
        |\mc{V}(t)|_{\mr{max}} = \exp(2\gamma t) |V|_{\mr{max}}
    \end{align}

    \item
    We require
    \begin{align}
        \max_{0 \le t \le T} \lambda_{\max}(t)\, \Delta t \le 1.7\,,
        \label{eq:rk4-cfl}
    \end{align}
    where $\lambda_{\max}(t) \le |H(t)|_{\mr{max}} = |K(t)|_{\mr{max}} + |\mathcal{V}(t)|_{\mr{max}}$ is the largest eigenvalue of the discretized Hamiltonian, computed by power iteration. Since the discretized Hamiltonian is Hermitian, RK4 is stable for $\lambda_{\max} \Delta t < 2\sqrt{2} \approx 2.83$.
    $\lambda_{\max}(t)$ is largest either at $t = 0$, where the kinetic term dominates, or at $t = T$, where the potential term $\ee^{2\gamma T} \eta V$ dominates. This gives $\Delta t = 3.7 \times 10^{-5}$ for $\gamma \le 0.1$ and $\Delta t = 1.1 \times 10^{-5}$ for $\gamma = 0.3$ (Tab.~\ref{tab:numerical-parameters}).
\end{enumerate}

\subsection{Numerical accuracy and convergence checks}
\label{app:accuracy}

\begin{figure}[t]
  \centering
  \includegraphics[width=9cm]{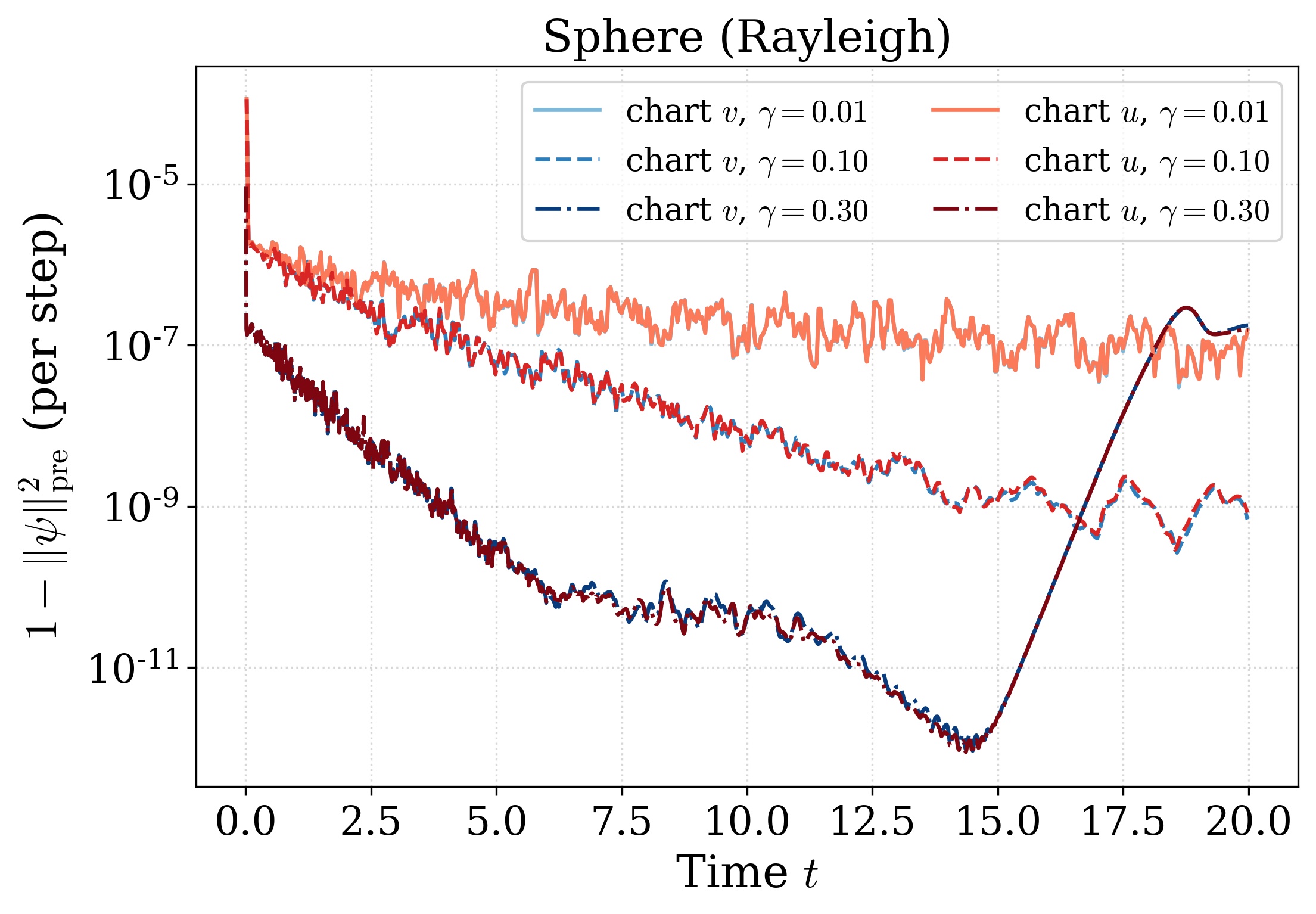} 
  \caption{
        Change of the norm per time step before the renormalization, $1 - \|\psi\|^2_{\rm pre}$, in the sphere solver (App.~\ref{app:setup}) in the charts $v$ and $u$. Block maxima over $1000$ steps are shown.
        In total, $4\%$ ($\gamma = 0.1$, $0.3$) and $11\%$ ($\gamma = 0.01$) of the initial norm are removed over the run.
        The flat-space runs use exact Dirichlet walls and conserve the norm to round-off (App.~\ref{app:setup}).
  }
  \label{fig:numerics-norm}
\end{figure}

\begin{figure}[t]
  \centering
  \includegraphics[width=16cm]{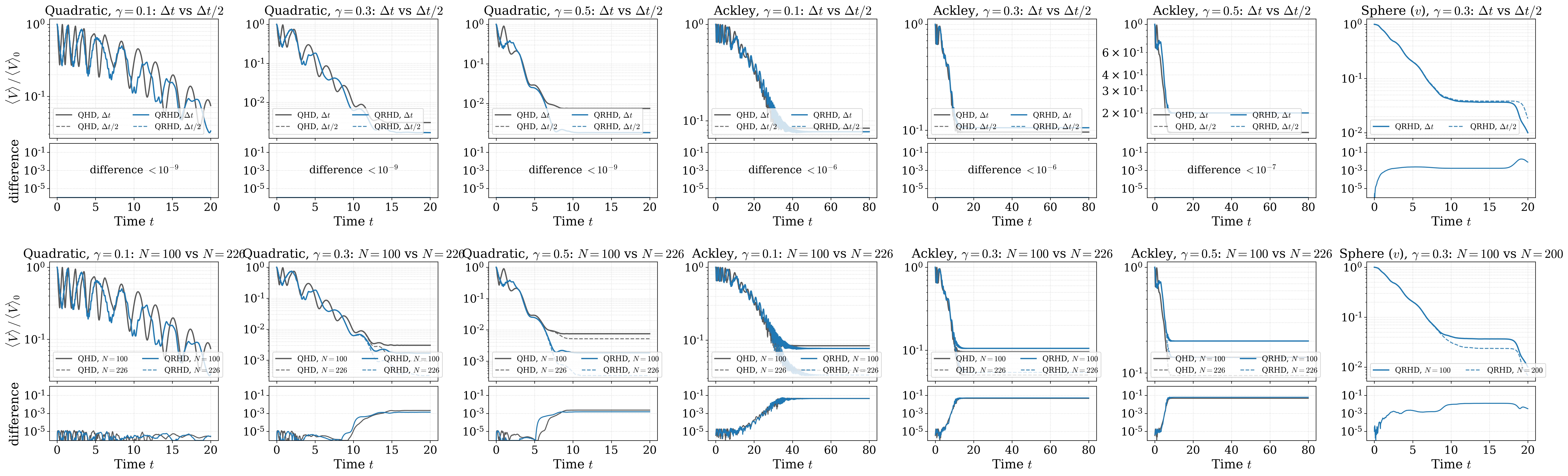}
  \caption{
   Convergence checks of the cost function $\braket{V(t)}/\braket{V(0)}$.
   The top row compares the production time step $\Delta t$ (solid) with $\Delta t/2$ (dashed), and the bottom row compares $N = 100$ (solid) with $N = 226$ (flat case) or $N = 200$ (sphere) (dashed), for the quadratic and Ackley potentials at the values of $\gamma$ shown in Figs.~\ref{fig:flat-quad-wavefunc}--\ref{fig:flat-ackley-costfunc} and for the sphere (chart $v$) with $\gamma = 0.3$.
   All runs start from the same initial state, Eq.~\eqref{eq:initial-state}, and the lower panels show the absolute difference on a common scale. Where the difference stays below the range of the axis, its upper bound is stated instead.
   In the flat case the two grids agree until the $N = 100$ curve levels off and then settle on different plateaus (see text).
  }
  \label{fig:numerics-convergence}
\end{figure}

\begin{table}[t]
    \centering
    \begin{tabular}{llccc}
        \hline\hline
        \textbf{Case} & \textbf{Method} & $\gamma$ & $\Delta t \to \Delta t/2$ & $N = 100 \to N'$ \\
        \hline
        Quadratic & QHD  & $0.1$ & $2.4 \times 10^{-10}$ & $8.4 \times 10^{-6}$ \\
        Quadratic & QRHD & $0.1$ & $1.0 \times 10^{-10}$ & $6.0 \times 10^{-6}$ \\
        Quadratic & QHD  & $0.3$ & $4 \times 10^{-12}$ & $2.1 \times 10^{-3}$ \\
        Quadratic & QRHD & $0.3$ & $3 \times 10^{-12}$ & $1.4 \times 10^{-3}$ \\
        Quadratic & QHD  & $0.5$ & $1.3 \times 10^{-10}$ & $2.3 \times 10^{-3}$ \\
        Quadratic & QRHD & $0.5$ & $6.4 \times 10^{-11}$ & $1.5 \times 10^{-3}$ \\
        Ackley    & QHD  & $0.1$ & $1.5 \times 10^{-7}$ & $4.7 \times 10^{-2}$ \\
        Ackley    & QRHD & $0.1$ & $2.1 \times 10^{-8}$ & $4.7 \times 10^{-2}$ \\
        Ackley    & QHD  & $0.3$ & $8.9 \times 10^{-8}$ & $4.6 \times 10^{-2}$ \\
        Ackley    & QRHD & $0.3$ & $1.0 \times 10^{-7}$ & $5.1 \times 10^{-2}$ \\
        Ackley    & QHD  & $0.5$ & $2.6 \times 10^{-8}$ & $4.6 \times 10^{-2}$ \\
        Ackley    & QRHD & $0.5$ & $7.9 \times 10^{-9}$ & $6.1 \times 10^{-2}$ \\
        Sphere ($v$) & QRHD & $0.3$ & $1.8 \times 10^{-2}$ & $1.3 \times 10^{-2}$ \\
        \hline\hline
    \end{tabular}
    \caption{Convergence checks of the runs of Fig.~\ref{fig:numerics-convergence}: $\max_{t > T/2} |\Delta \braket{V(t)}| / \braket{V(0)}$ between the production run and the run with the halved time step (column 4) or with the finer grid (column 5, with $N' = 226$ in the flat case and $200$ on the sphere).}
    \label{tab:numerical-accuracy}
\end{table}

We assess numerical accuracy by monitoring norm changes and examining the sensitivity of the results to temporal and spatial discretization. 
For the latter two checks, we repeat the flat-space simulations of Figs. 2–5 and the spherical simulation at $\gamma=0.3$ with a halved time step and a finer spatial grid. 
We increase the number of grid points per axis from $N=100$ to $226$ for the flat cases and to $200$ for the sphere. All comparisons use the same initial state defined in Eq.~\eqref{eq:initial-state}. 
The results are summarized in Fig.~\ref{fig:numerics-convergence} and Tab.~\ref{tab:numerical-accuracy}.

\paragraph{Norm.}
The flat runs use exact Dirichlet walls and conserve the norm to $3 \times 10^{-10}$ as mentioned in App.~\ref{app:setup}.
On the sphere the norm changes by about $10^{-7}$ per step before the renormalization while the wave function is extended over the domain, and by less as it contracts toward the minimum, except for a rise to $3 \times 10^{-7}$ per step at the end of the run with $\gamma = 0.3$ as in Fig.~\ref{fig:numerics-norm}.
In the sphere runs, $4$--$11\%$ of the initial norm is removed by the boundary layer over a whole run.

\paragraph{Time step.}
The time step is converged for the flat runs: halving $\Delta t$ changes $\braket{V(t)}$ by less than $2 \times 10^{-7}$ of $\braket{V(0)}$ (Fig.~\ref{fig:numerics-convergence}, top row).
On the sphere the change is $1.8 \times 10^{-2}$ of $\braket{V(0)}$, which is comparable to the late-time value of $\braket{V}$ itself ($\approx 10^{-2}$ of $\braket{V(0)}$).
The late-time value of $\braket{V}$ on the sphere is therefore uncertain by a factor of about two, whereas its decrease by two orders of magnitude from the initial value is robust.

\paragraph{Grid.}
The quantities used in Sec.~\ref{sec:numerical-simulation} are converged in the grid, whereas the late-time plateaus are not.
In the flat case, the cost-function expectation values computed with $N=100$ and $N=226$ agree to within $10^{-3}$ of $\braket{V(0)}$ as in Figs.~\ref{fig:numerics-convergence} before the coarser-grid result reach their late-time plateaus.
The widths also exhibit consistent transient behavior on the tow girds, as shown in Fig.~\ref{fig:flat-quad-costfunc}.
The plateau values, however, depend on the grid resolution.
The comparison of QHD and QRHD in Sec.~\ref{sec:numerical-simulation} is based on the evolution before the plateaus, i.e.\ on how fast $\braket{V}$ and the widths decrease, and not on the plateau values (their origin is not analyzed here).
For the quadratic potential with $\gamma = 0.1$ no plateau is reached within $T = 20$, and the two grids agree to $10^{-5}$ over the whole run.
For the Ackley potential the local wells are separated by six grid points at $N = 100$ and by $14$ at $N = 226$, and the plateau of $\braket{V}$ is about half as large at $N = 226$ (Tab.~\ref{tab:numerical-accuracy}, last column).
The basin probability $P_{\rm basin}(T)$, in contrast, is insensitive to the grid: refining $N$ from $100$ to $226$ changes it by at most $0.035$ (for $\gamma = 0.1$, $0.3$, $0.5$: from $0.948$, $0.923$, $0.772$ to $0.957$, $0.940$, $0.778$ for QHD and from $0.944$, $0.914$, $0.773$ to $0.960$, $0.920$, $0.808$ for QRHD).
The difference in $P_{\rm basin}(T)$ between QHD and QRHD does not exceed $0.03$ on either grid, which supports the statement of Sec.~\ref{sec:numeric-flat} that the two methods perform on a par.
On the sphere the difference between the two grids is $1.3 \times 10^{-2}$ of $\braket{V(0)}$, comparable to the late-time value of $\braket{V}$ (see above).

\newcommand{\arxivfont}{\rmfamily}
\bibliographystyle{yautphalpm}
\bibliography{ref-QRHD}

\newcommand{\etalchar}[1]{$^{#1}$}
\providecommand{\bysame}{\leavevmode\hbox to3em{\hrulefill}\thinspace}
\providecommand{\href}[2]{#2}
\providecommand{\arxivfont}{\ttfamily}
\begingroup\raggedright
\begin{thebibliography}{CLVBY24}

\bibitem[AGFM81]{Alvarez-Gaume:1981exa}
L.~Alvarez-Gaume, D.~Z. Freedman, and S.~Mukhi, {\itshape {The Background Field
  Method and the Ultraviolet Structure of the Supersymmetric Nonlinear Sigma
  Model}}, \href{http://dx.doi.org/10.1016/0003-4916(81)90006-3}{Annals Phys.
  {\bfseries 134} (1981) 85}.

\bibitem[Ama96]{1996-Amari}
S.-i. Amari, {\itshape Neural learning in structured parameter spaces - natural
  riemannian gradient}, in Advances in Neural Information Processing Systems,
  M.~Mozer, M.~Jordan, and T.~Petsche, eds., vol.~9.
\newblock MIT Press, 1996.
\newblock
  \url{https://proceedings.neurips.cc/paper\_files/paper/1996/file/39e4973ba3321b80f37d9b55f63ed8b8-Paper.pdf}.

\bibitem[AOBL19]{2019-Alimisis-et-al}
F.~Alimisis, A.~Orvieto, G.~B'ecigneul, and A.~Lucchi, {\itshape A
  continuous-time perspective for modeling acceleration in riemannian
  optimization}, in International Conference on Artificial Intelligence and
  Statistics.
\newblock 2019.
\newblock \href{http://arxiv.org/abs/1910.10782}{{\arxivfont 1910.10782
  [math.OC]}}.
\newblock \url{https://api.semanticscholar.org/CorpusID:204852166}.

\bibitem[ATS03]{2003-Aharonov-TaShma}
D.~Aharonov and A.~Ta-Shma,
  \href{http://dx.doi.org/10.1145/780542.780546}{{\itshape {Adiabatic quantum
  state generation and statistical zero knowledge}},} in {35th annual ACM
  symposium on Theory of computing}.
\newblock 1, 2003.
\newblock \href{http://arxiv.org/abs/quant-ph/0301023}{{\arxivfont
  quant-ph/0301023}}.

\bibitem[BACS07]{2007-Berry-et-al}
D.~W. Berry, G.~Ahokas, R.~Cleve, and B.~C. Sanders, {\itshape {Efficient
  Quantum Algorithms for Simulating Sparse Hamiltonians}},
  \href{http://dx.doi.org/10.1007/s00220-006-0150-x}{Commun. Math. Phys.
  {\bfseries 270} no.~2, (2007) 359--371}
  [\href{http://arxiv.org/abs/quant-ph/0508139}{{\arxivfont
  quant-ph/0508139}}].

\bibitem[BC12]{2009-Berry-Childs}
D.~W. Berry and A.~M. Childs, {\itshape {Black-box Hamiltonian simulation and
  unitary implementation}},
  \href{http://dx.doi.org/10.26421/QIC12.1-2-4}{Quant. Inf. Comput. {\bfseries
  12} no.~1-2, (2012) 0029--0062}
  [\href{http://arxiv.org/abs/0910.4157}{{\arxivfont 0910.4157 [quant-ph]}}].

\bibitem[BCV17]{Bastianelli:2017wsy}
F.~Bastianelli, O.~Corradini, and E.~Vassura, {\itshape {Quantum mechanical
  path integrals in curved spaces and the type-A trace anomaly}},
  \href{http://dx.doi.org/10.1007/JHEP04(2017)050}{JHEP {\bfseries 04} (2017)
  050} [\href{http://arxiv.org/abs/1702.04247}{{\arxivfont 1702.04247
  [hep-th]}}].

\bibitem[BRS{\etalchar{+}}16]{2015-Brown-Roberts-Susskind-Swingle-Zhao}
A.~R. Brown, D.~A. Roberts, L.~Susskind, B.~Swingle, and Y.~Zhao, {\itshape
  {Holographic Complexity Equals Bulk Action?}},
  \href{http://dx.doi.org/10.1103/PhysRevLett.116.191301}{Phys. Rev. Lett.
  {\bfseries 116} no.~19, (2016) 191301}
  [\href{http://arxiv.org/abs/1509.07876}{{\arxivfont 1509.07876 [hep-th]}}].

\bibitem[CHW{\etalchar{+}}25]{2025-Chakrabarti-et-al}
S.~Chakrabarti, D.~Herman, J.~Watkins, E.~Fontana, B.~Augustino, J.~L. Kim, and
  M.~Pistoia, {\itshape {On Speedups for Convex Optimization via Quantum
  Dynamics}}, \href{http://arxiv.org/abs/2503.24332}{{\arxivfont 2503.24332
  [quant-ph]}}.

\bibitem[CK11]{2019-Childs-Kothari}
A.~M. Childs and R.~Kothari, {\itshape {Simulating sparse Hamiltonians with
  star decompositions}},
  \href{http://dx.doi.org/10.1007/978-3-642-18073-6_8}{Lect. Notes Comput. Sci.
  {\bfseries 6519} (2011) 94--103}
  [\href{http://arxiv.org/abs/1003.3683}{{\arxivfont 1003.3683 [quant-ph]}}].

\bibitem[CLVBY24]{2022-Camps-Lin-VanBeeumen-Yang}
D.~Camps, L.~Lin, R.~Van~Beeumen, and C.~Yang, {\itshape {Explicit Quantum
  Circuits for Block Encodings of Certain Sparse Matrices}},
  \href{http://dx.doi.org/10.1137/22M1484298}{SIAM J. Matrix Anal. Appl.
  {\bfseries 45} no.~1, (2024) 801--827}
  [\href{http://arxiv.org/abs/2203.10236}{{\arxivfont 2203.10236 [quant-ph]}}].

\bibitem[CLW{\etalchar{+}}25]{2023-Chen-et-al}
Z.~Chen, Y.~Lu, H.~Wang, Y.~Liu, and T.~Li, {\itshape Quantum langevin dynamics
  for optimization}, Communications in Mathematical Physics {\bfseries 406}
  no.~3, (2025) 52.

\bibitem[CSW25]{2025-Catli-Simon-Wiebe}
A.~B. Catli, S.~Simon, and N.~Wiebe, {\itshape Exponentially better bounds for
  quantum optimization via dynamical simulation},
  \href{http://arxiv.org/abs/2502.04285}{{\arxivfont 2502.04285 [quant-ph]}}.

\bibitem[DeW64]{DeWitt:1964mxt}
B.~S. DeWitt, {\itshape {Dynamical theory of groups and fields}}, Conf. Proc. C
  {\bfseries 630701} (1964) 585--820.

\bibitem[DeW67]{DeWitt:1967ub}
\bysame, {\itshape {Quantum Theory of Gravity. 2. The Manifestly Covariant
  Theory}}, \href{http://dx.doi.org/10.1103/PhysRev.162.1195}{Phys. Rev.
  {\bfseries 162} (1967) 1195--1239}.

\bibitem[DT88]{1988-DOlivo-Torres}
J.~C. D'Olivo and M.~Torres, {\itshape The canonical formalism and path
  integrals in curved spaces},
  \href{http://dx.doi.org/10.1088/0305-4470/21/16/012}{Journal of Physics A:
  Mathematical and General {\bfseries 21} no.~16, (Aug, 1988) 3355}.

\bibitem[EAS98]{1998-Edelman-Arias-Smith}
A.~Edelman, T.~A. Arias, and S.~T. Smith, {\itshape The geometry of algorithms
  with orthogonality constraints}, SIAM journal on Matrix Analysis and
  Applications {\bfseries 20} no.~2, (1998) 303--353
  [\href{http://arxiv.org/abs/physics/9806030}{{\arxivfont physics/9806030}}].

\bibitem[Esc25]{2025-Escalante}
J.~A.~M. Escalante, {\itshape {Quantum Stochastic Gradient Descent in its
  continuous-time limit based on the Wigner formulation of Open Quantum
  Systems}}, \href{http://arxiv.org/abs/2510.25910}{{\arxivfont 2510.25910
  [quant-ph]}}.

\bibitem[FHS10]{Feynman:2010}
R.~P. Feynman, A.~R. Hibbs, and D.~F. Styer, {\itshape {Quantum Mechanics and
  Path Integrals: Emended Edition}}.
\newblock Dover Publications, 2010.

\bibitem[Gab82]{1982-Gabay}
D.~Gabay, {\itshape Minimizing a differentiable function over a differential
  manifold}, \href{http://dx.doi.org/10.1007/BF00934767}{Journal of
  Optimization Theory and Applications {\bfseries 37} (1982) 177--219}.
  \url{https://api.semanticscholar.org/CorpusID:120623235}.

\bibitem[GAW19]{2019-Gilyen-et-al-grad}
A.~Gily{\'e}n, S.~Arunachalam, and N.~Wiebe,
  \href{http://dx.doi.org/10.1137/1.9781611975482.87}{{\itshape Optimizing
  quantum optimization algorithms via faster quantum gradient computation},} in
  Proceedings of the Thirtieth Annual ACM-SIAM Symposium on Discrete
  Algorithms, pp.~1425--1444, SIAM.
\newblock 2019.
\newblock \href{http://arxiv.org/abs/1711.00465}{{\arxivfont 1711.00465
  [quant-ph]}}.

\bibitem[Gil75]{Gilkey:1975iq}
P.~B. Gilkey, {\itshape {The Spectral geometry of a Riemannian manifold}},
  \href{http://dx.doi.org/10.4310/jdg/1214433164}{J. Diff. Geom. {\bfseries 10}
  no.~4, (1975) 601--618}.

\bibitem[GJ76]{Gervais:1976ws}
J.-L. Gervais and A.~Jevicki, {\itshape {Point Canonical Transformations in
  Path Integral}}, \href{http://dx.doi.org/10.1016/0550-3213(76)90422-3}{Nucl.
  Phys. B {\bfseries 110} (1976) 93--112}.

\bibitem[GSLW18]{2019-Gilyen-Su-Low-Wiebe}
A.~Gily{\'e}n, Y.~Su, G.~H. Low, and N.~Wiebe,
  \href{http://dx.doi.org/10.1145/3313276.3316366}{{\itshape {Quantum singular
  value transformation and beyond: exponential improvements for quantum matrix
  arithmetics}},} in {51st Annual ACM SIGACT Symposium on Theory of Computing}.
\newblock 6, 2018.
\newblock \href{http://arxiv.org/abs/1806.01838}{{\arxivfont 1806.01838
  [quant-ph]}}.

\bibitem[Ita13]{2013-Itami}
T.~Itami, {\itshape
  エネルギー散逸を許す実時間量子ダイナミクスによる大域最適化},
  \href{http://dx.doi.org/10.1541/ieejeiss.133.985}{電気学会論文誌Ｃ（電子・情報・システム部門誌）
  {\bfseries 133} no.~5, (2013) 985--993}.

\bibitem[KAL{\etalchar{+}}25]{2025-Kharazi-et-al}
T.~Kharazi, A.~M. Alkadri, J.-P. Liu, K.~K. Mandadapu, and K.~B. Whaley,
  {\itshape Explicit block encodings of boundary value problems for many-body
  elliptic operators}, Quantum {\bfseries 9} (2025) 1764
  [\href{http://arxiv.org/abs/2407.18347}{{\arxivfont 2407.18347 [quant-ph]}}].

\bibitem[KBB15]{2015-Krichene-Bayen-Bartlett}
W.~Krichene, A.~Bayen, and P.~L. Bartlett, {\itshape Accelerated mirror descent
  in continuous and discrete time}, in Advances in Neural Information
  Processing Systems, C.~Cortes, N.~Lawrence, D.~Lee, M.~Sugiyama, and
  R.~Garnett, eds., vol.~28.
\newblock Curran Associates, Inc., 2015.

\bibitem[Kel65]{Keldysh:1964ud}
L.~V. Keldysh, {\itshape {Diagram Technique for Nonequilibrium Processes}},
  \href{http://dx.doi.org/10.1142/9789811279461_0007}{Sov. Phys. JETP
  {\bfseries 20} (1965) 1018--1026}.

\bibitem[KLP{\etalchar{+}}24]{2025-Kushnir-Leng-Peng-Fan-Wu}
S.~Kushnir, J.~Leng, Y.~Peng, L.~Fan, and X.~Wu, {\itshape {QHDOPT: A Software
  for Nonlinear Optimization with Quantum Hamiltonian Descent}},
  \href{http://arxiv.org/abs/2409.03121}{{\arxivfont 2409.03121 [quant-ph]}}.

\bibitem[KSB19]{2018-Dyson}
M.~Kieferov{\'a}, A.~Scherer, and D.~W. Berry, {\itshape {Simulating the
  dynamics of time-dependent Hamiltonians with a truncated Dyson series}},
  \href{http://dx.doi.org/10.1103/PhysRevA.99.042314}{Phys. Rev. A {\bfseries
  99} no.~4, (2019) 042314} [\href{http://arxiv.org/abs/1805.00582}{{\arxivfont
  1805.00582 [quant-ph]}}].

\bibitem[LC17]{2016-Low-Chuang-QSP}
G.~H. Low and I.~L. Chuang, {\itshape {Optimal Hamiltonian Simulation by
  Quantum Signal Processing}},
  \href{http://dx.doi.org/10.1103/PhysRevLett.118.010501}{Phys. Rev. Lett.
  {\bfseries 118} no.~1, (2017) 010501}
  [\href{http://arxiv.org/abs/1606.02685}{{\arxivfont 1606.02685 [quant-ph]}}].

\bibitem[LHLW23]{2023-Leng-Hickan-Li-Wu}
J.~Leng, E.~Hickman, J.~Li, and X.~Wu, {\itshape {Quantum Hamiltonian
  Descent}}, \href{http://arxiv.org/abs/2303.01471}{{\arxivfont 2303.01471
  [quant-ph]}}.

\bibitem[LLPW24]{2024-Leng-Li-Peng-Wu}
J.~Leng, J.~Li, Y.~Peng, and X.~Wu, {\itshape Expanding hardware-efficiently
  manipulable hilbert space via hamiltonian embedding}, arXiv preprint
  arXiv:2401.08550 (2024) .

\bibitem[LS25]{Leng:2025msd}
J.~Leng and B.~Shi, {\itshape {Quantum Optimization via Gradient-Based
  Hamiltonian Descent}}, in {42nd International Conference on Machine
  Learning}.
\newblock 5, 2025.
\newblock \href{http://arxiv.org/abs/2505.14670}{{\arxivfont 2505.14670
  [quant-ph]}}.

\bibitem[Lue72]{1972-Luenberger}
D.~G. Luenberger, {\itshape The gradient projection method along geodesics},
  Management Science {\bfseries 18} no.~11, (1972) 620--631.
  \url{http://www.jstor.org/stable/2629156}.

\bibitem[LW18]{2018-Dyson-IntPic}
G.~H. Low and N.~Wiebe, {\itshape {Hamiltonian Simulation in the Interaction
  Picture}}, \href{http://arxiv.org/abs/1805.00675}{{\arxivfont 1805.00675
  [quant-ph]}}.

\bibitem[LWWZ25]{2025-Leng-et-al}
J.~Leng, K.~Wu, X.~Wu, and Y.~Zheng, {\itshape {(Sub)Exponential Quantum
  Speedup for Optimization}},
  \href{http://arxiv.org/abs/2504.14841}{{\arxivfont 2504.14841 [quant-ph]}}.

\bibitem[LZW23]{2023-Leng-Zhang-Wu}
J.~Leng, Y.~Zheng, and X.~Wu, {\itshape {A quantum-classical performance
  separation in nonconvex optimization}},
  \href{http://arxiv.org/abs/2311.00811}{{\arxivfont 2311.00811 [quant-ph]}}.

\bibitem[Mac99]{MacKenzie:1999pu}
R.~MacKenzie, {\itshape {Path integral methods and applications}}, in {6th
  Vietnam International School on Physics}.
\newblock 12, 1999.
\newblock \href{http://arxiv.org/abs/quant-ph/0004090}{{\arxivfont
  quant-ph/0004090}}.

\bibitem[Miz75]{1975-Mizrahi}
M.~M. Mizrahi, {\itshape {The Weyl Correspondence and Path Integrals}},
  \href{http://dx.doi.org/10.1063/1.522468}{J. Math. Phys. {\bfseries 16}
  (1975) 2201--2206}.

\bibitem[Muk86]{Mukhi:1985vy}
S.~Mukhi, {\itshape {The Geometric Background Field Method, Renormalization and
  the Wess-Zumino Term in Nonlinear Sigma Models}},
  \href{http://dx.doi.org/10.1016/0550-3213(86)90502-X}{Nucl. Phys. B
  {\bfseries 264} (1986) 640--652}.

\bibitem[OS72]{1972-Omote-Sato}
M.~Omote and H.~Sato, {\itshape {Quantum mechanics of a nonlinear system}},
  \href{http://dx.doi.org/10.1143/PTP.47.1367}{Prog. Theor. Phys. {\bfseries
  47} (1972) 1367--1377}.

\bibitem[PCSZ25]{2025-Peng-et-al}
S.~Peng, S.~Chen, X.~Sun, and H.~Zhou, {\itshape {Stochastic Quantum
  Hamiltonian Descent}}, \href{http://arxiv.org/abs/2507.15424}{{\arxivfont
  2507.15424 [quant-ph]}}.

\bibitem[Pod28]{1928-Podolsky}
B.~Podolsky, {\itshape {Quantum-Mechanically Correct Form of Hamiltonian
  Function for Conservative Systems}},
  \href{http://dx.doi.org/10.1103/PhysRev.32.812}{Phys. Rev. {\bfseries 32}
  (1928) 812--816}.

\bibitem[PS95]{Peskin:1995ev}
M.~E. Peskin and D.~V. Schroeder,
  \href{http://dx.doi.org/10.1201/9780429503559}{{\itshape {An Introduction to
  quantum field theory}}}.
\newblock Addison-Wesley, Reading, USA, 1995.

\bibitem[Ros60]{1960-Rosen}
J.~B. Rosen, {\itshape The gradient projection method for nonlinear
  programming. part i. linear constraints},
  \href{http://dx.doi.org/10.1137/0108011}{Journal of the Society for
  Industrial and Applied Mathematics {\bfseries 8} no.~1, (1960) 181--217}.

\bibitem[Ros61]{1961-Rosen}
\bysame, {\itshape The gradient projection method for nonlinear programming.
  part ii. nonlinear constraints},
  \href{http://dx.doi.org/10.1137/0109044}{Journal of the Society for
  Industrial and Applied Mathematics {\bfseries 9} no.~4, (1961) 514--532}.

\bibitem[Ros97]{1997-Rosenberg}
S.~Rosenberg, {\itshape The Laplacian on a Riemannian Manifold: An Introduction
  to Analysis on Manifolds}.
\newblock London Mathematical Society Student Texts. Cambridge University
  Press, 1997.

\bibitem[Sak85]{Sakita:1985exh}
B.~Sakita, {\itshape {Quantum theory of many variable systems and fields}},
  vol.~1.
\newblock 1985.

\bibitem[Sat77]{Sato:1976hy}
M.-a. Sato, {\itshape {Operator Ordering and Perturbation Expansion in the Path
  Integration Formalism}}, \href{http://dx.doi.org/10.1143/PTP.58.1262}{Prog.
  Theor. Phys. {\bfseries 58} (1977) 1262}.

\bibitem[SBC15]{2015-Su-Boyd-Candes}
W.~Su, S.~Boyd, and E.~J. Candes, {\itshape A differential equation for
  modeling nesterov's accelerated gradient method: Theory and insights},
  \href{http://arxiv.org/abs/1503.01243}{{\arxivfont 1503.01243 [stat.ML]}}.

\bibitem[Sch61]{Schwinger:1960qe}
J.~S. Schwinger, {\itshape {Brownian motion of a quantum oscillator}},
  \href{http://dx.doi.org/10.1063/1.1703727}{J. Math. Phys. {\bfseries 2}
  (1961) 407--432}.

\bibitem[SGS{\etalchar{+}}25]{2025-Suzuki-et-al}
Y.~Suzuki, M.~Gluza, J.~Son, B.~H. Tiang, N.~H.~Y. Ng, and Z.~Holmes, {\itshape
  {Grover's algorithm is an approximation of imaginary-time evolution}},
  \href{http://arxiv.org/abs/2507.15065}{{\arxivfont 2507.15065 [quant-ph]}}.

\bibitem[SIKC20]{2020-Stokes-et-al}
J.~Stokes, J.~Izaac, N.~Killoran, and G.~Carleo, {\itshape Quantum natural
  gradient}, \href{http://dx.doi.org/10.22331/q-2020-05-25-269}{Quantum
  {\bfseries 4} (2020) 269} [\href{http://arxiv.org/abs/1909.02108}{{\arxivfont
  1909.02108 [quant-ph]}}].

\bibitem[SKY{\etalchar{+}}25]{2025-Sato-et-al}
Y.~Sato, J.~Kato, H.~Yano, K.~Ito, and N.~Yamamoto, {\itshape {Explicit
  block-encoding for partial differential equation-constrained optimization}},
  \href{http://arxiv.org/abs/2511.14420}{{\arxivfont 2511.14420 [quant-ph]}}.

\bibitem[Smi14]{1994-Smith}
S.~T. Smith, {\itshape Optimization techniques on riemannian manifolds},
  \href{http://arxiv.org/abs/1407.5965}{{\arxivfont 1407.5965 [math.OC]}}.

\bibitem[SS25]{2025-Sturm-Schillo}
A.~Sturm and N.~Schillo, {\itshape {Efficient and Explicit Block Encoding of
  Finite Difference Discretizations of the Laplacian}},
  \href{http://arxiv.org/abs/2509.02429}{{\arxivfont 2509.02429 [quant-ph]}}.

\bibitem[SSZ21]{2021-Sanz-et-al}
J.~M. Sanz-Serna and K.~C. Zygalakis, {\itshape The connections between
  lyapunov functions for some optimization algorithms and differential
  equations}, \href{http://dx.doi.org/10.1137/20M1364138}{SIAM Journal on
  Numerical Analysis {\bfseries 59} no.~3, (2021) 1542--1565}
  [\href{http://arxiv.org/abs/2009.00673}{{\arxivfont 2009.00673 [math.NA]}}].

\bibitem[WRJ16]{2016-Wilson-Recht-Jordan}
A.~C. Wilson, B.~Recht, and M.~I. Jordan, {\itshape A lyapunov analysis of
  momentum methods in optimization},
  \href{http://arxiv.org/abs/1611.02635}{{\arxivfont 1611.02635 [math.OC]}}.

\bibitem[WRJ21]{2021-Wilson-Recht-Jordan}
A.~C. Wilson, B.~Recht, and M.~I. Jordan, {\itshape A lyapunov analysis of
  accelerated methods in optimization}, Journal of Machine Learning Research
  {\bfseries 22} no.~113, (2021) 1--34.
  \url{http://jmlr.org/papers/v22/20-195.html}.

\bibitem[WW15]{2015-Wibisono-Wilson}
A.~Wibisono and A.~C. Wilson, {\itshape On accelerated methods in
  optimization}, \href{http://arxiv.org/abs/1509.03616}{{\arxivfont 1509.03616
  [math.OC]}}.

\bibitem[WWJ16]{2016-Wibisono-Wilson-Jordan}
A.~Wibisono, A.~C. Wilson, and M.~I. Jordan, {\itshape A variational
  perspective on accelerated methods in optimization},
  \href{http://dx.doi.org/10.1073/pnas.1614734113}{Proceedings of the National
  Academy of Sciences {\bfseries 113} no.~47, (2016) E7351--E7358}
  [\href{http://arxiv.org/abs/1603.04245}{{\arxivfont 1603.04245 [math.OC]}}].

\bibitem[Yam19]{2019-Yamamoto}
N.~Yamamoto, {\itshape On the natural gradient for variational quantum
  eigensolver}, \href{http://arxiv.org/abs/1909.05074}{{\arxivfont 1909.05074
  [quant-ph]}}.

\bibitem[ZS16]{2016-Zhang-Suvrit}
H.~Zhang and S.~Sra, {\itshape First-order methods for geodesically convex
  optimization}, in Conference on learning theory, pp.~1617--1638, PMLR.
\newblock 2016.
\newblock \href{http://arxiv.org/abs/1602.06053}{{\arxivfont 1602.06053
  [math.OC]}}.

\bibitem[ZS18]{2018-Zhang-Suvrit}
\bysame, {\itshape Towards riemannian accelerated gradient methods},
  \href{http://arxiv.org/abs/1806.02812}{{\arxivfont 1806.02812 [math.OC]}}.

\end{thebibliography}
\endgroup

\end{document}